\documentclass[aps,prx,longbibliography,twocolumn,superscriptaddress,floatfix]{revtex4-2}
\usepackage{amsmath,amssymb,bm,graphicx,color,gensymb,bbold,hyperref,keyval,url,latexsym}
\usepackage[usenames,dvipsnames]{xcolor}
\usepackage[normalem]{ulem} 
\usepackage{CJK}
\usepackage{float}

\usepackage{multirow}
\usepackage{hyperref}
\hypersetup{
    colorlinks=true,    
    linkcolor=NavyBlue, 
    citecolor=NavyBlue,   
    filecolor=NavyBlue, 
    urlcolor=NavyBlue   
}

\begin{document}

\title{The Saga of $\alpha$-RuCl$_3$: Parameters, Models, and Phase Diagrams}

\begin{CJK*}{UTF8}{}
\author{Marius M\"{o}ller}
\affiliation{Institut f\"{u}r Theoretische Physik, Goethe-Universit\"{a}t
  Frankfurt,  Max-von-Laue Strasse 1, 60438 Frankfurt, Germany}
\author{P. A. Maksimov}
\affiliation{Bogolyubov Laboratory of Theoretical Physics, Joint Institute for Nuclear Research, 
Dubna, Moscow region 141980, Russia}
\author{Shengtao Jiang (\CJKfamily{gbsn}蒋晟韬)}
\affiliation{Stanford Institute for Materials and Energy Sciences, SLAC National Accelerator Laboratory and Stanford University, Menlo Park, California 94025, USA}
\author{Steven R. White}
\affiliation{Department of Physics and Astronomy, University of California, Irvine, California 92697, USA}
\author{Roser Valent\'{i}}
\affiliation{Institut f\"{u}r Theoretische Physik, Goethe-Universit\"{a}t
  Frankfurt,  Max-von-Laue Strasse 1, 60438 Frankfurt, Germany}
\author{A. L. Chernyshev}
\affiliation{Department of Physics and Astronomy, University of California, Irvine, California 92697, USA}
\date{\today} 
\begin{abstract}
RuCl$_3$ was likely the first ever deliberately synthesized ruthenium compound,  following  the discovery of the $_{44}$Ru element in 1844. For a long time it was known as an oxidation catalyst, with its physical properties being discrepant and confusing, until a decade ago when its  allotropic form $\alpha$-RuCl$_3$ rose to  exceptional prominence.  This ``re-discovery'' of $\alpha$-RuCl$_3$ has not only reshaped the hunt for a material manifestation of the Kitaev spin liquid, but it has  opened the floodgates  of  theoretical and experimental research in the  many unusual phases and excitations that the anisotropic-exchange magnets as a class of compounds have to offer. Given its importance for the field of Kitaev materials, it is astonishing that the low-energy  spin model that describes this compound and its possible proximity to the much-desired spin-liquid state is still a subject of significant debate ten years later. In the present study, we argue that the existing key phenomenological observations put strong natural constraints on the effective microscopic spin model of $\alpha$-RuCl$_3$, and specifically on its spin-orbit-induced anisotropic-exchange parameters that are responsible for the non-trivial physical properties of this material. These constraints allow one to focus on the relevant region of the multi-dimensional phase diagram of the $\alpha$-RuCl$_3$ model, suggest an intuitive description of it via a different parametrization of the exchange matrix, offer a unifying view on the earlier assessments of its parameters, and bring closer together several approaches to the derivation of anisotropic-exchange models. We explore extended phase diagrams relevant to the  $\alpha$-RuCl$_3$ parameter space using quasi-classical, Luttinger-Tisza, exact diagonalization, and density-matrix renormalization group methods, demonstrating a remarkably close quantitative accord between them on the general structure and hierarchy of the phases, with the zigzag, ferromagnetic, and incommensurate phases that are proximate to each other. One of the  highlights is the detailed agreement on the nature of the incommensurate phases that realize two  distinct counter-rotating helical states.
\end{abstract}
\maketitle
\end{CJK*}
\section{Introduction}
\label{Sec_intro}

As every story needs a hero~\cite{hero}, every condensed matter  field needs a champion material, be it silicon for semiconductors~\cite{Si}, yttrium-iron garnet for spintronics~\cite{YIGsaga}, or graphene for 2D materials~\cite{grapheneRMP}.  With its 50-year-long quest for a spin-liquid compound, the field of quantum magnetism is, arguably, still on the search for one~\cite{savary_balents17,SL_RMP}. 

For some time, the mineral herbertsmithite, a copper-zinc hydroxychloride with its five-element composition and  elaborate crystal structure, seemed to be destined for that 
title~\cite{Norman_RMP}. Made of kagom\'{e} layers of  corner-sharing triangles of  nearly isotropically coupled spin-$\frac12$  Cu$^{2+}$ ions---a motif that provides  maximum frustration of magnetic orders and promotes spin-singlets---it was fitting all theoretical stereotypes for the leading spin-liquid paradigm, 
suggesting a true realization of the singlet-soup state of resonating valence bonds~\cite{PWA73}.  That is, until the issues of site disorder mimicking  spin-liquid features rendered this hope fruitless, although efforts continue to find its cleaner incarnation~\cite{Norman_RMP,Zorko17}.  

But then something happened that the archetypal construct of the isotropic spin models on triangular-motif lattices did not anticipate. The fallen champion's title 
was picked up by one of the most unlikely materials imaginable. Also a chloride, but a mere binary compound with a geometrically unfrustrated bipartite arrangement of magnetic ions in  stacked honeycomb-lattice sheets. When Karl Ernst Claus, a 19th century chemist at Kazan University, synthesized ruthenium trichloride in 1845~\cite{Claus}, little did he know that nearly two hundred years later physicists and material scientists alike would pin their hopes on his humble creation as being a close realization of an exotic state of matter---a Kitaev spin liquid---that could shape the fortunes of not just quantum magnetism, but the broader fields of quantum materials and topological quantum computing~\cite{KITAEV2006}.

Historically, $\alpha$-RuCl$_3$ has always been puzzling~\cite{Hyde65}. With its layered honeycomb-lattice composition of  edge-sharing RuCl$_6$ octahedra~\cite{Fletcher67}, the actual space group symmetry,   amount of the octahedral distortion, and  propensity toward interlayer stacking faults  have all been the subject of debate~\cite{WinterReview}. Its electronic characterization has evolved from being a small-gap magnetic semiconductor to a narrow-band Mott-Hubbard insulator that combines spin-orbit coupling and electronic correlations~\cite{Binotto71,Rojas83,Pollini96,Plumb14}.  By analogy with  other halides of transition metals~\cite{McGuire2017}, it was initially suggested to have ferromagnetic Ru-planes ordered antiferromagnetically~\cite{Fletcher67,Kobayashi92}, but this was proven incorrect later~\cite{Plumb15,Coldea15}. 

This was the state of understanding of $\alpha$-RuCl$_3$ about ten years ago, when the search for the  Kitaev spin liquid in the initially more promising $5d$ iridium-oxide compounds was waning~\cite{WinterReview}. The focus of that search has moved to the structurally similar materials of the $4d$ elements with spin-orbit coupling, in which $\alpha$-RuCl$_3$ stood out because of its supposed nearly ideal edge-sharing octahedral structure in the honeycomb-lattice planes and uncertain electronic and magnetic properties~\cite{Plumb14}.

$\alpha$-RuCl$_3$'s ascent to fame was not without  theoretical input. The celebrated spin-liquid solution of the compass-like~\cite{KK_82} honeycomb-lattice model by Kitaev~\cite{KITAEV2006}, featuring topological fermionic excitations, was initially perceived as a largely theoretical construct. It was propelled to the domain of  physical reality by the seminal microscopic insight of Jackeli and Khaliullin~\cite{Jackeli,chaloupka2010}, which laid out that Mott insulators of  transition metal ions in the edge-sharing octahedral environment of ligands can lead to that very model in the limit of strong spin-orbit coupling. The anticipated nearly perfect edge-sharing RuCl$_6$ octahedra, weak magnetic ordering, subsequent convincing evidence of  continuum-like excitations  \cite{Plumb15,Cao16,Sandilands15,nagler16,Zvyagin17,Banerjee17,banerjee18}, and observations of a significant and potentially anomalous thermal Hall effect~\cite{Matsuda18,Matsuda18a,Matsuda18b,Ong21,Balents18,Rosch18,Takagi22}  have all fueled hopes for  $\alpha$-RuCl$_3$'s singular place in history as the key material realizing this model.

While inspiring an enormous amount of theoretical and experimental research~\cite{Takagi2019,kaibValenti19,Knolle18,Motome20,Trebst22,Kee24,Kee_matsuda2025}, the initial expectations of   becoming a Rosetta stone for the Kitaev spin liquid have likely not played out for $\alpha$-RuCl$_3$. The longer-range exchanges and other significant interactions in its low-energy spin model have arguably moved it away from the pure-Kitaev limit and its phenomenology~\cite{WinterReview}.  Various spectroscopic signatures of the broad continuum, combined with the sharper excitations at  lower energies, have received a more natural explanation as a combination of strongly coupled magnons and their multi-magnon states~\cite{rethinking,Balz19,Smit20}, while substantial contribution of phonons has not left much room for exotic explanations of the field-induced thermal transport  in $\alpha$-RuCl$_3$~\cite{Taillefer21,Taillefer23,Taillefer23a,Okamoto22,Takagi22a,Mangeolle22,Nevidomskyy23,dhakal2024}.  

Nevertheless,  the research effort devoted to this material has already laid a prominent keystone in revealing exceptional richness of phenomena offered by anisotropic-exchange magnets on the honeycomb and other lattices. Thus, it is rather remarkable that the agreement on the low-energy model describing this compound has not yet been reached,  its place in the phase diagram of that model has not yet been settled,  and relevant proximate phases have not yet been uniquely identified. 

In the following few pages, we provide a brief digest describing the origin of the problem, suggesting a  path to its resolution, and giving an overview of such an approach, with the detailed expos\'{e} of the results and their  cross-examination presented in the main text.

\vspace{-0.3cm}
\subsection{Overall Summary}
\label{Sec_intro_summary}
\vskip -0.2cm

While the problem of finding a definitive set of parameters for an effective model is common to many materials, it has been particularly difficult for $\alpha$-RuCl$_3$ because of its fluctuating ground and field-induced states and complicated interactions in its low-energy description. The main result of this work is a systematic path to a narrower parameter space for the $\alpha$-RuCl$_3$ effective model.

The key message is the approach itself. It consists of breaking the problem into stages. First is the task of identifying experimental observables that would not occur without the anisotropic-exchange parameters of the model, which originate from spin-orbit interaction. Such observables can be shown to work as effective phenomenological constraints for the most challenging parameters of the model. Then, any set of remaining parameters that are not fixed at this stage from such a partially restricted manifold is expected to meet the imposed constraints. The remaining, more traditional isotropic exchanges can be either fixed with additional constraints, or left for further adjustment. We demonstrate the success of such a staged approach in systematically narrowing the allowed parameter space of $\alpha$-RuCl$_3$.

Another important general message of our study is the deep insight provided by the use of an alternative crystallographic parameterization of the effective model of $\alpha$-RuCl$_3$. It offers a unifying view of most prior efforts to determine its underlying model and also gives a fresh perspective on it as a member of a coherent group of other materials and models.

The numerical approaches, such as exact diagonalization (ED) and density-matrix renormalization group (DMRG), allow us to verify that this entire approach, which involves quasiclassical approximations, actually holds up quantitatively. The other main utility of numerical explorations, specifically using DMRG, is the study of incommensurate phases. These phases are consistently found to be proximate to the relevant parameter space of $\alpha$-RuCl$_3$, but very little research has been done on their nature. The present study closes this gap by demonstrating detailed quantitative agreement on their structure, which corresponds to counter-rotating helical states.

\vspace{-0.3cm}
\subsection{Parameters' drama}
\label{Sec_intro_param}
\vskip -0.2cm

\begin{figure}[t]
\centering
\includegraphics[width=\linewidth]{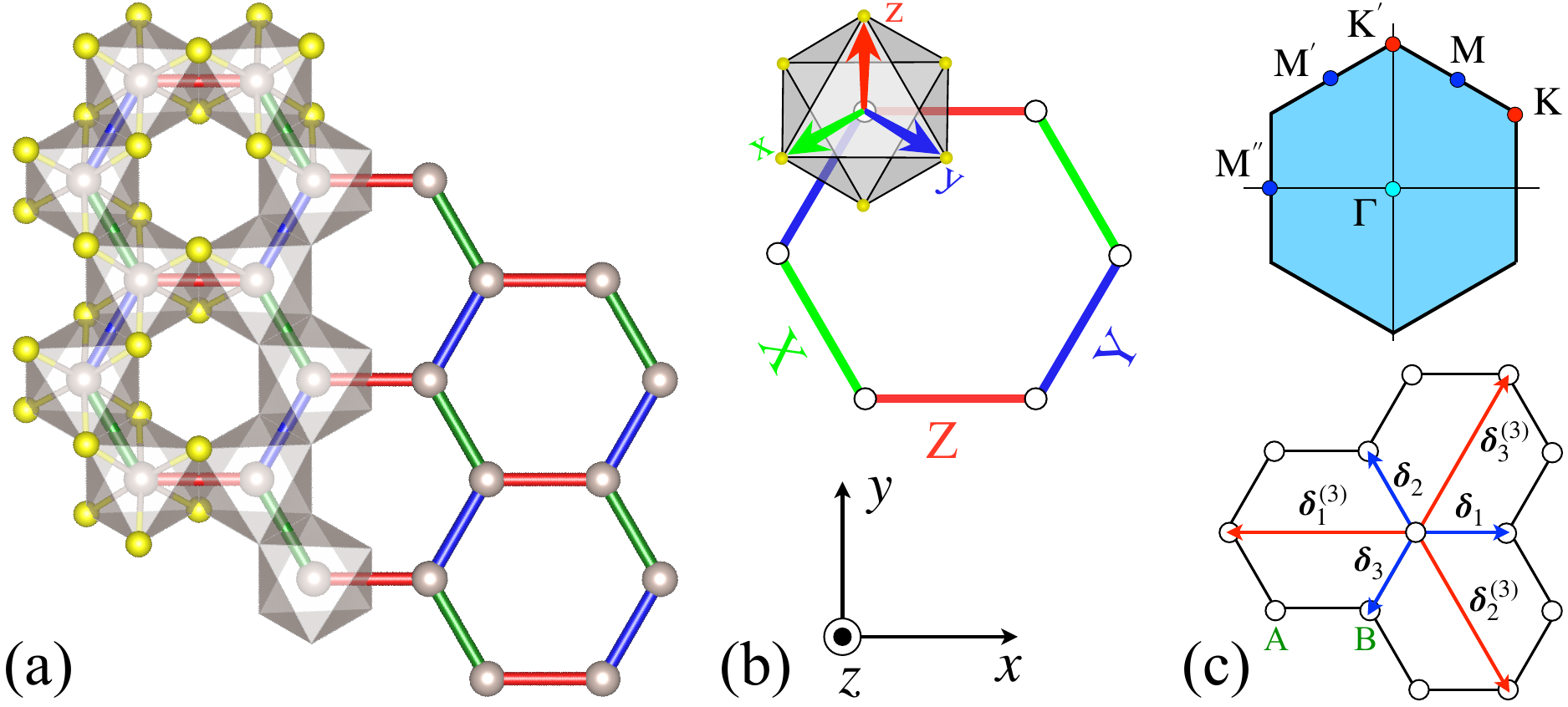}
\vskip -0.2cm
\caption{(a) Ru$^{3+}$ ions in the octahedral cages of Cl$^{-}$ and the nearest-neighbor bonds. (b) $\{{\rm X,Y,Z}\}$ bonds with the cubic $\{\rm x,y,z\}$ and crystallographic $\{x,y,z\}$ axes. (c) A and B sublattices, nearest-  and third-nearest-neighbor primitive vectors, $\bm{\delta}_\alpha$ and $\bm{\delta}^{(3)}_\alpha$,  and the Brillouin zone (BZ) of the honeycomb lattice with the  high-symmetry points.}
\label{fig_axes}
\vskip -0.4cm
\end{figure}

Generally,  the anisotropic-exchange Hamiltonians lack spin-rotational symmetries. For  the effective low-energy spin-orbit-coupled  spin degrees of freedom, it is the discrete symmetry of  the honeycomb lattice of magnetic ions with edge-sharing octahedral ligand environment  that allows {\it four}  terms in the nearest-neighbor exchange matrix
~\cite{WinterReview}. In a  parametrization within the reference frame of the cubic axes of the idealized ligand bonds, see Fig.~\ref{fig_axes}(a) and \ref{fig_axes}(b), these four are  Kitaev, Heisenberg, and  off-diagonal  $\Gamma$ and $\Gamma'$ exchanges~\cite{RauG,RauGp}, with  three anisotropic terms stemming from the spin-orbit coupling within the electronic states of  Ru$^{3+}$ ions. 

Arguably, deviations from the ideal Kitaev-only model are unavoidable in any realistic material~\cite{rethinking}, with all  these four  terms  present in $\alpha$-RuCl$_3$ for any reasonable choice of electronic parameters~\cite{WinterReview}. An important role is also played by the third-neighbor, more isotropic  $J_3$ exchange~\cite{WinterReview}, as the superexchange path for the second-nearest neighbor is more convoluted for a wide class of the honeycomb-lattice compounds~\cite{WinterReview,Winter_Co_2022,LP90}. This yields the minimal  spin $S_{\rm eff}\!=\!\frac12$  model of  the magnetic two-dimensional (2D)   honeycomb-lattice planes of $\alpha$-RuCl$_3$ 
\noindent
\begin{align}
\hat{\cal H}=\hat{\cal H}_1+\hat{\cal H}_3
=\sum_{\langle ij\rangle} \mathbf{S}^{\rm T}_i \hat{\rm \bf J}_{ij} \mathbf{S}_j
+J_3\sum_{\langle ij\rangle_3} \mathbf{S}_i \cdot \mathbf{S}_j \, ,
\label{eq_Hij}
\end{align}
\noindent
where ${\bf S}_i^{\rm T}\!=\!\left(S_i^{x},S_i^{y},S_i^{z}\right)$ and the nearest-neighbor part of the model 
in  the cubic-axes parametrization is
\noindent
\begin{align}
\mathcal{H}_1=&\sum_{\langle ij \rangle_\gamma} \Big[J {\bf S}_i \cdot {\bf S}_j 
+K S^\gamma_i S^\gamma_j +\Gamma \big(S^\alpha_i S^\beta_j +S^\beta_i S^\alpha_j\big)\nonumber \\
\label{H_JKGGp}
&\ \ \ \ \ +\Gamma' \big(S^\gamma_i S^\alpha_j+
S^\gamma_i S^\beta_j+S^\alpha_i S^\gamma_j +S^\beta_i S^\gamma_j\big)\Big],
\end{align}
\noindent
where the spin indices $\{\alpha,\beta,\gamma\}$ follow the  $\{{\rm X,Y,Z}\}$ bonds according to Fig.~\ref{fig_axes}(b): $\{{\rm y,z,x}\}$ for the X, $\{{\rm z,x,y}\}$ for the Y, and $\{{\rm x,y,z}\}$ for the Z bond, respectively.

There is a broad consensus that this generalized Kitaev-Heisenberg (KH), or {\it effective} $KJ\Gamma\Gamma'$--$J_3$ model  is the {\it minimal} microscopic model of $\alpha$-RuCl$_3$, which absorbs further-neighbor exchanges and their anisotropies into the fewer  parameters of the nearest-neighbor matrix without affecting the symmetry of the model. However, even the parameters of this minimal model have received  very wide ranges of estimates, see Ref.~\cite{rethinking} for the earlier overview and Sec.~\ref{Sec_constraints} below.

Because of the initial optimism regarding the near-proximity of $\alpha$-RuCl$_3$ to the ideal Kitaev case~\cite{Sandilands15,Cao16,Zvyagin17,nagler16}, early strategies have involved adding other symmetry-allowed terms to the pure Kitaev model in small quantities, with a hope for a reasonable phenomenology~\cite{motome16,motome17,K1K2}. Nonetheless, significant systematic efforts have also been  made to restrict parameters of the more complete versions of the generalized KH model using first-principles approaches, perturbative orbital model derivations, exact diagonalization, spin-wave calculations, and other methods, many in combination with the symmetry analysis and with the goal  to match various experimental observations~\cite{hozoi16,Kee,Chaloupka16,Winter18,Vojta1,NaglerVojta18,Mandrus18,%
kee16,winter16,gong17,li17,berlijn19,wen17,winter17,suga18,moore18,orenstein18,gedik19,kaib19,kim19,%
okamoto19,Keimer20,Kaib20,Andrade20,Janssen20,Li21,Ran22,Samarakoon22,Giniyat22,Pollet21,kaib_npj_22}. However, in spite of  its importance  for the field of Kitaev materials, this significant effort has not yet yielded an agreement on the parameters of the $\alpha$-RuCl$_3$ model, nor  are there sets of them that successfully describe the full set of experimental observations.

One objective reason for the difficulty with the convergence on the physical set of parameters for $\alpha$-RuCl$_3$ can be attributed to the model's multi-dimensional parameter space, which generally complicates the search for a unique set of strong microscopic constraints~\cite{RauReview16,WinterReview}. It can also be argued that the natural instinct to  reduce the dimensions of this space by neglecting some of the parameters only exacerbates the problem of convergence. This is not only due to artificial restrictions on the parameter space, possibly affecting one's ability to describe physical phenomena, but also because  this reductionist approach suffers from the lack of intuition regarding the outcomes of individual terms in the standard $KJ\Gamma\Gamma'$ parametrization of the exchange matrix. In a sense,  the model's own complexity is a problem that has been calling for a better, more intuitive parametrization. 

Another objective obstacle on the path to convergence to an ultimate parameter set are  strong quantum fluctuations in both ordered  and nominally polarized field-induced paramagnetic phases of $\alpha$-RuCl$_3$, the situation common to   anisotropic-exchange magnets~\cite{ColdeaEssler,Tsirlin_Review}. These fluctuations limit the access to the regime of the  truly polarized state with quenched quantum fluctuations, where a direct determination of  the ``bare'' model parameters can be made using spectroscopic measurements of spin excitations,  as is successfully done in the other, more fortunate quantum magnets~\cite{RaduCsCuCl,Mourigal1D,MM1,Ross11,RaduYbTiO,JeffYbTiO,CoNb,CsCeSe}. This limitation  leaves most of the  phenomenological analyses of $\alpha$-RuCl$_3$ dealing with the observables that can be strongly affected by  quantum effects. To extract model parameters  from such observables in a meaningful part of the parameter space requires a prohibitively demanding numerical effort. Alternatively, they are extracted using semi-classical approximations, resulting in the parameters that are themselves renormalized from the bare ones, making uncertain their use for the other phenomenologies and complicating their comparison with the other suggested sets~\cite{okamoto19,rethinking}. 

\vspace{-0.3cm}
\subsection{The solution}
\label{Sec_intro_sols}
\vskip -0.2cm

In the present study, we propose to have reached, if not the final scene of the $\alpha$-RuCl$_3$ parameters' drama, but perhaps its final act.

\vspace{-0.3cm}
\subsubsection{Anisotropic strategy}
\label{Sec_intro_anisotropic}
\vskip -0.2cm

We argue that it is precisely the  spin-orbit-induced anisotropic-exchange components of the effective spin model of $\alpha$-RuCl$_3$ that can be strongly constrained by the existing phenomenologies. The crucial ideological steps are to identify  phenomena that (i) would {\it not} have occurred if such terms were absent, 
(ii)     provide  {\it orthogonal}, i.e., not redundant, constraints that restrict different combinations of anisotropic exchanges, and (iii) have minimal effects of quantum fluctuations in them. Thus, the strategy is to focus on the bounds on the $K$, $\Gamma$, and $\Gamma'$ terms of the generalized KH model, leaving the isotropic ones, $J$ and $J_3$, to further phenomenological constraints and  explorations of the resultant lower-dimensional phase diagram in the restricted parameter subspace.  

We claim to have identified three nearly orthogonal phenomenological constraints that can fulfill such a mission: (a) The out-of-plane tilt angle $\alpha$ of the ordered moments  in the zero-field  zigzag state of $\alpha$-RuCl$_3$  \cite{Plumb15,Coldea15,Cao16,kim19,kim_angle24}, (b) the energy offset  $\Delta E_g$ of the lowest single-spin-flip excitation in the Raman,  terahertz (THz), and electron spin resonance (ESR) spectra at high fields~\cite{Zvyagin17,kaib19,Zvyagin20}, referred below as to the ``ESR gap,'' and (c) the difference of the critical fields $\Delta H_c$ for the transition to the paramagnetic phase in the two principal orthogonal in-plane  directions~\cite{NaglerVojta18,Mandrus18,Zvyagin20}. Each of these quantities is thoroughly discussed in Sec.~\ref{Sec_constraints}.

\begin{figure}[t]
\includegraphics[width=1\linewidth]{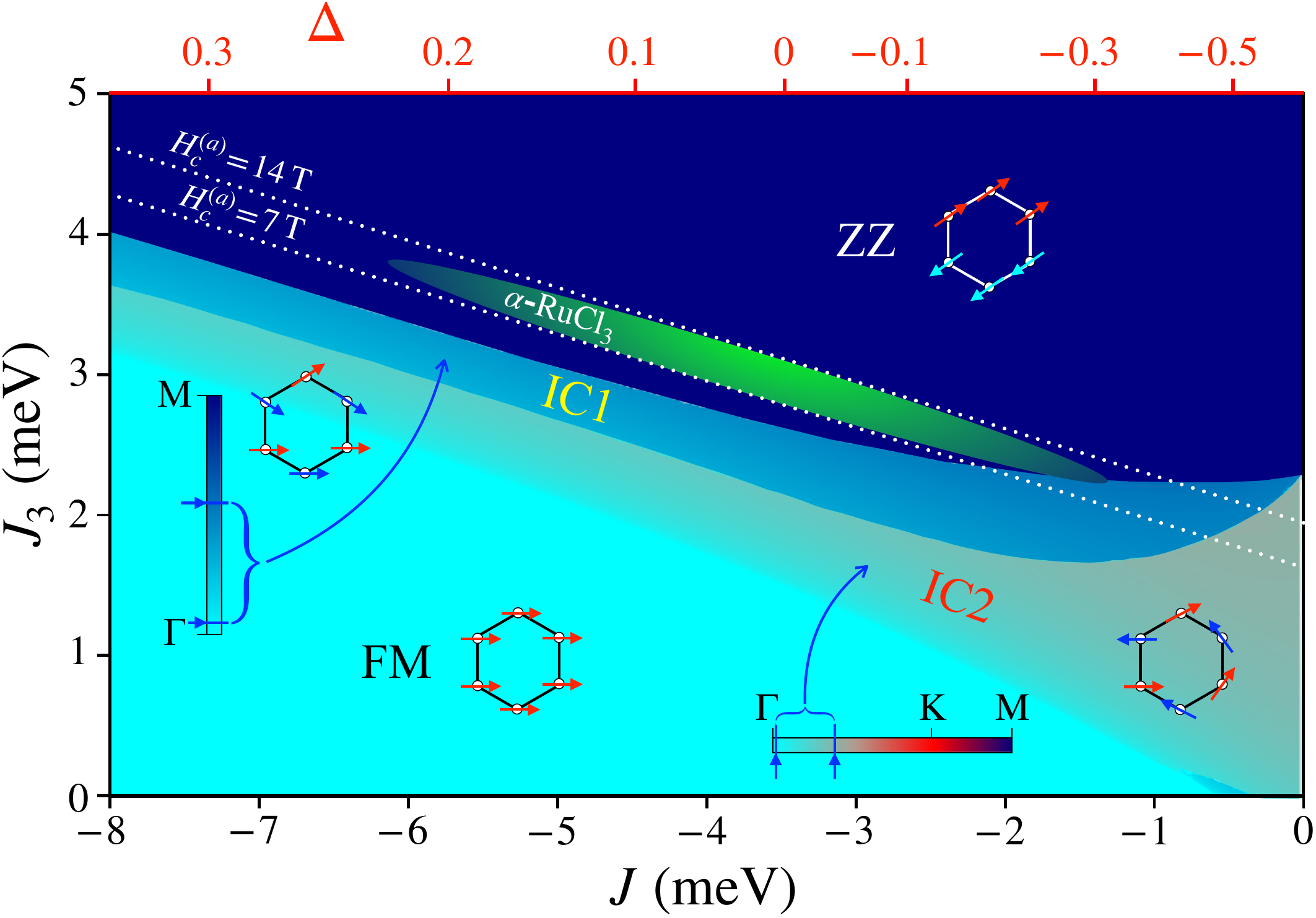} 
\vskip -0.2cm
\caption{The $J$--$J_3$ phase diagram of the model (\ref{eq_Hij}) for  $J\!<\!0$ and $J_3\!>\!0$ obtained by the LT approach  at fixed ``Point B'' set of $\{K,\Gamma,\Gamma'\}$, see the text. The ZZ, FM, and IC phases, the color bars indicating ranges of their ordering vectors, and the constraints from  the critical field $H_c^{(a)}$ are highlighted.}
\label{fig_phaseDJ1J3_B_LT}
\vskip -0.4cm
\end{figure}

By imposing physical bounds that are guided by  phenomenologies on the theoretical expressions for these quantities, we can infer the  bounds on the three anisotropic  terms of the model~(\ref{H_JKGGp}). To illustrate the result of this approach in very broad strokes, the small critical-field  difference $\Delta H_c$  strongly ties  $\Gamma'$ to $\Gamma$, with $\Gamma'$ necessarily positive and significant~\cite{rethinking}. The  physical limits on the ESR gap provide tight bounds  for $\Gamma$. Lastly, the out-of-plane tilt angle $\alpha$  binds Kitaev $K$-term in a significantly narrower range than the prior estimates, see Sec.~\ref{Sec_constraints} for further details. 

In this context,  two recent works  suggest a promising convergence of  very different approaches on the  parameter space relevant to $\alpha$-RuCl$_3$ of the same effective model. 

The first is Ref.~\cite{Giniyat22}, which has used the extensive  perturbative orbital model derivation in the spirit of the original work by Jackeli and Khaliullin~\cite{Jackeli}. By considering octahedral distortion pertinent to $\alpha$-RuCl$_3$ and using an exhaustive set of hopping channels for all relevant orbitals, their systematic microscopic derivation of the nearest-neighbor exchange parameters has yielded  qualitative and quantitative trends which are very similar to our purely phenomenological results, perhaps aside from the overall rescaling, and specifically pointed to a positive and sizable $\Gamma'$ term. Other close similarities between our results will be further highlighted below. 

The second is Ref.~\cite{Samarakoon22}, which has  extracted interaction parameters of $\alpha$-RuCl$_3$ from the neutron-scattering data using machine learning, but  restricted itself to an artificially abbreviated  model with no $\Gamma'$ term, underutilizing its parameter space. Despite this restriction,  their resultant  model is able to satisfy, even if for the wrong reasons, two of the phenomenological constraints suggested in this work, while failing  the more stringent one, as we show below. Our analysis suggests that a redo of the same effort with no artificial restrictions in the model and with an additional experimental input may open the full potential of this approach and lead to a convergence of our parameter sets. 

\begin{figure}[t]
\includegraphics[width=1\linewidth]{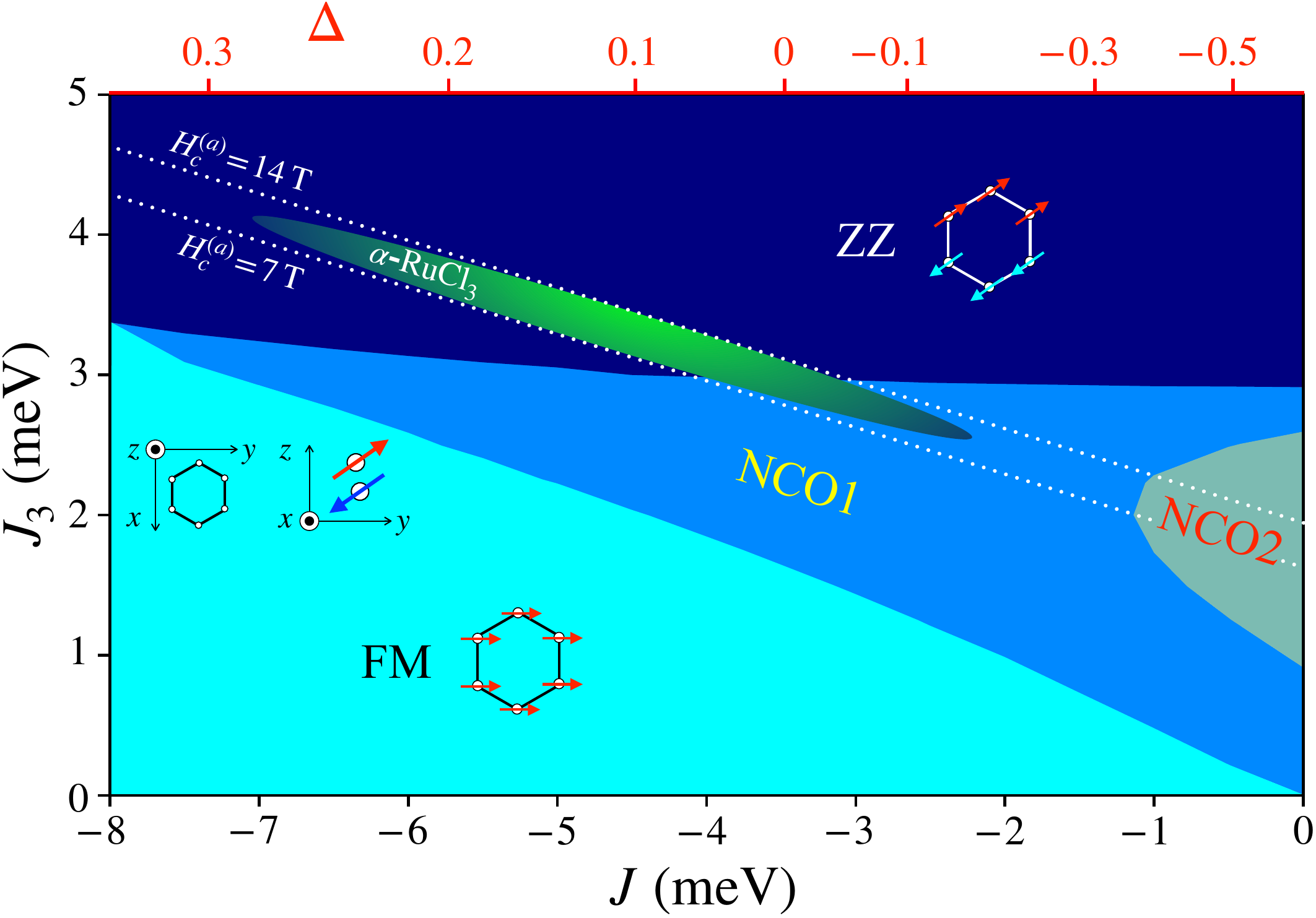} 
\caption{Same as Fig.~\ref{fig_phaseDJ1J3_B_LT} by ED.}
	\label{fig_phaseDJ1J3_B_ED}
\end{figure}

With the proposed constraints, one is able to focus on the relevant region of the multi-dimensional phase diagram of this model. In Figure~\ref{fig_phaseDJ1J3_B_LT}, we provide a representative example of such a phase diagram, in which $\{K,\Gamma,\Gamma'\}$ anisotropic terms are fixed to a set of values referred below as the ``Point B'' that belongs to a narrow 3D region bounded by the constraints, while the remaining ``free'' isotropic parameters of the model (\ref{eq_Hij}) are varied in the relevant  region of the $J$--$J_3$ plane. The Point~B selection, Luttinger-Tisza (LT) method~\cite{lt_original,Lyons_Kaplan_1960,Friedman_1974,Litvin_1974,Niggemann_2019}, which is used for obtaining Fig.~\ref{fig_phaseDJ1J3_B_LT}, and specific features of the spin configurations in the individual phases are  all thoroughly described in Secs.~\ref{Sec_constraints} and \ref{Sec_PhDs} below. Here,  we point out  two ubiquitous features of this phase diagram. 

First, the zigzag (ZZ) and ferromagnetic (FM) phases are prominently present in the relevant phase space of $\alpha$-RuCl$_3$, in agreement with several prior works~\cite{winter17,winter16,Winter18,Keimer20},  and are separated by a layer of two incommensurate (IC) phases with varying ordering vectors~\cite{rethinking,Pollet21}. 

Second, while the difference of the critical fields for the transition to the paramagnetic phase $\Delta H_c$, already utilized as a constraint for the anisotropic terms, does not depend on the isotropic $J$ and $J_3$ exchanges,  the values of these critical fields themselves are strongly dependent on a linear combination of $J$ and $J_3$, providing another strong constraint on the $\alpha$-RuCl$_3$ parameters~\cite{rethinking}. While the critical fields can be strongly renormalized by quantum fluctuations, Fig.~\ref{fig_phaseDJ1J3_B_LT} demonstrates the narrow strip of the $J$--$J_3$ space that is carved by such a constraint, showing the boundaries that correspond to the bare value of one of the critical fields $H_c^{(a)}\!\approx\!7$~T~\cite{Winter18,Cao16,NaglerVojta18}, and the value  twice as high, allowing for its strong quantum renormalization. The acceptable range of $|J|$ is also broadly restricted from below by $\alpha$-RuCl$_3$ being in the zigzag phase and from above by the overall  widths of the magnetic excitation spectra~\cite{rethinking}.  

We also demonstrate that the chosen strategy allows us to rein in on the otherwise prohibitively demanding numerical exploration of the $\alpha$-RuCl$_3$ parameter space. With the phase diagram in Fig.~\ref{fig_phaseDJ1J3_B_LT} obtained by a quasiclassical Luttinger-Tisza approach~\cite{lt_original,Lyons_Kaplan_1960,Friedman_1974,Litvin_1974,Niggemann_2019}, one may be skeptical of its  applicability to the quantum $S\!=\!1/2$ limit of the model (\ref{eq_Hij}). In Figure~\ref{fig_phaseDJ1J3_B_ED}, we present such a quantum phase diagram obtained by ED, see  Sec.~\ref{Sec_PhDs} for further details. While the nature of the  intermediate phases between the ZZ and FM cannot be fully characterized in the ED approach, hence the ``non-commensurate'' (NCO) nomenclature, the close {\it quantitative} accord between the two diagrams in Fig.~\ref{fig_phaseDJ1J3_B_LT} and Fig.~\ref{fig_phaseDJ1J3_B_ED} on the general structure and hierarchy of the proximate phases is rather impressive. These numerical explorations are also  furthered by the DMRG approach, which  provides various selfconsistency checks on our phenomenological constraints, see Sec.~\ref{Sec_DMRGchecks}, and  yields unambiguous insights into the nature of the incommensurate phases,  see  Sec.~\ref{Sec_IC}.

\vspace{-0.3cm}
\subsubsection{Alternative parametrization}
\label{Sec_intro_alternative}

Generally, the exchange matrix $\hat{\rm \bf J}_{ij}$ in the model~(\ref{eq_Hij}) is not invariant under   axes transformations if anisotropic terms are present. While the choice of the cubic axes in  Fig.~\ref{fig_axes}(b) is designed to make explicit the Kitaev-like structure of the model (\ref{H_JKGGp}) which may  not be obvious otherwise, this may not be the optimal  choice when  other significant terms are present.

The most natural physical alternative is the crystallographic reference frame associated with the planes of magnetic ions, with $x$ and $y$ axes corresponding  to the principal in-plane directions of the honeycomb lattice and $z$ axis pointing out of this plane, see Figs.~\ref{fig_axes}(b) and \ref{fig_axes}(c). Several virtues of this reference frame, which include making explicit  the symmetries of the model and bond directionality of interactions while also  elucidating some of the enigmatic duality relations of the cubic-axis representation of the model (\ref{H_JKGGp}), have been discussed in the past~\cite{RauGp,ChKh15,Maksimov22,Giniyat22,rethinking}. 

With the nomenclature for the $\{x,y,z\}$ frame inherited from the related spin-ice models~\cite{Tsirlin_Review,Zhu19,Ross11,RauYb},  the nearest-neighbor Hamiltonian  (\ref{H_JKGGp})  is recast into the $XXZ$--${\sf J_{\pm\pm}}$--${\sf J_{z\pm}}$ form 
\begin{align}
\label{HJpm}
{\cal H}_1=&\sum_{\langle ij\rangle} \!\Big\{
{\sf J_1} \big(S^{x}_i S^{x}_j+S^{y}_i S^{y}_j+\Delta S^{z}_i S^{z}_j\big)\\
&\ \ +\!2 {\sf J_{\pm \pm}} 
\Big( \big( S^{x}_i S^{x}_j \!- \!S^{y}_i S^{y}_j \big) \tilde{c}_\alpha 
\!-\!\big( S^{x}_i S^{y}_j\!+\!S^{y}_i S^{x}_j\big)\tilde{s}_\alpha \Big)\nonumber\\ 
&\ \ +\!{\sf J_{z\pm}}\Big(\big( S^{x}_i S^{z}_j \!+\!S^{z}_i S^{x}_j \big) \tilde{s}_\alpha 
 \!-\!\big( S^{y}_i S^{z}_j\!+\!S^{z}_i S^{y}_j\big)\tilde{c}_\alpha \Big)\Big\}\nonumber,
\end{align}
where we use the shorthand notations $\tilde{c}_\alpha\!=\!\cos\tilde{\varphi}_\alpha$ and  $\tilde{s}_\alpha\!=\!\sin\tilde{\varphi}_\alpha$, with the phases $\tilde{\varphi}_\alpha\!=\!\{0,2\pi/3,-2\pi/3\}$  for the $\{{\rm Z,X,Y}\}$ bonds  in Fig.~\ref{fig_axes} being the bond angles of the primitive vectors ${\bm \delta}_\alpha$ with the $x$ axis. 

The  matrix of the transformation  from the cubic $\{\rm x,y,z\}$ to the crystallographic $\{x,y,z\}$ reference frame, ${\bf S}_{\rm cryst}\!=\! \hat{\mathbf{R}}_c{\bf S}_{\rm cubic}$, is given in Appendix~\ref{app_A} together with the translation of the  parameters of the generalized KH model in the form (\ref{H_JKGGp})   to that of   (\ref{HJpm}) and back: $\{J,K,\Gamma,\Gamma'\}\!\Leftrightarrow\! \{ {\sf J_1},\Delta{\sf J_1},{\sf J}_{\pm \pm},{\sf J}_{z\pm}\}$. The isotropic third-neighbor ${\cal H}_3$ in (\ref{eq_Hij}) is unchanged by this transformation.

One immediate advantage of the crystallographic form of the model (\ref{HJpm}) is in having fewer bond-dependent terms, with its first line being a conventional  $XXZ$ Hamiltonian. As we argue in Sec.~\ref{Sec_constraints}, the  $XXZ$--${\sf J_{\pm\pm}}$--${\sf J_{z\pm}}$ model (\ref{HJpm}) also provides a much more intuitive understanding of the outcomes of its individual terms, allowing one to tie them directly to  specific observables. One such example is the out-of-plane tilt angle $\alpha$ discussed above, which can only be caused by the ${\sf J}_{z\pm}$-term, the sole term in the model that connects the in-plane and the out-of-plane spin components.  Another example is the ESR gap $\Delta E_g$  that depends on an uninformative combination of $\Gamma$ and $\Gamma'$ in the $KJ\Gamma\Gamma'$ language (\ref{H_JKGGp}), but corresponds to a simple $XXZ$ anisotropy, ${\sf J_1}(1-\Delta)$, upon translation to the crystallographic frame (\ref{HJpm}), see Sec.~\ref{Sec_constraints}. 
 
Moreover,   significant additional progress in understanding  $\alpha$-RuCl$_3$ can be made by reforging  phenomenological constraints on  the anisotropic parameters  of the generalized KH model  (\ref{H_JKGGp}) $\{K,\Gamma,\Gamma'\}$ into  similar constraints on the $\{ {\sf J_1}(1-\Delta),{\sf J}_{\pm \pm},{\sf J}_{z\pm}\}$ terms of the model (\ref{HJpm}). This conversion results in a clear hierarchy of the parameters of this model and, ultimately, in a simpler and more intuitive version of it.

For the advocated parameter ranges  of $\alpha$-RuCl$_3$, the nearest-neighbor part of the spin model  in the crystallographic language can be shown to have two leading terms, ${\sf J_1}$ and ${\sf J}_{z\pm}$, and two secondary ones, the $XXZ$  parameter $\Delta$ and one of the bond-dependent anisotropic terms, ${\sf J_{\pm\pm}}$.  This hierarchy is also in  very close accord with the results of the orbital model expansion of Ref.~\cite{Giniyat22} discussed above, and is  in  broader agreement with the majority of the earlier assessments of the $\alpha$-RuCl$_3$ parameters,  offering a unifying view of them as all pointing in a similar direction, although often in a limited way. 

\begin{figure}[t]
\centering
\includegraphics[width=\linewidth]{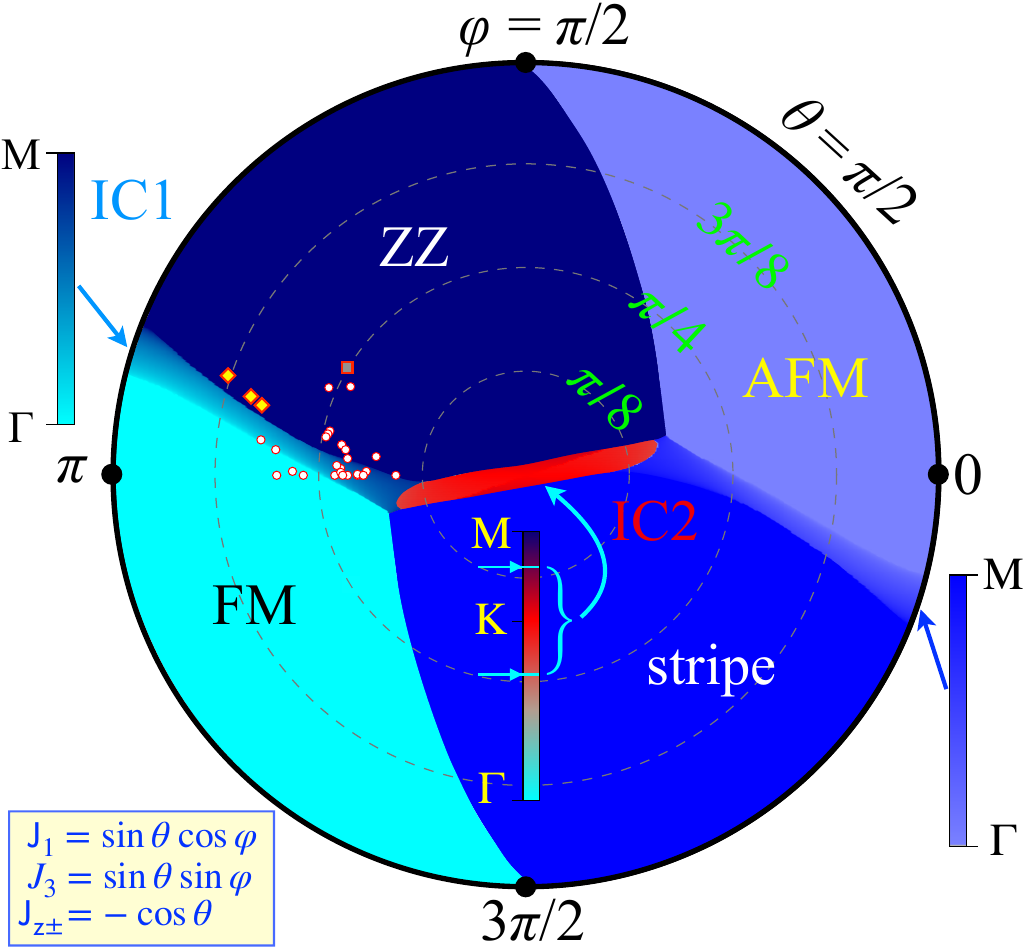}
\caption{Polar phase diagram of the model (\ref{HJpm}) for $\Delta\!=\!{\sf J_{\pm\pm}}\!=\!0$, with ${\sf J_1}$ and $J_3$ encoding the polar and ${\sf J_{z\pm}}$ the radial coordinates in units of ${\sf J^{\rm 2}_1}+{\sf J^{\rm 2}_{z\pm}}+J_3^2\!=\!1$, see the legend. Phases are discussed below and the small symbols near the ZZ-FM boundary are  projections of  prior $\alpha$-RuCl$_3$ parameter searches onto the $\Delta\!=\!{\sf J_{\pm\pm}}\!=\!0$ plane,  see Sec.~\ref{Sec_polarPD} for details.}
\label{fig_phaseDpolar}
\end{figure}

It transpires that the resultant minimal model that closely describes $\alpha$-RuCl$_3$ is dominated by the $XY$ ($\Delta\!\approx\!0$) ferromagnetic ${\sf J_1}$ and   sizable anisotropic ${\sf J_{z\pm}}$  terms, which are complemented by the isotropic antiferromagnetic $J_3$ exchange.  The neglected  $\Delta$ and  ${\sf J_{\pm\pm}}$ can  be verified  as playing only secondary roles in the physical outcomes compared to the leading terms.

Needless to say, this revelation suggests a considerable simplification of the $\alpha$-RuCl$_3$ parameter problem, as it allows a significantly more focused and detailed exploration of the relevant and physically justified three-dimensional parameter space of the resultant ${\sf J^{\it XY}_1}$--${\sf J_{z\pm}}$--$J_3$  model. 

In Figure~\ref{fig_phaseDpolar}, we provide an example of such an exploration in the form of its polar phase diagram, with  details given in Sec.~\ref{Sec_PhDs}. In this phase diagram, obtained by the same LT  method as Fig.~\ref{fig_phaseDJ1J3_B_LT},  ${\sf J_1}$ and $J_3$ encode the polar coordinate and ${\sf J_{z\pm}}$ is the radial one, as  indicated in the legend. While covering the entire parameter space of the  ${\sf J^{\it XY}_1}$--${\sf J_{z\pm}}$--$J_3$  model, which contains additional antiferromagnetic (AFM) and stripe phases, the relevant segment of it is populated by the zigzag and ferromagnetic phases, separated by  the incommensurate phase. This proximity is  already familiar from the Cartesian phase diagrams of Figs.~\ref{fig_phaseDJ1J3_B_LT} and \ref{fig_phaseDJ1J3_B_ED}, which offer a  different slice through the same higher-dimensional parameter space. In addition, the small symbols in Fig.~\ref{fig_phaseDpolar} are the  projections of the majority of the previously proposed $\alpha$-RuCl$_3$ parameter sets onto the $\Delta\!=\!{\sf J_{\pm\pm}}\!=\!0$ plane of the phase diagram, demonstrating the commonality of the trends and phenomenologies that all of them were  trying to capture.
   
It is also important to point out that the model in the crystallographic frame (\ref{HJpm}) and its abbreviated ${\sf J_1}$--${\sf J_{z\pm}}$--$J_3$  version offer a direct connection to the broader spectrum of the paradigmatic  models in frustrated magnetism. In fact, the circumference of the phase diagram in Fig.~\ref{fig_phaseDpolar} corresponds to the $J_1$--$J_3$ FM-AFM model that has been studied since long ago~\cite{italians} and has attracted significant attention very recently in the context of the other material candidates of the Kitaev model realization~\cite{J1J3us,Arun23,Hickey23,Halloran23}. This model and its ubiquitous characteristic sequence of the FM, ZZ, and intermediate phases also provide a wider context to the studies of $\alpha$-RuCl$_3$.

\subsection{The plan}
\label{Sec_intro_plan}

With the brief   synopsis of our findings given in the previous pages, the rest of the paper is organized as follows. 

The  empirical constraints  and representation of them in the cubic and crystallographic reference frames, the outline of the advocated parameter space, and selected  representative choices of the anisotropic exchanges are discussed in Section~\ref{Sec_constraints} together with the systematic analysis of the prior efforts at the $\alpha$-RuCl$_3$ parameters in the context of the proposed  phenomenologies.  

In Section~\ref{Sec_PhDs}, we provide a detailed discussion of the phase diagrams for the relevant region of the multi-dimensional parameter space of the $\alpha$-RuCl$_3$ model, such as the ones in Figs.~\ref{fig_phaseDJ1J3_B_LT} and \ref{fig_phaseDJ1J3_B_ED} in the generalized KH, and in Fig.~\ref{fig_phaseDpolar} in the crystallographic axes. Here we also elaborate on the technical details of the approaches that are used for  their derivation. 

Further insights and selfconsistency checks from the unbiased numerical methods on the phenomenological constraints and other assumptions of our approach   are provided in Section~\ref{Sec_DMRGchecks}.
 
A comparative analysis uncovering the nature of the incommensurate phases in these phase diagrams using DMRG is presented in Section~\ref{Sec_IC}. 

We discuss possible future directions  as an outlook in Section~\ref{Sec_Outlook} and present a summary in Section~\ref{Sec_conclusions}. Technical details are delegated to Appendices. 

\section{Constraints and parameters}
\label{Sec_constraints}

In this Section, we demonstrate that the  available  phenomenologies are powerful enough to significantly restrict the  physically  reasonable parameter space of the generalized KH model (\ref{eq_Hij}) for $\alpha$-RuCl$_3$. With the help of the translation of the model using the alternative parametrization described in Sec.~\ref{Sec_intro_alternative}, we also emphasize  important physical insights and intuition associated with the crystallographic reference frame.  The 3D boundaries of the proposed parameter space in both parametrizations are shown, the representative choices of the model parameters that are used for the phase diagrams in Secs.~\ref{Sec_PhDs}, \ref{Sec_IC}, and \ref{Sec_intro_anisotropic} are suggested, and a  comparison of the $\alpha$-RuCl$_3$ parameter sets with  recent works  is highlighted.

\subsection{Constraints}
\label{Sec_constraints_details}

According to our discussion in Sec.~\ref{Sec_intro_anisotropic}, the observables that are effective in restricting anisotropic terms of the spin model of $\alpha$-RuCl$_3$  are the ones that (i) should occur only if such terms are present, (ii) should preferably depend on a distinct set of such terms to avoid redundancy of the empirical constraints, and (iii) should contain minimal or controllable quantum effects.

We  identify three phenomenological constraints that can fulfill such a mission and provide a thorough discussion of them.  In the end of  Section~\ref{Sec_constraints},  Table~\ref{table3} offers a comprehensive  compilation of the previously proposed parameter sets for the anisotropic exchanges  $\{K,\Gamma,\Gamma'\}$ of $\alpha$-RuCl$_3$~\cite{kee16,winter16,gong17,li17,berlijn19,wen17,winter17,suga18,moore18,orenstein18,gedik19,%
kaib19,kim19,okamoto19,Keimer20,Kaib20,Andrade20,Janssen20,Li21,Ran22,Samarakoon22,Giniyat22}, with the only general restriction of  $K\!<\!0$, which is consistently demonstrated by the first-principles guidance and spectroscopic measurements~\cite{winter16,kim19}.  Table~\ref{table3} also shows the ability of these parameter sets to match the proposed phenomenological constraints. Since the number of  entries in this compilation is significant, we also use histogram distributions of the data in these sets to showcase the proposed physical limits on the observables and the effect of the constraints on the possible ranges of the model parameters. 

\subsubsection{Tilt angle}
\label{Sec_tilt_angle}

The first proposed phenomenological constraint is the substantial tilt of the ordered moments  out of the crystallographic plane of magnetic ions in the low-temperature zigzag state of $\alpha$-RuCl$_3$~\cite{Plumb15,Coldea15}, see inset of Fig.~\ref{fig_ESRfit}. This tilt angle has been measured by  the neutron and  resonant elastic x-ray scattering spectroscopies, which put its value to $\alpha\!\approx \! 35\degree$ and $\alpha\!\approx \! 32\degree$, respectively~\cite{Cao16,kim19,kim_angle24}. The sign of this angle was also important for the determination of the sign of the Kitaev $K$-term in $\alpha$-RuCl$_3$~\cite{kim19}. 

Needless to say, the isotropic exchanges cannot be responsible for the tilt, with the classical energy minimization connecting the tilt angle to a rather non-intuitive mix of all three anisotropic-exchange terms of the generalized KH model (\ref{H_JKGGp})~\cite{ChKh15,Chaloupka16,rethinking}
\begin{align}
\tan 2\alpha=  4\sqrt{2}\cdot\frac{\Gamma-K-\Gamma'}{7\Gamma+2K+2\Gamma'}.
\label{eq_alphaKG}
\end{align}
In the crystallographic notations of the model (\ref{HJpm}), this expression becomes significantly more telling 
\begin{align}
\tan 2\alpha=  \frac{4{\sf J_{z\pm}}}{{\sf J_1}\big(1-\Delta\big)+4{\sf J_{\pm\pm}}},
\label{eq_alpha}
\end{align}
showing plainly that the tilt can only be induced by the term that explicitly connects the in-plane and the out-of-plane spin components, ${\sf J}_{z\pm}$.

One can question the use of the classical expressions for the tilt angle to match the actual observations in $\alpha$-RuCl$_3$. According to Ref.~\cite{Chaloupka16}, there are also possible differences between the calculated direction of the pseudospin in (\ref{eq_alphaKG}) and the experimentally measured direction of the magnetic moment. 

For the first concern, prior comparisons to exact diagonalization for several representative sets of parameters have shown only insignificant  quantum corrections to the tilt angle~\cite{Chaloupka16}. Similarly small deviation from the classical value of the out-of-plane tilt of the magnetic moments was recently observed in a strongly-fluctuating ground state  of a related anisotropic-exchange model of the 1D compound CoNb$_2$O$_6$, where it was rationalized as due to a natural compensation  of the contributions of different terms~\cite{CoNb2O6}.  The conclusive DMRG check  of the numerical correspondence of the tilt angle in the quantum and classical models for the  parameters proposed in this work will be   given in Sec.~\ref{Sec_DMRGchecks}.

In order to account for these effects, and instead of fixing $\alpha$ to a particular value, we take a relatively broad span of  $30^\circ\!\leq\!\alpha\!\leq\!37^\circ$ as the  physically allowed range of the tilt angle.

\subsubsection{ESR gap}
\label{Sec_ESR_gap}

While the tilt angle criterion has been previously discussed~\cite{Chaloupka16}, the second phenomenological constraint has received less attention, despite its clear advantages and being close in spirit to the standard approach to the problem of determining parameters in fluctuating magnets~\cite{rethinking}. As was mentioned earlier, significant quantum fluctuations in the nominally polarized paramagnetic phase of $\alpha$-RuCl$_3$ limit the insights that one can draw from such a state, compared to the case when fluctuations are fully quenched by the field~\cite{RaduCsCuCl,Mourigal1D,MM1,Ross11,RaduYbTiO,JeffYbTiO,CoNb,CsCeSe}.

\begin{figure}[t]
\centering
\includegraphics[width=\linewidth]{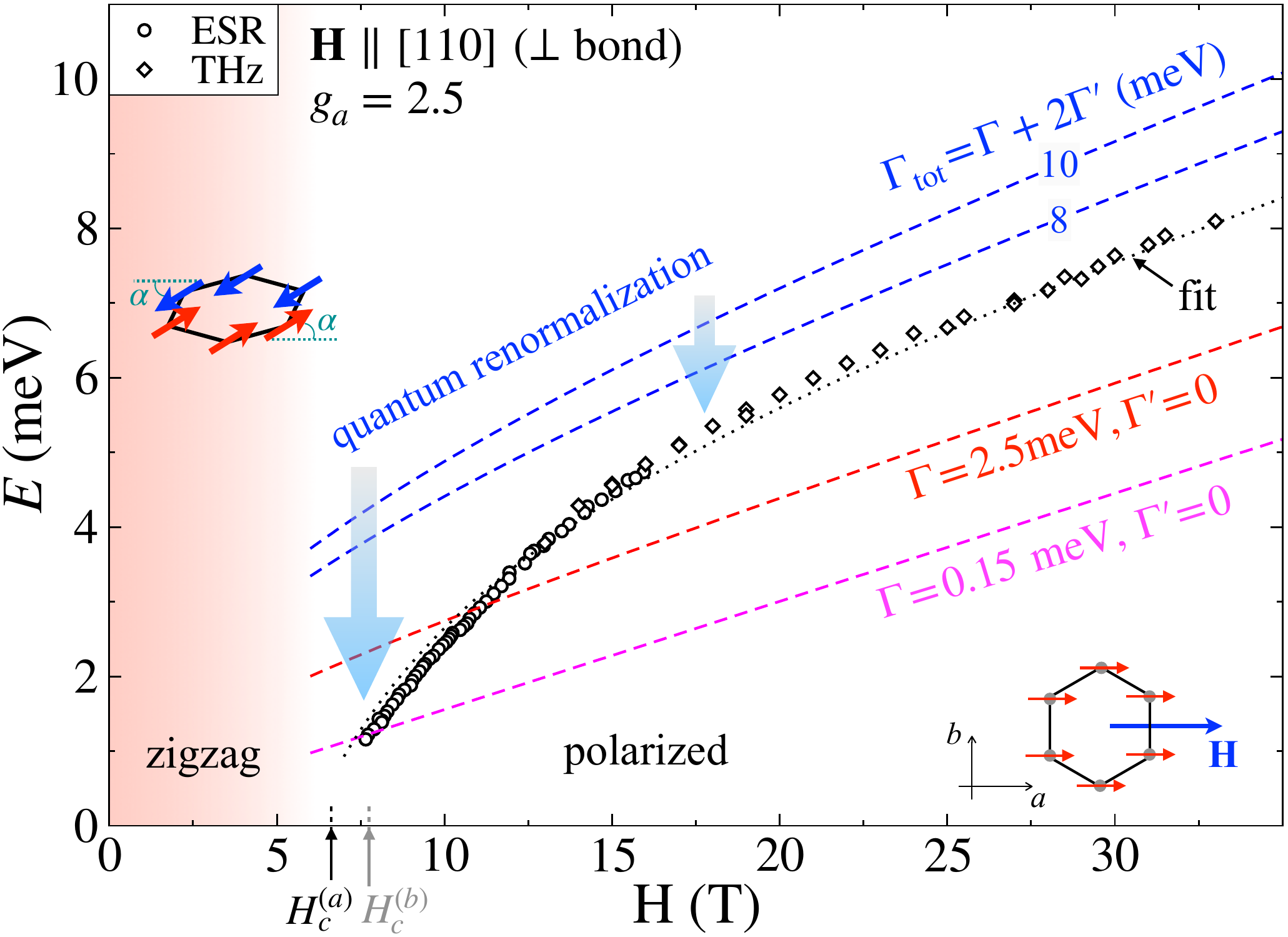}
\caption{ESR~\cite{Zvyagin17} and THz~\cite{kaib19} data for the magnon energy, $E_0$, 
at $\mathbf{k}\!=\!0$ vs field in the $a$-direction. Results from Eq.~(\ref{Ek0}) 
for $\Gamma_{\rm tot}$ from Refs.~\cite{Samarakoon22} and~\cite{winter17} and for 8~meV and 10~meV.  Arrows emphasize the downward renormalization by quantum fluctuations.
Insets: sketches of the zigzag and polarized states and  in-plane $a$ and $b$ directions.}
\label{fig_ESRfit}
\end{figure}

Although limiting, they do not completely prohibit such insights, because some spectroscopies can be utilized in the fields that are much higher than the ones available to neutron scattering. Specifically, the field-dependence of the lowest-energy spin excitation at ${\bf k}\!=\!0$ has been probed up to the high fields of 35~T using Raman and THz spectroscopies~\cite{kaib19}, with the ESR results essentially coinciding with them up to its  feasibility range of  17~T~\cite{Zvyagin17}, see Fig.~\ref{fig_ESRfit}. For the fields $\agt\!20$~T, quantum fluctuations in $\alpha$-RuCl$_3$ are substantially quenched, leading to   $\alt$20\% suppression of the ordered moment~\cite{Coldea15,Winter18}, the number that can be used to gauge their overall effect. 

All three techniques are consistent with each other and in their interpretation of the lowest excitation in their spectra as that of the single magnon-like spin-flip in the partially polarized ferromagnetic background~\cite{kaib19,Zvyagin20}, to which we will refer to as the lowest ``ESR mode.'' The behavior of this mode differs markedly from the Zeeman-like linear field-dependence, typical of the isotropic magnets, with the high-field asymptote of $E_0$ vs $H$ having a large positive shift. The latter follows straightforwardly from the quasiclassical result~\cite{rethinking} 
\begin{align}
\label{Ek0}
E_0=\sqrt{h\big(h+\Delta E_g\big)}\,,
\end{align}
with $h\!=\!g\mu_B H$ and the ``ESR gap'' $\Delta E_g$ 
\begin{align}
\label{ESRgap}
\Delta E_g=3S \big(\Gamma+2\Gamma'\big)=-3S{\sf J_1}(1-\Delta),
\end{align}
that depends on a  subset of terms of the model (\ref{eq_Hij}), underscoring its utility as of the ``orthogonal'' constraint. 

The ESR gap depends on a simple, but uninformative combination of $\Gamma$ and $\Gamma'$ in the generalized KH parametrization  (\ref{H_JKGGp}). In contrast, its explicit relation to the departure of the $XXZ$ part of the model in the crystallographic frame (\ref{HJpm}) from the isotropic Heisenberg limit, $(1-\Delta)$, is not only simple, but   also explains the physical origin and generic character of the positive shift of the ESR mode. Given the large observed value of $\Delta E_g$, it provides the most direct evidence of the strong deviation of the KH model of $\alpha$-RuCl$_3$ from the Kitaev limit and a compelling argument for its natural description as an easy-plane $J_1$--$J_3$ ferro-antiferromagnet.
 
In Fig.~\ref{fig_ESRfit}, we reproduce the data for the energy of the lowest ESR mode from Refs.~\cite{Zvyagin17,kaib19} for the in-plane field  perpendicular to the  bond ($a$-direction),  varying from the critical field $H^{(a)}_{c}\!\approx\! 7$~T  to about 35~T, with the data for the $b$-direction, along the bond, being very similar. The high-field expansion of Eq.~(\ref{Ek0}), $E_0\!\approx\!h + a_0 + a_1/h$ provides the fit to the data. In  this study, we use $g_a\!=\!g_b\!=\!2.5$, in accord with the earlier estimates~\cite{hozoi16,Winter18,Kubota15}.

Qualitatively, quantum fluctuations should produce a downward shift of the ``bare'' quasiclassical mode in (\ref{Ek0})  due to a repulsion from the two-magnon states, which are also observed in Refs.~\cite{Zvyagin17,kaib19}. This effect is  naturally stronger near the critical field, as is emphasized schematically in Fig.~\ref{fig_ESRfit}, where these states may overlap, according to the analysis of Ref.~\cite{kaib19}. The downward renormalization has also been  confirmed by exact  diagonalization and selfconsistent spin-wave theory for representative parameter sets~\cite{Winter18,rethinking}. 

This downward trend implies a logical {\it lower} limit on the value of the ``bare'' ESR gap, or $\Gamma_{\rm tot}\!=\!\Gamma+2\Gamma'$, as the unrenormalized theory result in (\ref{Ek0}) must be at least above the experimental curve in order to be able to reach it upon a downward renormalization. 

To demonstrate the power of this criterion as a test for the existing parameter sets from Table~\ref{table3}, we use Fig.~\ref{fig_ESRfit} to show $E_0$ vs $H$ from Eq.~(\ref{Ek0}) for two representative  $\Gamma_{\rm tot}\!=\!\Gamma+2\Gamma'$, with the lowest dashed line from Ref.~\cite{Samarakoon22} of the machine-learning approach and the second lowest from Refs.~\cite{winter17,moore18}, which is successful in describing the low-field phenomenology of  $\alpha$-RuCl$_3$.  Both miss the high-field results by a significant margin, with the data suggesting substantially larger values of $\Delta E_g$ and the corresponding combination of $\Gamma$ and $\Gamma'$. With the residual quantum fluctuations still present up to 35~T, our analysis suggests  that $\Gamma_{\rm tot}$ cannot be less than $\approx\!7.5$~meV. 

Importantly,  quantum fluctuations are reduced in higher fields, which also  puts a logical {\it upper} limit on $\Delta E_g$  and $\Gamma_{\rm tot}$.  As is made clear in Fig.~\ref{fig_ESRfit} with the help of the quasiclassical results for $E_0$ vs $H$ curves for $\Gamma_{\rm tot}\!=$8 and 10~meV,  with  quantum effects in magnetization being  $\alt$20\%, it would be very hard to justify $\Gamma_{\rm tot}$ to be larger than $\approx\!10$ meV, as this would imply unphysically large fluctuations in a strongly gapped high-field state. 

Altogether, the high-field results for the ESR gap provide the lower and the upper bounds of the  physically allowed range  directly to the linear combination of the model parameters,   7.5~meV$\leq\!\Gamma+2\Gamma'\!\leq\!10$~meV, which are used in our analysis below. The quantitative verification of the  downward-renormalization effects in the field-dependence of the ESR gap is presented in Sec.~\ref{Sec_DMRGchecks}, using ED for the quantum model with one of the representative parameter sets proposed in this work. This also serves as a validation of our strategy for constraining the anisotropic terms of the $\alpha$-RuCl$_3$ effective model.

\subsubsection{Critical fields}
\label{Sec_dHc}

The third  phenomenological constraint is counterintuitive. If the  bond-dependent  terms in the generalized KH model of $\alpha$-RuCl$_3$ are significant, it is natural to inquire why the experimentally observed in-plane  critical fields for the transition to the partially polarized paramagnetic phase in the two principal field directions are so close, $\Delta H^{\rm exp}_c\!=\!H_c^{(b)}\!-\!H_c^{(a)}\!\approx\!0.8$~T \cite{NaglerVojta18,Mandrus18,Zvyagin20};  both fields are indicated in  Fig.~\ref{fig_ESRfit}.  Such a small difference would be natural for the nearly isotropic Heisenberg or pure KH models, but with  the ESR gap suggesting substantial $\Gamma$ and $\Gamma'$  terms,  small $\Delta H_c$ condition should be able to bind them in a way that is distinct from the other constraints. 

One can use the quasiclassical expressions for the critical fields given below to calculate such  $\Delta H_c$ for the parameter sets proposed in prior works, most of which put strong emphasis on the anisotropic terms in $\alpha$-RuCl$_3$. It is intriguing  that {\it all} but one of them yield values of these quasiclassical estimates of $\Delta H_c$ that are not just larger, but typically an order of magnitude larger than the observed value---of the same magnitude as the $H_c^{(b)}$ and $H_c^{(a)}$ critical fields themselves, see Table~\ref{table3}. The one exception, Ref.~\cite{Samarakoon22}, fails the ESR gap criterion as it  suggests a model which is close to the pure KH limit.
 
One concern for the critical fields being a phenomenological constraint is their expected strong quantum effects.  However, while quantum fluctuations should considerably suppress the critical fields from their classical values~\cite{Vojta2020_NLSWT}, it would be very unnatural for their {\it difference} to  change drastically~\cite{rethinking}. This suggested lack of  quantum effects in $\Delta H_c$ can be taken as a falsifiable prediction of our strategy, with the representative parameter sets proposed in this work given a numerical verification by DMRG   in Sec.~\ref{Sec_DMRGchecks}.

The transition from the  field-induced paramagnetic   to the zigzag phase corresponds to a closing of a magnon gap at the ordering vector associated with zigzag order~\cite{sears17,orenstein18,Janssen20}. This condition yields quasiclassical expressions for the transition fields  in the two principal directions, $H\!\parallel\! a$  and $H\!\parallel\! b$~\cite{rethinking}, perpendicular and parallel to the nearest-neighbor bond, respectively,  see Fig.~\ref{fig_axes}(b)
\noindent
\begin{align}
\label{Hca}
h_c^{(a)}&=J+3J_3+\frac{1}{12}\big(5K-5\Gamma-16\Gamma'\big)+\frac{1}{12}R_a,\\
&R_a=\sqrt{\big(K+5\Gamma+4\Gamma'\big)^2+24 \big( K-\Gamma+\Gamma'\big)^2},\nonumber\\
\label{Hcb}
h_c^{(b)}&=J+3J_3+\frac{1}{4}\big(2K-\Gamma-6\Gamma'\big)+\frac{1}{12}R_b,\\
&R_b=\sqrt{\big(2K+7\Gamma+2\Gamma'\big)^2+32 \big( K-\Gamma+\Gamma'\big)^2}, \nonumber
\end{align}
\noindent
where $h_c^{(\alpha)}\!=\!g_\alpha \mu_B H_c^{(\alpha)}$. Consistent with our proposed anisotropic strategy, the difference of the critical fields in (\ref{Hca}) and (\ref{Hcb}) is a function of only anisotropic terms of the model, $\{K,\Gamma,\Gamma'\}$. Although not leading to significant new insights, rewriting the critical fields in the crystallographic language yields   more compact expressions
\noindent
\begin{align}
\label{Hca_alt}
h_c^{(a)}&={\sf J_1}+3J_3+ \frac{1}{4} {\sf J_1}(1-\Delta)-\frac{1}{2}{\sf J_{\pm\pm}}+\frac{1}{4}R_a,\\
&R_a=\sqrt{\big( {\sf J_1}(1-\Delta) +2{\sf J_{\pm\pm}}\big)^2+12 {\sf J_{z\pm}}^2}, \nonumber\\
\label{Hcb_alt}
h_c^{(b)}&={\sf J_1}+3J_3+ \frac{1}{4} {\sf J_1}(1-\Delta)-{\sf J_{\pm\pm}} +\frac{1}{4}R_b,\\
&R_b=\sqrt{\big( {\sf J_1}(1-\Delta) +4{\sf J_{\pm\pm}}\big)^2+16 {\sf J_{z\pm}}^2},\nonumber
\end{align}
\noindent 
 suggesting that some subtle near-cancellation of the anisotropic terms is needed to yield   $\Delta H_c\!\approx\!0$. 

Aside from the quantum fluctuations suppressing $H_c^{(a)}$  and $H_c^{(b)}$ from their quasiclassical values, the interplane 3D couplings in $\alpha$-RuCl$_3$ should also alter them, as  discussed in Ref.~\cite{Janssen20}. However, since these 3D couplings are mostly isotropic~\cite{Janssen20},  they are not expected to modify the functional expression for  the field difference $\Delta H_c$, thus, effectively,  redefining only the combination of the isotropic exchanges of the  2D model (\ref{eq_Hij}), $J+3J_3$ in Eqs.~(\ref{Hca}) and (\ref{Hcb}) or ${\sf J_1}+3J_3$ in Eqs.~(\ref{Hca_alt}) and (\ref{Hcb_alt}), and can be absorbed into them. These couplings may also play an additional role in stabilizing the zigzag phase~\cite{Janssen20}.

In $\alpha$-RuCl$_3$, the in-plane critical field $H_c^{(a)}$ for the transition from the zigzag to the paramagnetic state found experimentally is  $\approx\!7$T and $H_c^{(b)}\!\approx\!7.8$T~\cite{Winter18,Cao16,NaglerVojta18}, see  Fig.~\ref{fig_ESRfit}.  There is an additional transition for $H\!\parallel\! a$ at $\approx\!6$T, which has been identified with an interplane ordering of the zigzag planes~\cite{NaglerVojta18,Balz19,Janssen20} that is unrelated to the  2D physics of $\alpha$-RuCl$_3$ discussed in this work. In the following analysis, the physical range of $0\!\leq\!\Delta H_c\!\leq\!1.5$T is assumed to allow for  modest quantum effects. 

\subsection{Parameter space}
\label{Sec_parameter_space}

\begin{figure}[t]
\centering
\includegraphics[width=\linewidth]{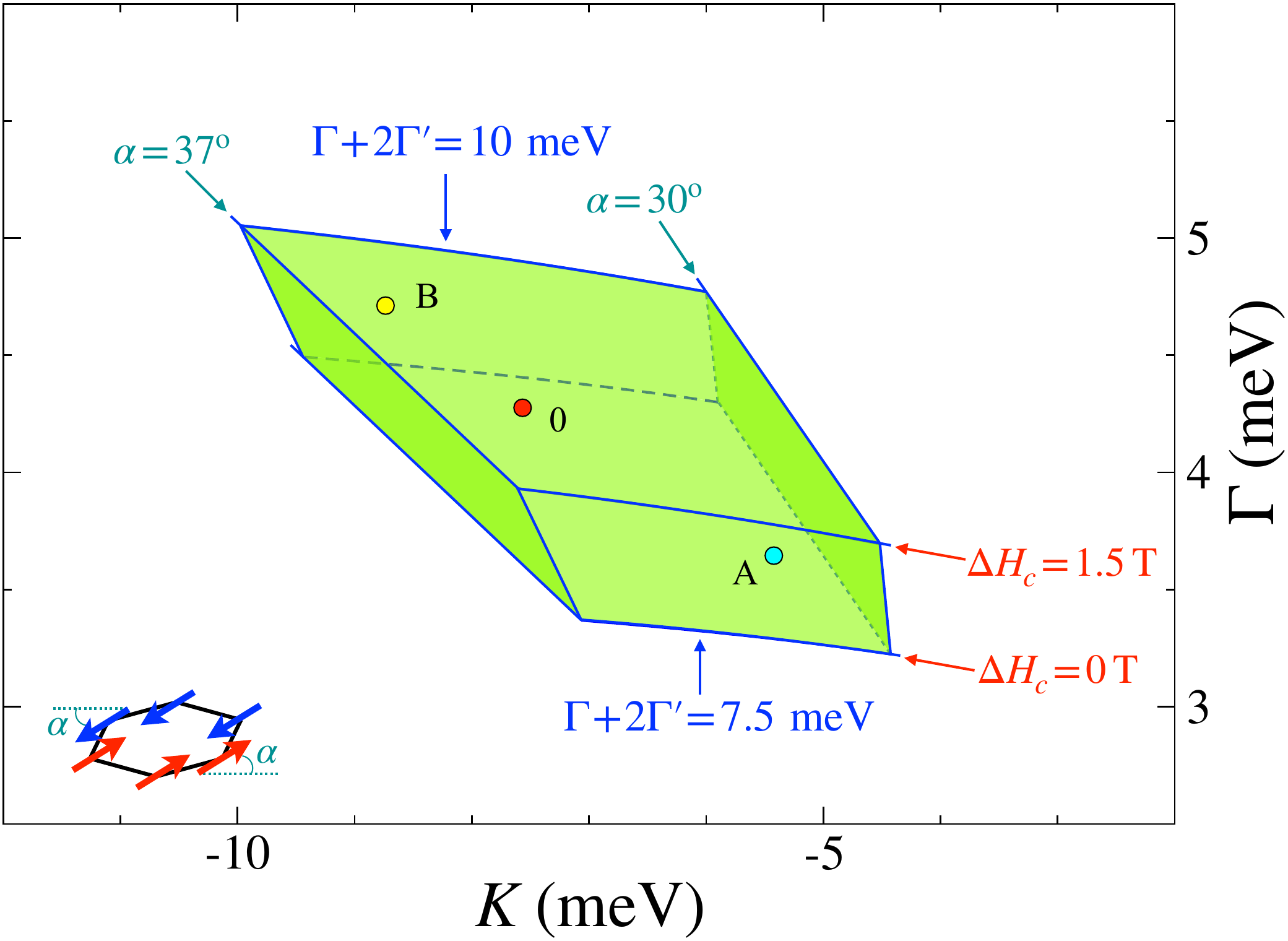}
\caption{The projection of the $\{K,\Gamma,\Gamma'\}$ parameter space on the $K$--$\Gamma$ plane. The colored region is the subspace  bounded by the phenomenological constraints for $\alpha$-RuCl$_3$. 
Boundaries due to tilt angle $\alpha$ (\ref{eq_alphaKG}), critical fields difference $\Delta H_c$ (\ref{Hca}) and (\ref{Hca}), and $\Gamma+2\Gamma'$  from the ESR gap (\ref{ESRgap}), and representative parameter sets, Points 0, A, and B, are indicated.}
\label{fig_GvsK}
\end{figure}

\begin{figure}[t]
\centering
\includegraphics[width=\linewidth]{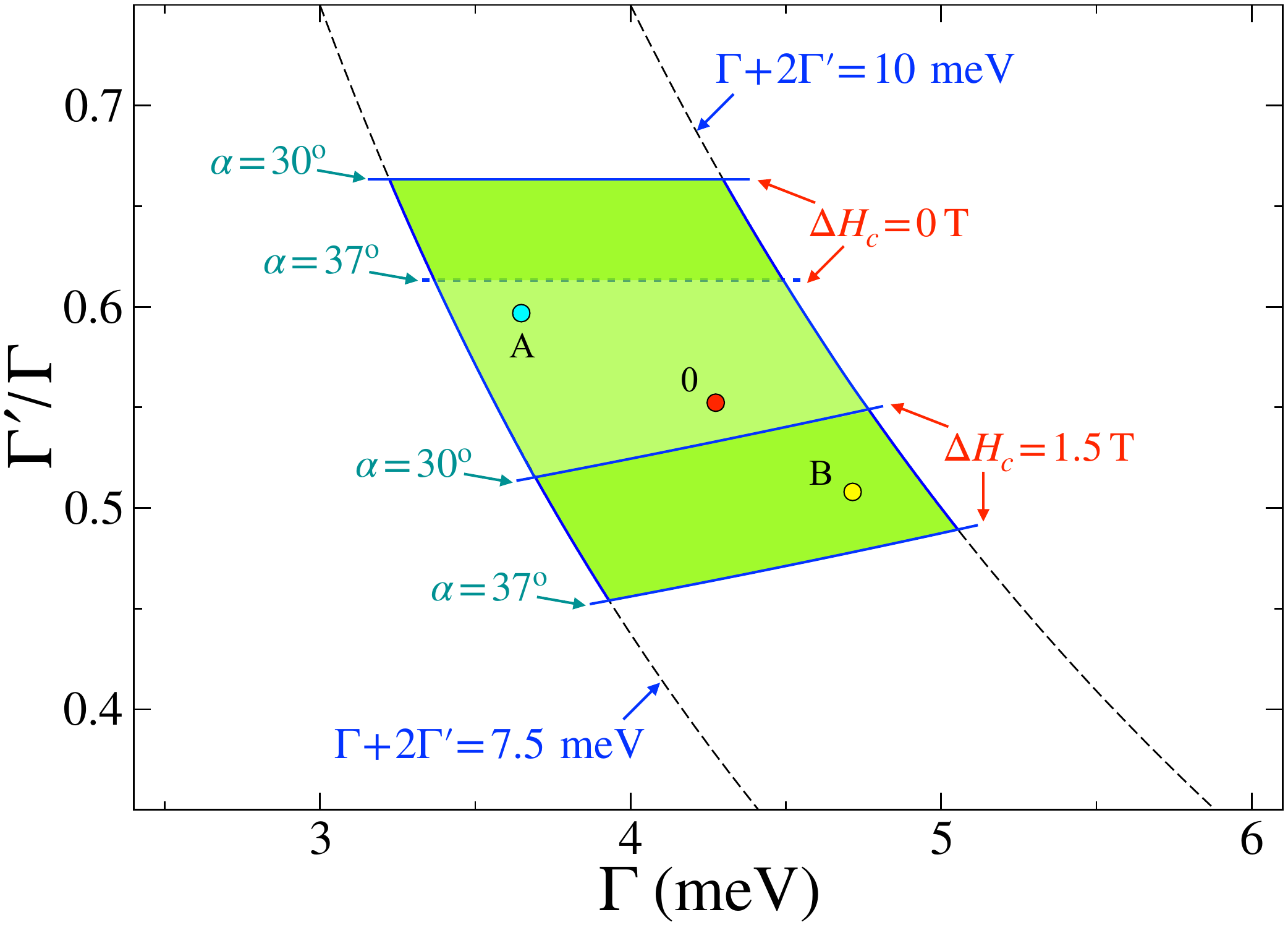}
\caption{Same as Fig.~\ref{fig_GvsK} with the projection on the $\Gamma$--$\Gamma'$ plane.}
\label{fig_G'vsG}
\end{figure}

We now turn to restricting the three anisotropic terms of the generalized KH model  (\ref{H_JKGGp}) by applying phenomenological constraints proposed in Sec.~\ref{Sec_constraints_details}. Our Figures \ref{fig_GvsK} and \ref{fig_G'vsG} show the 2D projections of the 3D $\{K,\Gamma,\Gamma'\}$ parameter space on the $K$--$\Gamma$ and $\Gamma$--$\Gamma'$ planes, respectively, with the colored region corresponding to the physical subspace of $\alpha$-RuCl$_3$ that is bounded by the constraints. In the figures, we also indicate the ranges of the physical observables responsible for the specific boundaries of this subspace, with the out-of-plane tilt angle $\alpha$ from Eq.~(\ref{eq_alphaKG}) and the field difference $\Delta H_c$ from Eqs.~(\ref{Hca}) and (\ref{Hcb}). For the ESR gap $\Delta E_g$ in Eq.~(\ref{ESRgap}), we apply the constraint directly to the $\Gamma+2\Gamma'$ combination.   

The constraints can be seen as sufficiently, if not nearly ``orthogonal,'' in the sense of being nonredundant and resulting in a closed compact physical region in the 3D parameter space. While all three constraints are essential for the physical bounds on all three parameters of the model, the rough account of their roles, summarized in Sec.~\ref{Sec_intro_anisotropic}, can be recounted here. The main role of the $\Delta H_c$ constraint can be seen in strongly tying $\Gamma'$ to $\Gamma$, with the resultant overall parameter range $0.45\!\alt\!\Gamma'/\Gamma\!\alt\!0.66$, followed by the ESR gap strongly limiting the physical values of $\Gamma$ to the 3.2~meV$\alt\!\Gamma\!\alt\! 5.0$~meV range. The tilt angle can be seen as responsible for binding $K$ to the range of 4.5~meV$\alt\!|K|\!\alt\!  10$~meV.   Obviously, these limits do not do the full justice to the picture as the constrained space is not orthorhombic,  so fixing one of the parameters would narrow the allowed ranges for the others.  

\subsubsection{Representative parameters}
\label{Sec_rep_parameters}

In Figs.~\ref{fig_GvsK} and \ref{fig_G'vsG}, one can see  three points within the colored  region. These are  projections of the representative $\{K,\Gamma,\Gamma'\}$ sets from the advocated physical parameter subspace, chosen to span it along its longer axis. 

These sets will be referred to as  Point~0, roughly at the center of the physical subspace, and Point~A and Point~B, located toward the opposite ends of it, respectively. Their $\{K,\Gamma,\Gamma'\}$ coordinates and the values of the   observables, $\alpha$, $\Gamma_{\rm tot}$, and $\Delta H_c$,  to which these parameter sets correspond,  are given by
\noindent
\begin{center}
\begin{tabular}{lllr | ccc}
     & $\{K,$ \ & $\Gamma,$   & $\Gamma'\ \ \ \ \}$   \  \ &  $\alpha$\  \ &  \ $\Gamma_{\rm tot}$ \   & $\Delta H_c$ \\ 
{Point 0:} & $\{$-7.567, & 4.276, & 2.362$\}$   \  \ & \ \ 35$\degree$\  \ & 9.0 & \ 0.8 T \\ 
{Point A:} & $\{$-5.427, & 3.647, & 2.176$\}$   \  \ &  \ \ 32$\degree$\  \ & 8.0 & \ 0.5 T \\ 
{Point B:} & $\{$-8.733, & 4.714, & 2.393$\}$   \  \ & \ \ 36$\degree$\  \ & 9.5 & \ 1.3 T \\ 
\end{tabular}
\end{center}
\noindent
All parameters are in meV unless indicated otherwise. 

The phase diagrams  for  Points~0 and A in the remaining $J$--$J_3$ parameter space of the model (\ref{eq_Hij}), such as the ones shown in Sec.~\ref{Sec_intro_anisotropic} for Point~B, will be explored in detail In Sec.~\ref{Sec_PhDs}, with their similarities and differences highlighted. The stability of the quasiclassical values of the observables to quantum effects for the Point~0  is verified by DMRG in Sec.~\ref{Sec_DMRGchecks}.

\subsubsection{Alternative space}
\label{Sec_Jpp_Jzp_space}

\begin{figure}[t]
\centering
\includegraphics[width=\linewidth]{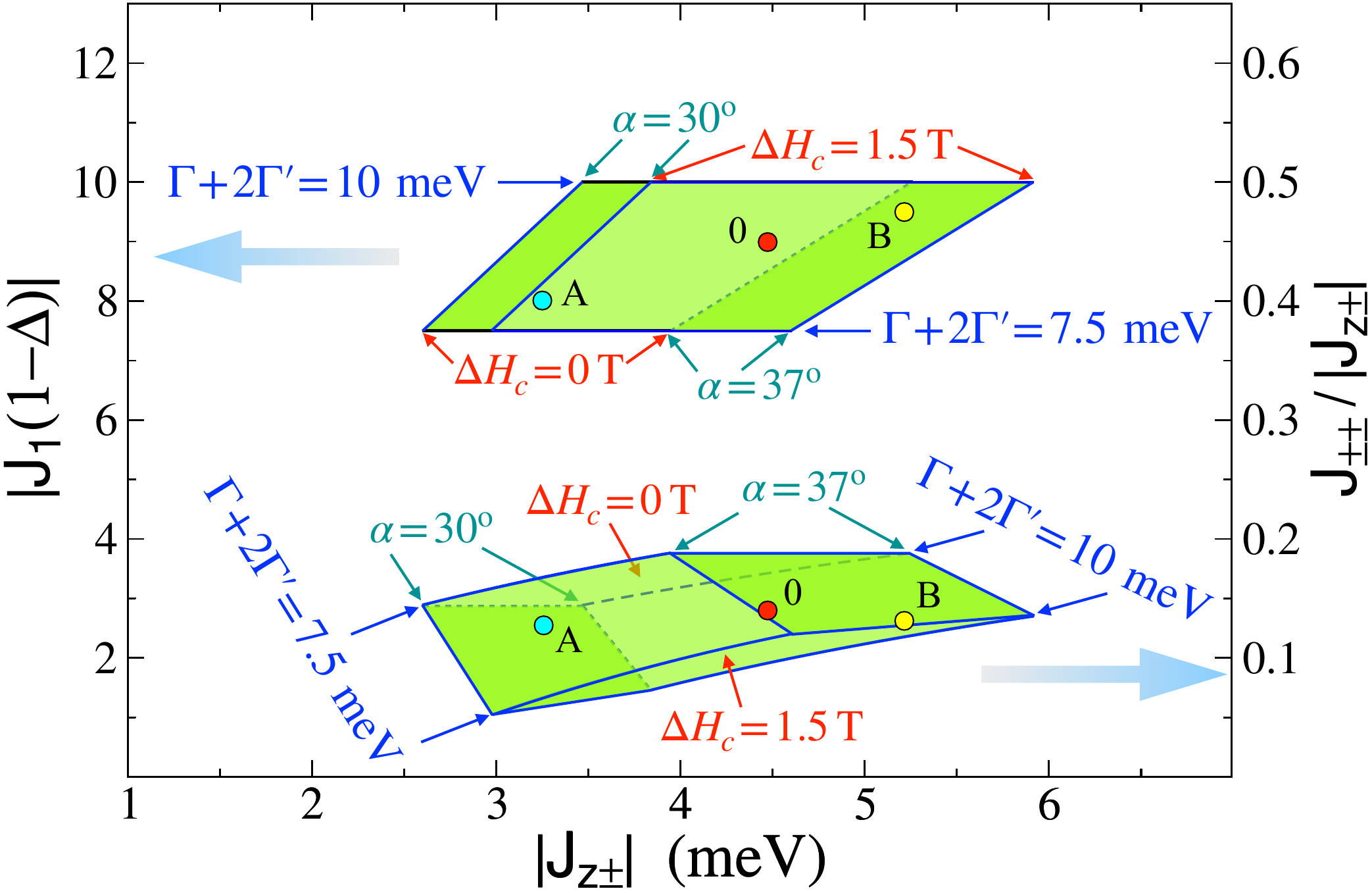}
\caption{Same as Figs.~\ref{fig_GvsK} and \ref{fig_G'vsG}  for the crystallographic parametrization of the model (\ref{HJpm}):  the ${\sf J_1}(1-\Delta)$--${\sf J}_{z\pm}$ plane (left axis) and the ${\sf J}_{\pm \pm}$--${\sf J}_{z\pm}$ plane (right axis) are shown.}
\label{fig_JppJzp}
\end{figure}

Our Figure~\ref{fig_JppJzp} presents  projections of the physical parameter space of $\alpha$-RuCl$_3$ in the alternative crystallographic parametrization of the $XXZ$--${\sf J_{\pm\pm}}$--${\sf J_{z\pm}}$ model (\ref{HJpm}). The corresponding anisotropic parameter space is also three-dimensional, with the  $\{ {\sf J_1}(1-\Delta),{\sf J}_{\pm \pm},{\sf J}_{z\pm}\}$ coordinates.  Fig.~\ref{fig_JppJzp} shows projections of this 3D space on the ${\sf J_1}(1-\Delta)$--${\sf J}_{z\pm}$ and ${\sf J}_{\pm \pm}$--${\sf J}_{z\pm}$ planes. 

Given the discussion in Sec.~\ref{Sec_constraints_details}, some of the  phenomenological constraints proposed in this work have much more direct and instructive relation to the parameters of the model in the crystallographic reference frame. For the ESR gap, Eq.~(\ref{ESRgap}), the $XXZ$ anisotropy parameter, ${\sf J_1}(1-\Delta)$, is constrained directly, hence the horizontal boundaries of the respective region in Fig.~\ref{fig_JppJzp}. Since the tilt angle, Eq.~(\ref{eq_alpha}), is induced by the coupling of the in-plane to the out-of-plane spin components from the ${\sf J}_{z\pm}$ term, its  physical range is largely dictated by the constraints on the angle.  In turn, while somewhat more complicated, the critical field difference $\Delta H_c$ is responsible for the bounds on ${\sf J}_{\pm \pm}$. 

An  important insight provided by the crystallographic  form of the model  is the clear hierarchy in the magnitude of the physical parameters  in this alternative space, which is seen in the representative  Points~0, A, and B,
\noindent
\begin{center}
\begin{tabular}{lllr | ccc}
     & $\{{\sf J_1}(1-\Delta),$ & ${\sf J}_{\pm \pm},$   & ${\sf J}_{z\pm}\ \ \ \}$   \ &  $\alpha$\  & \ $\Gamma_{\rm tot}$  & $\Delta H_c$ \\ 
{Point 0:} & $\{$-9.0, & 0.623, & -4.469$\}$    \ & \  35$\degree$\   & 9.0 & \ 0.8 T \\ 
{Point A:} & $\{$-8.0, & 0.414, & -3.252$\}$    \ &   \ 32$\degree$\  & 8.0 & \ 0.5 T \\ 
{Point B:} & $\{$-9.5, & 0.682, & -5.211$\}$    \ &  \ 36$\degree$\   & 9.5 & \ 1.3 T \\ 
\end{tabular}
\end{center}
\noindent
Specifically, the largest is the (negative) ${\sf J_1}(1-\Delta)$, followed by ${\sf J}_{z\pm}$, see Fig.~\ref{fig_JppJzp}. On the other hand, the physical range of ${\sf J}_{\pm \pm}$ is nearly an order of magnitude smaller than that of the leading terms. While its small positive values will be important for the details of the incommensurate phases in the phase diagrams discussed in Sec.~\ref{Sec_PhDs}, one expects a secondary role of this term to that of ${\sf J}_{z\pm}$~\cite{Zhu19}.  

\begin{figure}[t]
\centering
\includegraphics[width=\linewidth]{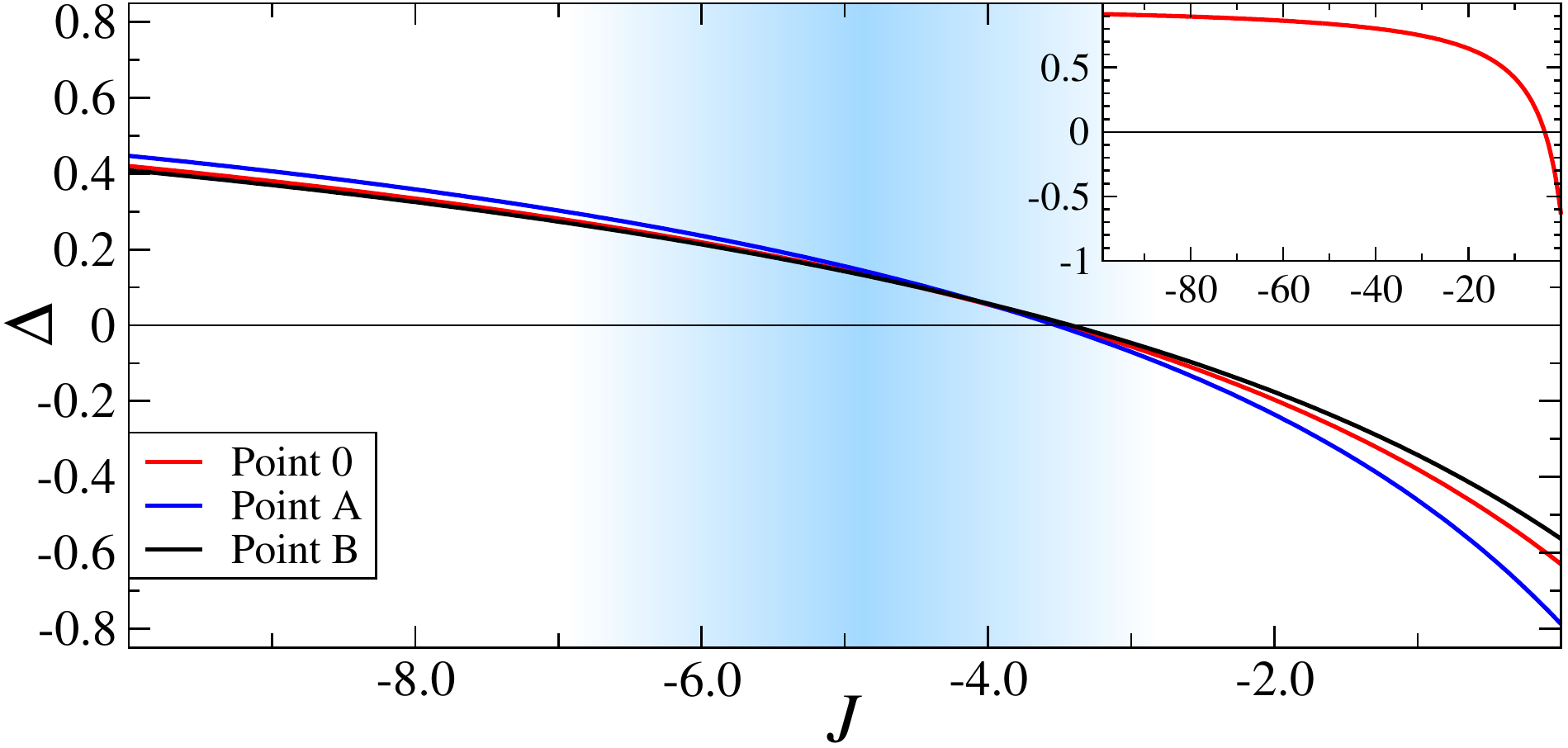}
\caption{The $XXZ$  $\Delta$ of the model (\ref{HJpm}) vs  $J$ of the KH model (\ref{H_JKGGp}) for Points 0, A, and B. Inset:  Point 0 in a larger range.}
\label{fig_DeltavsJ}
\end{figure}

A less obvious, but   arguably more dramatic element of this hierarchy is the strongly easy-plane character of the model (\ref{HJpm}) in the physical space of $\alpha$-RuCl$_3$. Since ${\sf J_1}(1-\Delta)$ is the largest parameter, it already implies that the $XXZ$ term cannot be close to the Heisenberg limit ($\Delta\!=\!1$) for any physically reasonable values of ${\sf J_1}$. 

In order to determine a plausible range of $\Delta$, one can rely on the broad expectations for the isotropic term $J$ in the KH representation to be negative and not to exceed $|K|$~\cite{Keimer20,winter16,rethinking}. Using  relations between the crystallographic and KH frames in Appendix~\ref{app_A} and taking $|K|\!\approx\!\Gamma_{\rm tot}$ for the $\alpha$-RuCl$_3$ range of parameters from Sec.~\ref{Sec_parameter_space}, one can find that $\Delta$ should vary from about $-0.5$ at $J\!=\!0$ to 1  for  $J\!\rightarrow\!-\infty$, crossing zero at $J\!\approx\!-\Gamma_{\rm tot}/3$. These trends are verified for the Points 0, A, and B in Fig.~\ref{fig_DeltavsJ}. One can see that for the values of $J$  relevant to $\alpha$-RuCl$_3$ (shaded region), the $XXZ$ anisotropy parameter  $\Delta$ varies between -0.1 and 0.3.

Having ${\sf J}_{\pm \pm}$ and $\Delta$ near zero  suggests a simpler and more intuitive description of $\alpha$-RuCl$_3$ by the easy-plane ${\sf J^{\it XY}_1}$--$J_3$ FM-AFM model with strong bond-dependent  ${\sf J_{z\pm}}$  term, the perspective   further explored below.

\subsubsection{Comparison with Ref.~\cite{Giniyat22}}
\label{Sec_comparison_Giniyat}

\begin{figure}[t]
\centering
\includegraphics[width=\linewidth]{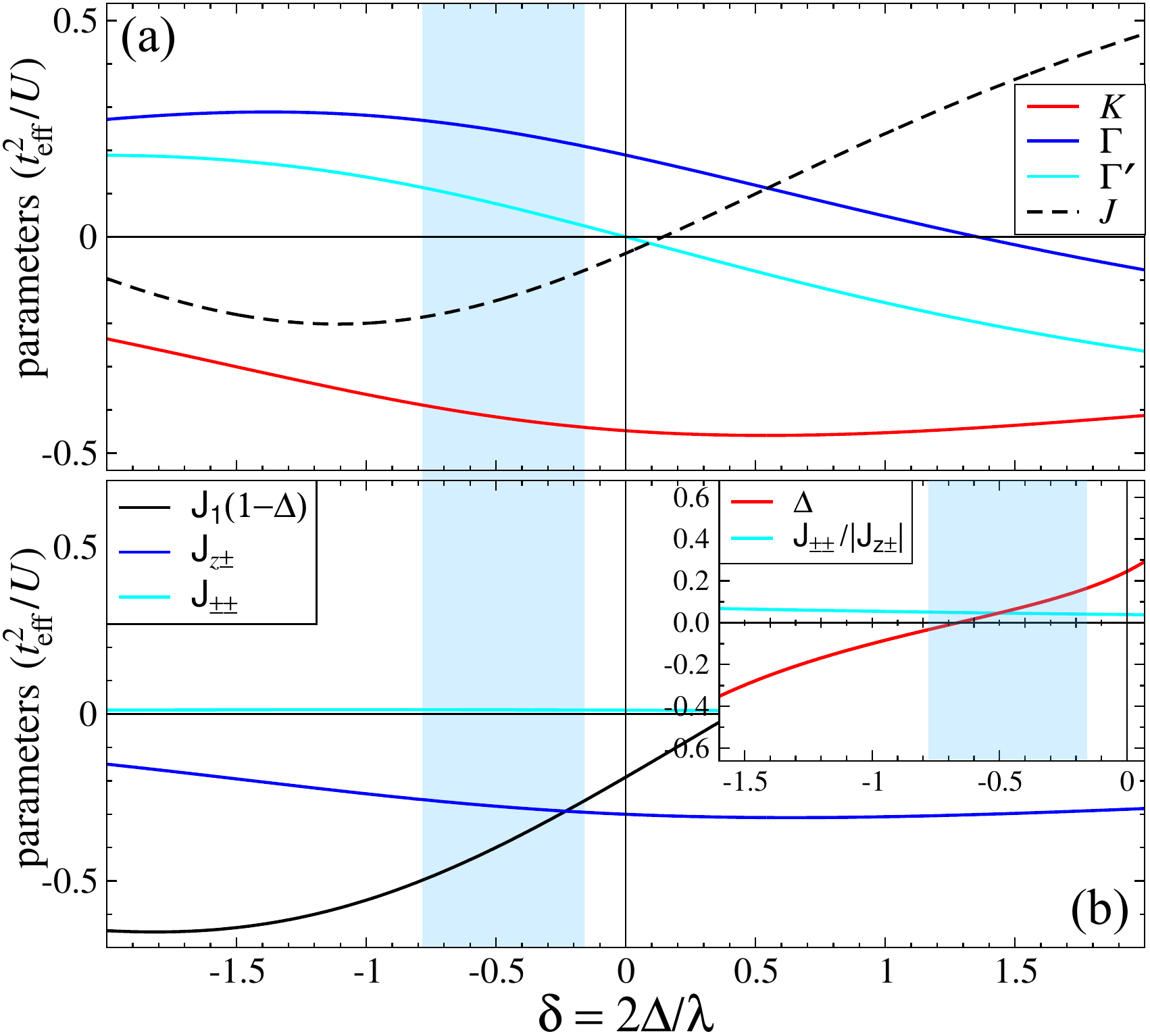}
\caption{(a) $J,K,\Gamma,$ and $\Gamma'$ from Ref.~\cite{Giniyat22},  $\delta$ is the octahedral distortion, shaded region is relevant to $\alpha$-RuCl$_3$. (b)  and inset: same data in the crystallographic parametrization (\ref{HJpm}).}
\label{fig_Giiyat}
\end{figure}

Here we  discuss in more detail recent results of Ref.~\cite{Giniyat22}, which has obtained nearest-neighbor exchanges for the $\alpha$-RuCl$_3$  model from a  completely different viewpoint from the phenomenological approach of our work. Their systematic perturbative derivation of the KH exchanges using microscopic treatment of the orbital model in the spirit of the original work by Jackeli and Khaliullin~\cite{Jackeli} has included  the most relevant orbitals and their hoppings together with the octahedral distortion of the ligand environment. Their results are reproduced in our Fig.~\ref{fig_Giiyat}(a). The horizontal axis is the dimensionless parametrization of the octahedral distortion and the shaded region marks the range relevant to $\alpha$-RuCl$_3$.

One can see a close accord of the hierarchy of  the  $\{K,\Gamma,\Gamma'\}$ terms in the shaded region of Fig.~\ref{fig_Giiyat}(a) with our suggested  ranges  in Sec.~\ref{Sec_rep_parameters}:  the leading $K$-term, $\Gamma$ close next, and, specifically, positive and sizable $\Gamma'$ term, originally proposed in Ref.~\cite{rethinking}. Here and in Fig.~\ref{fig_Giiyat}(b), $t^2_{\rm eff}/U$ is the overall energy scale of the model parameters in this perturbative approach, see Ref.~\cite{Giniyat22}.

However, the most notable results are shown in Fig.~\ref{fig_Giiyat}(b). It is the same data from Ref.~\cite{Giniyat22}, but in the crystallographic parametrization of the $XXZ$--${\sf J_{\pm\pm}}$--${\sf J_{z\pm}}$ model (\ref{HJpm}). One can see all the aspects of the physical parameter space of $\alpha$-RuCl$_3$ that are discussed in Sec.~\ref{Sec_Jpp_Jzp_space} above: leading ${\sf J_1}(1-\Delta)$ that is followed by ${\sf J}_{z\pm}$, a subleading ${\sf J}_{\pm \pm}$ for all distortion values, with its ratio of ${\sf J}_{\pm \pm}/|{\sf J}_{z\pm}|\!\approx\!0.1$ matching nicely the physical space in Fig.~\ref{fig_JppJzp}, and, most importantly, the value of $\Delta$ varying from -0.1 to 0.2 in the relevant region of Fig.~\ref{fig_Giiyat}(b).  

Needless to say, such a close agreement on the qualitative and quantitative trends of the two very different approaches to the same problem is noteworthy. It also   provides a sign of the promising convergence on the physical parameter space of $\alpha$-RuCl$_3$  for the effective $KJ\Gamma\Gamma'$--$J_3$ model.

\begin{table*}[t]
  \begin{tabular}{| l | c || c | c | c | c | c | c |}
    \hline
\text{Reference} 
& \text{Method} 
& $K$ $(<0)$ & $\Gamma$ $(>0)$ & $\Gamma'$ & $\alpha$ ($\degree$)  & $\Gamma\!+\!2\Gamma'$ & $\Delta H_c$ (T) \\ \hline\hline
\multirow{3}{2.5cm}{Kim et al. \cite{kee16}} 
& DFT+$t/U$, $P3$  
& {-6.55} & {5.25} & -0.95                            & {\bf 36.6}  &      3.35  &       9.64      \\ \cline{2-8}
& DFT+SOC+$t/U$ 
& {-8.21} & {4.16} & -0.93                             & 40.9            & 2.3        &        7.03              \\ \cline{2-8}
& same+fixed lattice 
& -3.55                &  7.08         & -0.54              & 28.4         & 6.01  &      14.4           \\ \cline{2-8}
\hline   
Winter et al. \cite{winter16}
& DFT+ED, $C2$   
& {-6.67} & { 6.6}   & -0.87                              & {\bf 34.4 }           & 4.87        &     12.2       \\ \cline{2-8}
\hline
&  DFT+$t/U$, $U\!=\!2.5$eV  
& { -14.4}     & { 6.43} &                             & 41.1                & 6.43     &     7.96  \\ \cline{2-8}
Hou et al. \cite{gong17} 
& same, $U\!=\!3.0$eV 
& { -12.2}     & { 4.83} &                         & 42.2     &          4.83        &   5.74    \\ \cline{2-8}
& same, $U\!=\!3.5$eV 
& { -10.7}     & { 3.8}   &                         & 43.2      &            3.8        &    4.36     \\ \hline
\multirow{2}{2.5cm}{Wang et al. \cite{li17}} 
& DFT+$t/U$, $P3$  
& { -10.9}       & { 6.1}   &                           & 38.9     &            6.1         &   8.15  \\ \cline{2-8}
& same, $C2$       
& {-5.5}   & {7.6}          &                             & {\bf 30.2}             &        {\bf  7.6}         &      13.3      \\ \hline
Eichstaedt\! et\! al. \cite{berlijn19} 
& DFT+Wannier+$t/U$ 
& {-14.3}            & 9.8            &      -2.23         &     38.3             &       5.33          &           18.1              \\ \hline
Ran et al. \cite{wen17} & LSWT, INS fit 
& {-6.8}    &            9.5 &                              &     {\bf 30.1}         &  {\bf 9.5}            &     16.6      \\ \hline
Winter et al. \cite{winter17} 
& \textit{Ab initio}+INS fit 
& {-5.0}   &            2.5  &                  &     40.0              &            2.5         &     3.22      \\ \hline
Suzuki et al. \cite{suga18} 
& ED, $C_p$ fit 
& -24.4          & { 5.25} &     -0.95 &        47.3        &           3.35         &    6.76  \\ \hline
Cookmeyer\! et\! al. \cite{moore18} 
& thermal Hall fit 
& {-5.0}   &            2.5 &                    &             40.0    &             2.5       &     3.22  \\ \hline
 Wu et al. \cite{orenstein18} 
& LSWT, THz fit 
& -2.8                  &            2.4 &                    &           {\bf 34.6}   &              2.4    &     3.68           \\ \hline
Ozel et al. \cite{gedik19}
& same 
& -3.5                  & 2.35          &                    &           {\bf  37.0}        &    2.35       &        3.34                \\ \hline
Sahasrabudhe et al. \cite{kaib19} & ED, Raman fit & 
{-10.0}              & 3.75             &                       & 42.7                & 3.75  &  4.38           \\ \hline
\multirow{2}{2.5cm}{Sears et al. \cite{kim19}}
& \multirow{2}{2.9cm}{Magnetization fit}
& {-10.0}           & 10.6             & -0.9           &   {\bf 33.4}        & {\bf 8.8}         &     19.0           \\  \cline{3-8}
&& {-10.0}     & 8.8          &                   &    {\bf  34.3}             & {\bf 8.8}           &      13.6          \\ \hline
Laurell et al. \cite{okamoto19} 
& ED, $C_p$ fit 
& {-15.1}           & 10.1          & -0.12        &        {\bf 37.2}            & {\bf 9.86}         &      14.6         \\ \hline
Suzuki  et al. \cite{Keimer20} 
& RIXS 
& -5.0       & 2.5           & +0.1                  &     39.8                     & 2.7         &         3.03                \\ \hline
Kaib et al. \cite{Kaib20}
& GGA+U 
& -10.1   & 9.35        & -0.73                   &   {\bf  34.5}               & {\bf 7.89}             &       16.0            \\ \hline
Andrade et al. \cite{Andrade20}
& $\chi$ 
& -6.6        & 6.6           &                   &    {\bf 33.1}               & 6.6         &           10.6                 \\ \hline
Janssen et al. \cite{Janssen20}
& LSWT+3D     
& -10.0      & 5.0           &                    &             40.0               & 5.0            &          6.43             \\ \hline
Li et al. \cite{Li21}
& $C_m$, $\chi$
&-25.0         & 7.5          & -0.5            &       44.8                & 6.5         &            9.03                 \\ \hline
Ran et al. \cite{Ran22}
& polarized INS
& -7.2           & 5.6          &                    &         {\bf 35.6}               & 5.6         &           8.33                      \\ \hline
Samarakoon et al. \cite{Samarakoon22} \  \
&\  Machine learning, INS \ 
& -5.3             &  0.15      &                   &   {\bf 36.4}          &   0.15           &        {\bf  0.11}      \\ \hline
Liu et al. \cite{Giniyat22}
& downfolding
& -5.0            &  2.8        &         +0.7       &   {\bf 37.3}             &       4.2      &      2.37               \\ \hline
    \hline  
\multirow{4}{2.5cm}{ \ \ \ \ \  \bf This work}  
&  {\bf realistic range}  \ \ 
&\ {\bf [-10.0,-4.4]} \ &\  {\bf [3.2,5.0]} \   &\  {\bf [1.8,2.85]}\  & \ {\bf [30.0,37.0]} \ & \ {\bf [7.5,10.0]}\ & \ {\bf [0.0,1.5]} \  \\ \cline{2-8}
&  {\bf point 0}   & {\bf -7.57}  & {\bf 4.28} & {\bf 2.36}   & {\bf 35.0}  &  {\bf 9.0} & {\bf 0.8}  \\ \cline{2-8}
& {\bf point A}    & {\bf -5.43} & {\bf 3.65}  & {\bf 2.18}   & {\bf 32.0}  & {\bf 8.0}    & {\bf 0.5}  \\ \cline{2-8}
&   {\bf point B}  & {\bf  -8.73}  & {\bf 4.71}   & {\bf 2.39} & {\bf 36.0}  & {\bf 9.5} & {\bf 1.3}   \\ \hline
    \hline    
\end{tabular}
\caption{The proposed sets of $\{K,\Gamma,\Gamma'\}$ parameters for $\alpha$-RuCl$_3$ and values of the  angle $\alpha$,  $\Gamma+2\Gamma'$, and $\Delta H_c$  that follow from them  according to Eqs.~(\ref{eq_alphaKG}),  (\ref{ESRgap}), and (\ref{Hca}) and (\ref{Hcb}), respectively. For all the results, only the nearest-neighbor anisotropic exchanges from the proposed parameter sets were used. The acronyms are linear spin-wave theory (LSWT), spin-orbit coupling (SOC), inelastic neutron scattering (INS), resonant inelastic x-ray scattering (RIXS); ``$P3$'' and ``$C2$'' refer to the lattice symmetry.  Representative sets of Points 0, A, and B, and proposed ranges are from Sec.~\ref{Sec_rep_parameters}, see the text.}
\label{table3}
\end{table*}

\subsubsection{Compilation of prior results}
\label{Sec_compilation}

Our Table~\ref{table3} presents a comprehensive  compilation of the previously proposed $\{K,\Gamma,\Gamma'\}$ parameter sets of the generalized KH model (\ref{eq_Hij}) for $\alpha$-RuCl$_3$, which are based on  density-functional theory (DFT)~\cite{kee16,winter16,gong17,li17,berlijn19} and various phenomenological, analytical, and numerical analyses~\cite{wen17,winter17,suga18,moore18,orenstein18,gedik19,kaib19,kim19,okamoto19,%
Keimer20,Kaib20,Andrade20,Janssen20,Li21,Ran22,Samarakoon22,Giniyat22}. Entries with the positive $K$ values, which contradict broad phenomenologies~\cite{Cao16,kim19}, are omitted. The reference is in the first and the abbreviated details of the  approach are in the second column; all parameters are in meV unless indicated otherwise. For the few cases of the  lower symmetry of the model, the bond-averaged values of the exchange parameters were used.  

The last three columns of the Table~\ref{table3} list the values of the key physical observables, tilt angle $\alpha$, the $\Gamma+2\Gamma'$ combination for the ESR gap $\Delta E_g$, and critical field difference $\Delta H_c$,  which  follow from the proposed $\{K,\Gamma,\Gamma'\}$ sets according to Eqs.~(\ref{eq_alphaKG}),  (\ref{ESRgap}), and (\ref{Hca}) and (\ref{Hcb}), respectively. These values are highlighted in bold in cases where they fall within, or come close to the suggested realistic ranges. The latter are listed at the bottom of the table together with the physical bounds for $\{K,\Gamma,\Gamma'\}$ from  Figs.~\ref{fig_GvsK} and \ref{fig_G'vsG} and   representative sets of Point 0, A, and B discussed above, see  Sec.~\ref{Sec_constraints_details} and Sec.~\ref{Sec_parameter_space}. 

\begin{figure}[t]
\centering
\includegraphics[width=\linewidth]{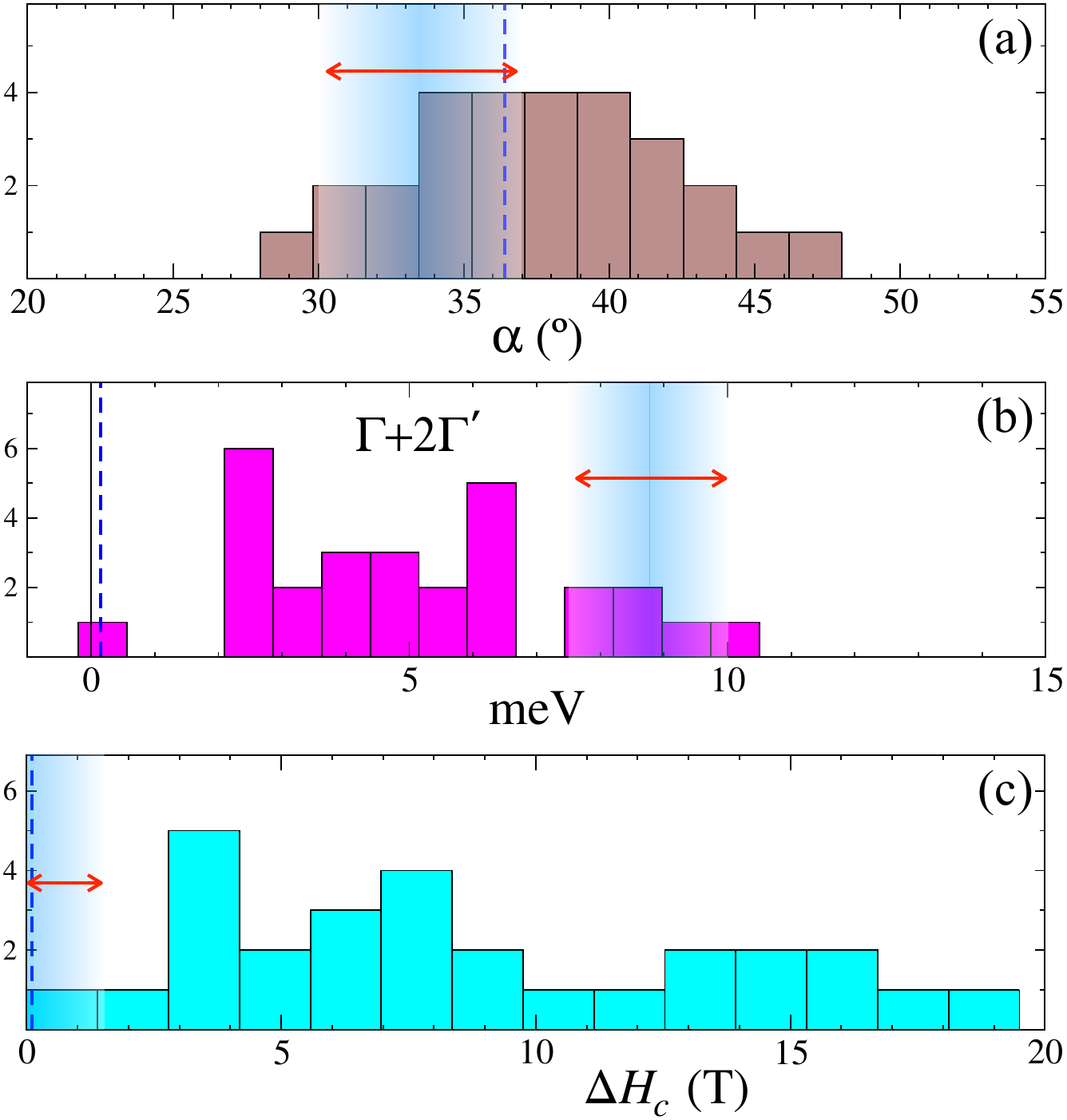}
\caption{Histograms of the  observables from Table~\ref{table3}, excluding present work.  Shaded regions are physical limits (see text) and  vertical dashed are for the data from  Ref.~\cite{Samarakoon22}.}
\label{fig_hist_exp}
\end{figure}

We note here again that the $KJ\Gamma\Gamma'$--$J_3$ model considered in this work is an effective one, with the nearest-neighbor exchange matrix encapsulating {\it all} anisotropic terms, while some of the microscopic studies listed in Table~\ref{table3} also allow such terms in the more extended exchanges~\cite{winter16,gong17,li17,berlijn19,wen17,winter17}. This may account for the lack of the significant nearest-neighbor $\Gamma'$ term in such models and lead to a better agreement with the proposed phenomenological constraints if the additional exchanges are included.

\begin{figure}[t]
\centering
\includegraphics[width=\linewidth]{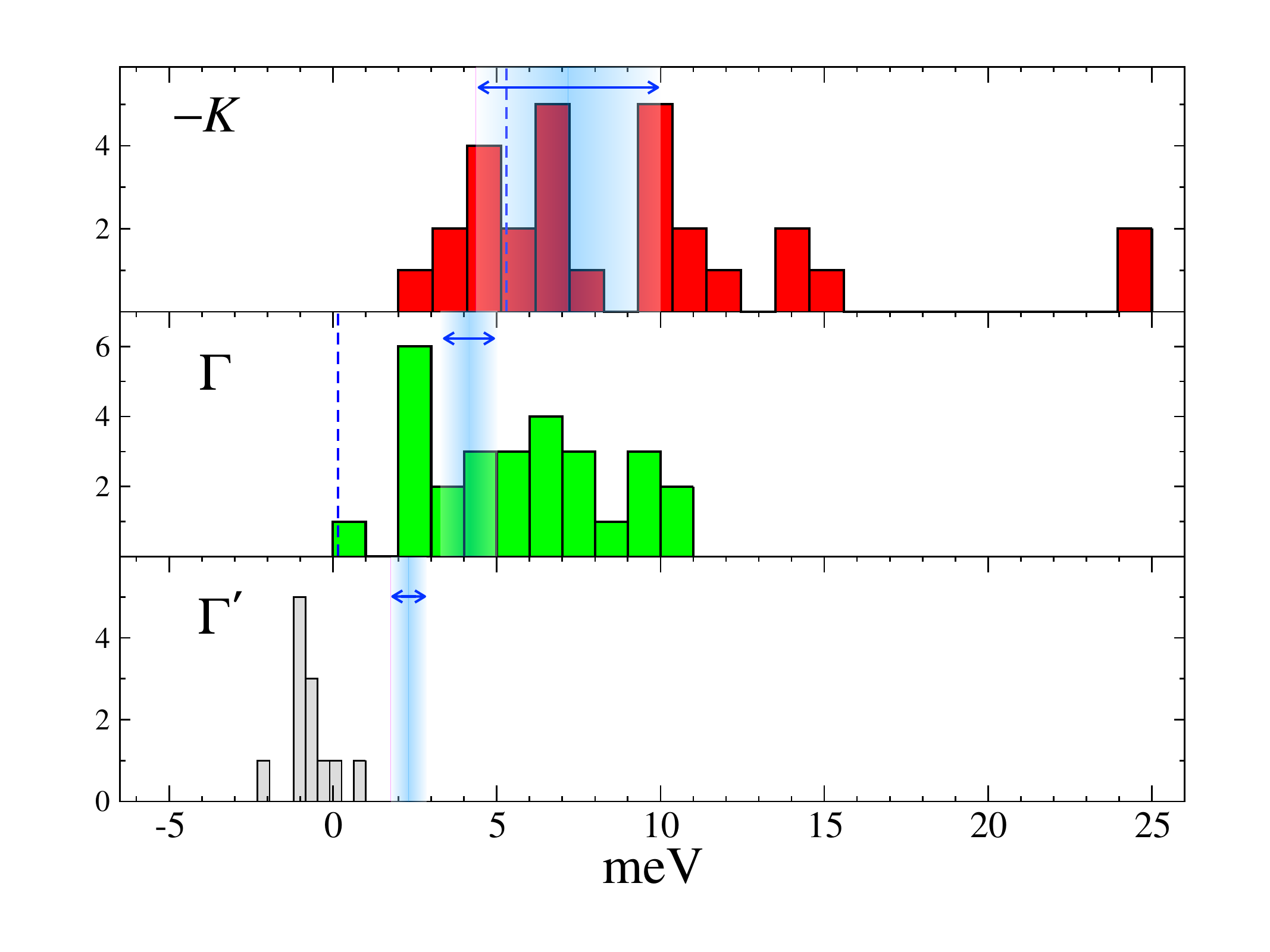}
\caption{Same as in Fig.~\ref{fig_hist_exp} for $\{K,\Gamma,\Gamma'\}$  from Table~\ref{table3}.}
\label{fig_histKGGp}
\end{figure}

\begin{figure}[t]
\centering
\includegraphics[width=\linewidth]{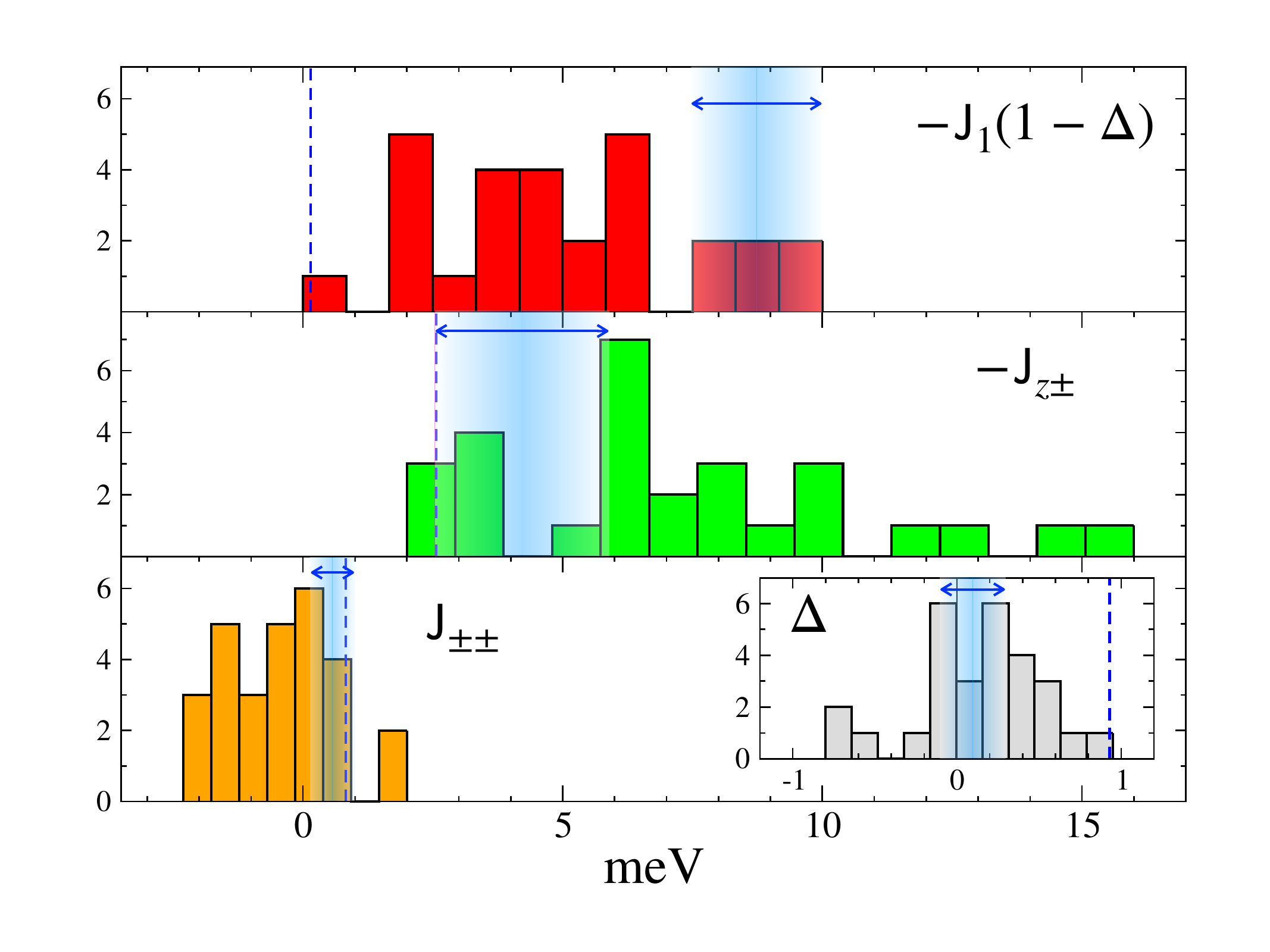}
\caption{Same as in Fig.~\ref{fig_histKGGp} for the $\{ {\sf J_1}(1-\Delta),{\sf J}_{\pm \pm},{\sf J}_{z\pm}\}$ parameters of the model (\ref{eq_Hij}) in the crystallographic parametrization (\ref{HJpm}), see also  Table~\ref{table2} in Appendix~\ref{app_A}.}
\label{fig_histJzpJpp}
\end{figure}

To provide a  graphical presentation of the variation of the entries in the data compilation of Table~\ref{table3} (28 entries, excluding our own), we  use the histograms  in Figures~\ref{fig_hist_exp}, \ref{fig_histKGGp}, and \ref{fig_histJzpJpp} to show the wide distribution of the prior attempts at the model parameters, highlight the  physical limits  on the observables proposed in our work, and emphasize the resultant effect of the constraints on the possible ranges of the model parameters. The physical limits discussed above are shown by shaded regions and  vertical dashed lines correspond to a representative set of this compilation from the second to last entry,  Ref.~\cite{Samarakoon22}, elaborated on below in more detail. 

As one can see in both Table~\ref{table3} and Fig.~\ref{fig_hist_exp}(a), about half of the entries correspond to the tilt angle range close to the physical one. However, for the ESR gap in Fig.~\ref{fig_hist_exp}(b) and in the second to last column of Table~\ref{table3}, only about 20\% of them hit within the physical bounds. For the last column and Fig.~\ref{fig_hist_exp}(c), only one parameter set matches the range of the empirical critical field difference, and, arguably, for a wrong reason. 

This set is from Ref.~\cite{Samarakoon22} that extracted effective model parameters of $\alpha$-RuCl$_3$ from the INS data using machine learning. It has neglected $\Gamma'$ and their resultant parameter set has near-zero $\Gamma$.  Because of that,  this  set fails the ESR gap criterion, which requires a large $\Gamma+2\Gamma'$ combination, while satisfying the remaining two of the phenomenological constraints put forward in our work.  Its ability to match $\Delta H_c$ range is because the pure $K$--$J$ model always yields $\Delta H_c\!\equiv\!0$, see  Eqs.~(\ref{Hca}) and (\ref{Hcb}).  

One can argue that a redo of the same analysis with no artificial restrictions in the model and with  additional experimental input, which can improve  the lack of orthogonality in their phenomenological constraints, may open the full potential of this approach and lead to a convergence of our parameter sets. 
 
The rest of the entries in Table~\ref{table3} that match the physical ranges of both   tilt angle {\it and}   ESR gap,  fail at the remaining criterion, all yielding  $\Delta H_c$  in excess of 13~T. 

Lastly, the orbital model expansion study of Ref.~\cite{Giniyat22}, highlighted  in Sec.~\ref{Sec_comparison_Giniyat} as showing great similarity to our results, suggested a representative set (the last entry in Table~\ref{table3}) that  has likely suffered from the choices of the smaller overall  $t^2/U$ scale for their parameters and  a somewhat conservative value of the octahedral distortion $\delta$, leading to a positive, but insufficiently large $\Gamma'$. A reasonable adjustment in both should produce  a set not far from the ones proposed in this work.

Following our discussions in Sec.~\ref{Sec_parameter_space}, the proposed constraints on the physical observables lead to  important bounds on the model parameters. In Fig.~\ref{fig_histKGGp}, we highlight these bounds with the backdrop of the distributions of  $K$, $\Gamma$, and $\Gamma'$  from Table~\ref{table3}. Needless to say, the phenomenological constraints lead to a significantly narrower physical parameter space, also visualized  in Figs.~\ref{fig_GvsK} and \ref{fig_G'vsG}. Our Fig.~\ref{fig_histKGGp} reinforces the statements that are already made above. For an adequate description of the $\alpha$-RuCl$_3$ phenomenology  within the effective model,  a positive $\Gamma'$-term is unavoidable,  a significant  $\Gamma$-term is necessary, and fairly tight limits on all three anisotropic exchanges can be established.

In Fig.~\ref{fig_histJzpJpp}, the same parameter sets from Table~\ref{table3} are recast into the alternative crystallographic parametrization, with the actual values of the $\{ {\sf J_1}(1-\Delta),{\sf J}_{\pm \pm},{\sf J}_{z\pm}\}$ parameters of the model  (\ref{HJpm}) given in Table~\ref{table2} in Appendix~\ref{app_A}.  Although  not listed in Table~\ref{table3}, the values of the $J$-term of the KH model (\ref{H_JKGGp}) were used to obtain the histogram of the $XXZ$ anisotropy $\Delta$, see inset.

Aside from demonstrating the effect of phenomenological constraints that lead to a significantly narrower physical parameter space in these alternative axes, the crystallographic language underscores  important common trends in most of the prior attempts  at the $\alpha$-RuCl$_3$ parameters. Whether they match the proposed physical parameter space or not, the majority of them respects the  hierarchy of terms proposed in this work: the two  leading ones, ${\sf J_1}(1-\Delta)$ and ${\sf J}_{z\pm}$, and a subleading ${\sf J}_{\pm \pm}$. 

Especially notable is another trend.  As is laid plain in  Fig.~\ref{fig_histJzpJpp},  {\it all} prior works with no exception suggest an easy-plane character of the ferromagnetic  nearest-neighbor exchange in their $\alpha$-RuCl$_3$ modeling, showing $|\Delta|\!<\!1$. In  retrospect, the strongly easy-plane character of the model (\ref{HJpm}) is one of the most direct arguments that the  parameter space of $\alpha$-RuCl$_3$ is far away from the Kitaev limit. 

In a sense, the crystallographic representation of the model offers a lens  that allows one to see the commonality of all earlier assessments of $\alpha$-RuCl$_3$ phenomenologies, suggesting a broader agreement between them as all pointing in a similar direction.

\subsection{Summary on physical parameters}
\label{Sec_parameters_summary}

In this Section, we have demonstrated  that the  available  phenomenologies should allow one to overcome the uncertainty in the model parameters for $\alpha$-RuCl$_3$ using  observables that are induced by the  anisotropic terms, see Sec.~\ref{Sec_constraints_details}. The  outline of the advocated parameter space and selected  representative choices of the anisotropic exchanges, which will be used in the next Section, have been proposed, see Sec.~\ref{Sec_rep_parameters}. A systematic analysis of the prior attempts at the $\alpha$-RuCl$_3$ parameters in the context of the proposed  phenomenologies has been provided, see Secs.~\ref{Sec_comparison_Giniyat} and \ref{Sec_compilation}. 

Important physical insights and intuition associated with the alternative crystallographic reference frame parametrization have been highlighted in Secs.~\ref{Sec_Jpp_Jzp_space} and \ref{Sec_compilation}. They will be expanded upon below within the discussion of the phase diagrams relevant to $\alpha$-RuCl$_3$ parameter space, also offering a  connection to the broader class of  paradigmatic  models in frustrated magnetism, thus providing a wider context to the studies of $\alpha$-RuCl$_3$.
 
\section{Phase diagrams}
\label{Sec_PhDs}
 
The purpose of the next two Sections is twofold. The first is to explore the remaining parameter space  of the model (\ref{eq_Hij}) for the representative  anisotropic parameter sets suggested above in order to pinpoint the ultimate parameter region for $\alpha$-RuCl$_3$ and to identify proximate phases that can be relevant to its properties. The  second is to provide a better understanding of such  phases and to characterize them with the help of numerical methods. We also use the opportunity to highlight the close agreement of various approaches to the derivation of these phase diagrams and to demonstrate that our anisotropic strategy outlined in Sec.~\ref{Sec_intro_anisotropic} allows one to reign in on the otherwise prohibitively costly numerical exploration of the $\alpha$-RuCl$_3$ model parameter space.
 
\subsection{$J$--$J_3$  phase diagrams of the KH model (\ref{eq_Hij})}
\label{Sec_cartesianPD}
 
With the logic for fixing   anisotropic exchanges articulated in Sec.~\ref{Sec_intro_anisotropic}  and executed in Sec.~\ref{Sec_constraints} above, the remaining two parameters of the effective model  (\ref{eq_Hij}) of $\alpha$-RuCl$_3$  in the generalized KH representation are the isotropic $J$ and $J_3$ exchanges. As  highlighted in Sec.~\ref{Sec_intro_anisotropic} for the representative set of $\{K,\Gamma,\Gamma'\}$ for Point~B from Sec.~\ref{Sec_rep_parameters} and Table~\ref{table3}, the relevant region of the parameter space resides in the $J\!<\!0$ and $J_3\!>\!0$ quadrant of the  2D $J$--$J_3$ phase diagram of the model,  conforming to the general expectations~\cite{winter16,winter17,okamoto19,rethinking}. 
 
\begin{figure*}
\centering
\includegraphics[width=1.0\linewidth]{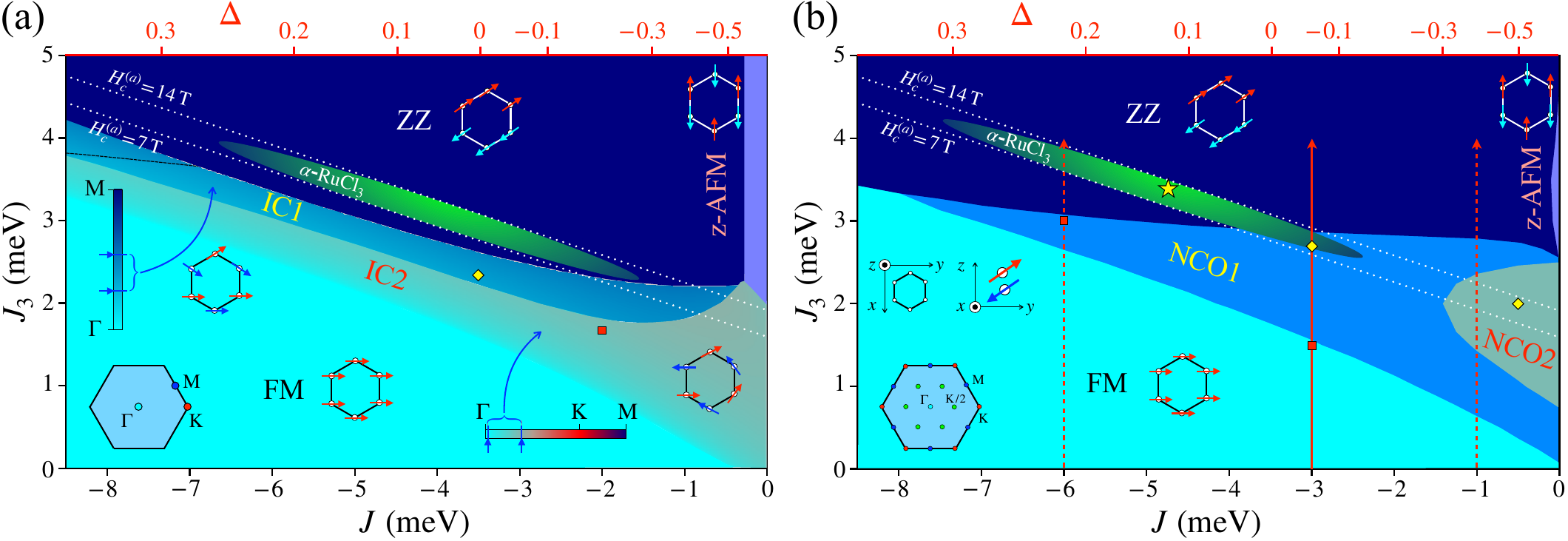}
\caption{The $J$--$J_3$ phase diagram of the model (\ref{eq_Hij}) for the Point~0 parameter set, see Sec.~\ref{Sec_rep_parameters} and Table~\ref{table3},  in the  $J\!<\!0$ and $J_3\!>\!0$ quadrant: (a)  by the LT and (b)  by ED. The ZZ, FM, AFM, IC, and  NCO  phases are designated and sketched using the $xy$ plane for the lattice and the $yz$ plane for the spin orientation. The upper horizontal axis is the $XXZ$ parameter $\Delta$ in the crystallographic axes (\ref{HJpm}).  Slanted dashed lines mark the boundaries of the parameter region for $\alpha$-RuCl$_3$ from the constraints on $H_c^{(a)}$ from Eq.~(\ref{Hca}), the star symbol in (b) is a representative point from that region, see Sec.~\ref{Sec_cartesianPD_aRuCl3}. Vertical solid and dashed arrows in (b) show the extent of the DMRG scans discussed in Sec.~\ref{Sec_helices}.  Squares and diamonds are representative points in the IC phases for (a) LT and (b) DMRG non-scans, see Sec.~\ref{Sec_helices}.  Insets: (a) color-coded bars for the ordering vector in the IC phases and the first BZ with the high-symmetry points,  (b) allowed momenta of the ED cluster in the first BZ.}
\label{fig_phaseDJ1J3_0}
\end{figure*}

\subsubsection{Phase diagrams by LT and ED methods}
\label{Sec_cartesianPD_methods}

\begin{figure*}[t]
\includegraphics[width=1.0\linewidth]{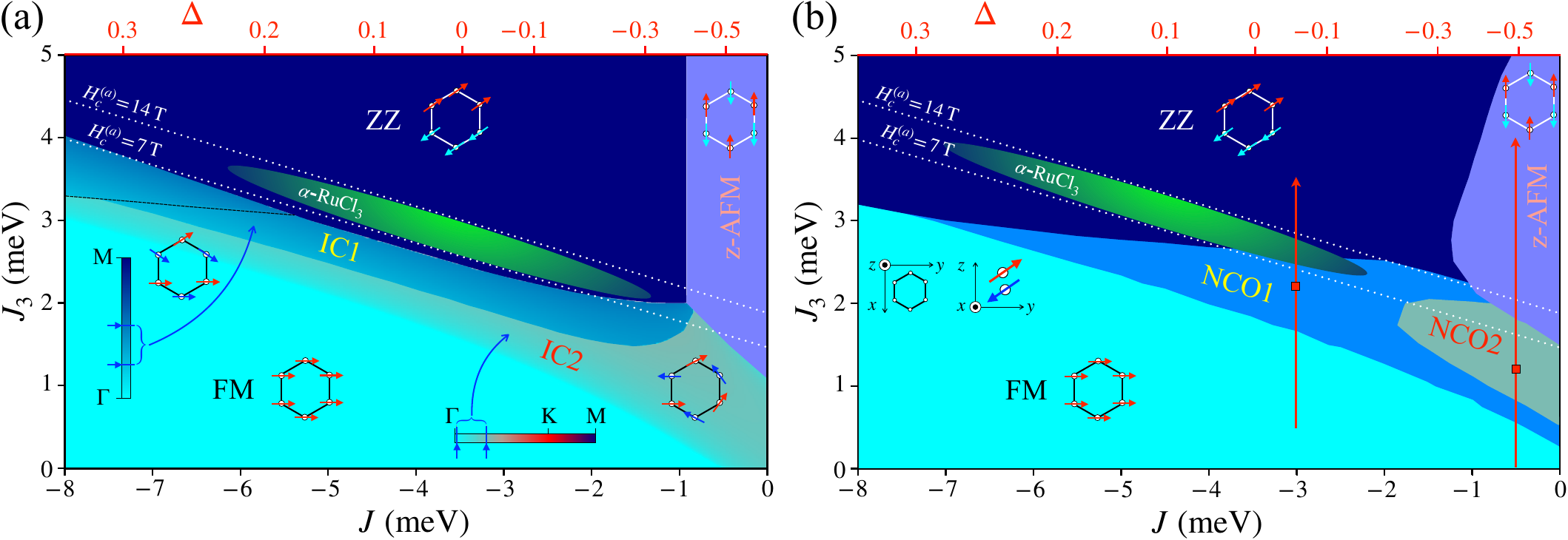}
\vskip -0.2cm
\caption{The $J$--$J_3$ phase diagrams, same as in Fig.~\ref{fig_phaseDJ1J3_0}, for the Point~A set, see Sec.~\ref{Sec_rep_parameters} and Table~\ref{table3}.}
\label{fig_phaseDJ1J3_A}
\vskip -0.4cm
\end{figure*}

 Here, we provide a detailed analysis  of such phase diagrams and elaborate on the technical details of the approaches that are used for their derivation. Setting $\{K,\Gamma,\Gamma'\}$ to Point~0, we show the phase diagram in the $J$--$J_3$ plane in Figure~\ref{fig_phaseDJ1J3_0}, and for Point~A in Figure~\ref{fig_phaseDJ1J3_A}. Both of these phase diagrams have regions consistent with the phenomenology of $\alpha$-RuCl$_3$.

In Fig.~\ref{fig_phaseDJ1J3_0}(a) and Fig.~\ref{fig_phaseDJ1J3_A}(a), the $J$--$J_3$ phase diagrams are obtained by the  quasiclassical  Luttinger-Tisza (LT) approach, which allows us to find spin arrangements of the classical spins that minimize their energy. The modern version of the original LT approach~\cite{lt_original,Lyons_Kaplan_1960,Friedman_1974,Litvin_1974}  involves diagonalization of the exchange matrix of the classical model in the momentum space, and the lowest eigenenergy, corresponding ordering vector, and associated types of spin arrangement are identified by a scan through  reciprocal space and a Fourier transform back into the real space. It  has been widely used in the studies of the anisotropic-exchange models~\cite{multiQ,Wang17,Zhu19,Niggemann_2019,K1K2,Maksimov22}, in which   classical groundstate spin configurations are often not obvious. The implementation of the LT method is computationally cheap and straightforward (see Appendix~\ref{app_B}), with the area shown in Fig.~\ref{fig_phaseDJ1J3_0}(a) and Fig.~\ref{fig_phaseDJ1J3_A}(a) containing a grid of several hundreds points in both the $J$ and $J_3$ directions.

In Figs.~\ref{fig_phaseDJ1J3_0}(b) and \ref{fig_phaseDJ1J3_A}(b) we show similar phase diagrams obtained from ED. The reduction of the parameter space is essential in making this feasible. To reduce the effort of mapping out the phase diagrams, sets of 1D sweeps along various lines through the plane are performed~\cite{Wang17,Trebst_tr,Winter18,Trebst22,Hickey23}. At individual points along a line, an ED  using a  24-site  cluster with all space-group symmetries of the lattice is performed. The second derivative of the ground-state energy with the sweep parameter, $\partial^2 E_0/\partial J_{(3)}^2$, is used to find phase boundaries, and the static spin structure factor, ${\cal S}({\bf q})$, is analyzed  to identify ordered states, see App.~\ref{app_C} for details. 

The FM, AFM, and ZZ phases were identified by the dominant peak of ${\cal S}({\bf q})$ at the corresponding ordering vector, with the insets in Figs.~\ref{fig_phaseDJ1J3_0}(b) and \ref{fig_phaseDJ1J3_A}(b) showing the allowed momenta of the 24-site cluster in the first Brillouin zone. In the cases with no definite dominance of a specific ${\bf q}$-point in ${\cal S}({\bf q})$, the states received   the ``non-commensurate'' (NCO) designation. Peaks in the energy derivatives $\partial^2 E_0/\partial J_{(3)}^2$ are associated with the phase boundaries and their relative sharpness can be suggestive of the order of the phase transition~\cite{Wang17,Trebst_tr}. 

The vertical dashed and solid lines in Figs.~\ref{fig_phaseDJ1J3_0}(b) and \ref{fig_phaseDJ1J3_A}(b)  show the direction and extent of the 1D DMRG ``scans,'' which will be discussed in Sec.~\ref{Sec_IC} together with the analysis  of the IC phases at the  representative points, marked by the squares and diamonds.

The resultant side-by-side comparison of the LT and ED phase diagrams in Fig.~\ref{fig_phaseDJ1J3_0} and Fig.~\ref{fig_phaseDJ1J3_A} is  revealing. Very different methods, having very  different limitations, one is classical and the other constrained by the finite-size effects, both point to a  very similar structure of the phase diagram with a rather  close {\it quantitative} correspondence of their boundaries and arrangements.  

In broad strokes,  large sectors of the phase diagrams are occupied by the ZZ and FM phases, both discussed as prominently present in  the relevant phase space of $\alpha$-RuCl$_3$ by prior analyses~\cite{winter17,winter16,Winter18,Keimer20}.

The z-AFM phase at low values of $|J|$ has  N\'{e}el order with spins pointing along the $z$ axis, not unlike the Ising-z phase discussed recently for the $J_1$--$J_3$ FM-AFM model~\cite{Arun23,Trebst22,J1J3us}, where it is stabilized by strong quantum fluctuations due to frustrating in-plane interactions. However, here its origin is simpler, as the out-of-plane order is also promoted by $\Delta\!<\!0$, as can be seen from the upper horizontal axis in Figs.~\ref{fig_phaseDJ1J3_0} and \ref{fig_phaseDJ1J3_A}. This means that the out-of-plane nearest-neighbor exchange in the crystallographic parametrization (\ref{HJpm}) is antiferromagnetic, cooperating with the $J_3$ term. As we discuss below, this region is likely irrelevant to $\alpha$-RuCl$_3$ for different reasons.

According to the LT phase diagrams in Figs.~\ref{fig_phaseDJ1J3_0}(a) and \ref{fig_phaseDJ1J3_A}(a),  the FM and ZZ phases are separated by the sequence  of two incommensurate phases, IC2 and IC1. While the IC phases are regularly found in the extended KH model~\cite{RauG,RauGp,rethinking,Pollet21}, they are rarely analyzed. In our case, the ordering in the  IC2 and IC1 phases  correspond to two variants of the deformed counter-rotating spin-helices,   discussed  in  detail in Sec.~\ref{Sec_IC}, with the ${\bf Q}$-vectors and corresponding periodicities varying according to the color maps shown in the insets. 

For the IC2 phase, the ${\bf Q}$-vector is along the $\Gamma K$ direction and it evolves from a value that is imperceptibly close to the $\Gamma$ point, suggesting a smooth evolution from the FM state, followed by a jump to the $\Gamma M$ direction when crossing to the IC1 phase, but with nearly the same value of $|{\bf Q}|$. In fact, according to LT, the two phases  at that transition correspond to very shallow energy minima and are close  to the other IC states with the intermediate ${\bf Q}$ directions. The transition from the IC1 to ZZ is by a finite jump in ${\bf Q}$ to the ordering vector  of the ZZ phase at the M-point and is significantly first order. Although the finite-size effects in ED do not allow it to contribute decisively to the discussion of the nature of the IC phases, the widths of the peaks in the second derivative of ED energy at the FM-NCO1 and NCO1-ZZ boundaries are not inconsistent with the former being second  and the latter being first order.
Further verifications of these traits are provided by DMRG in Sec.~\ref{Sec_IC}.

\begin{figure*}[t]
\centering
\includegraphics[width=1.0\linewidth]{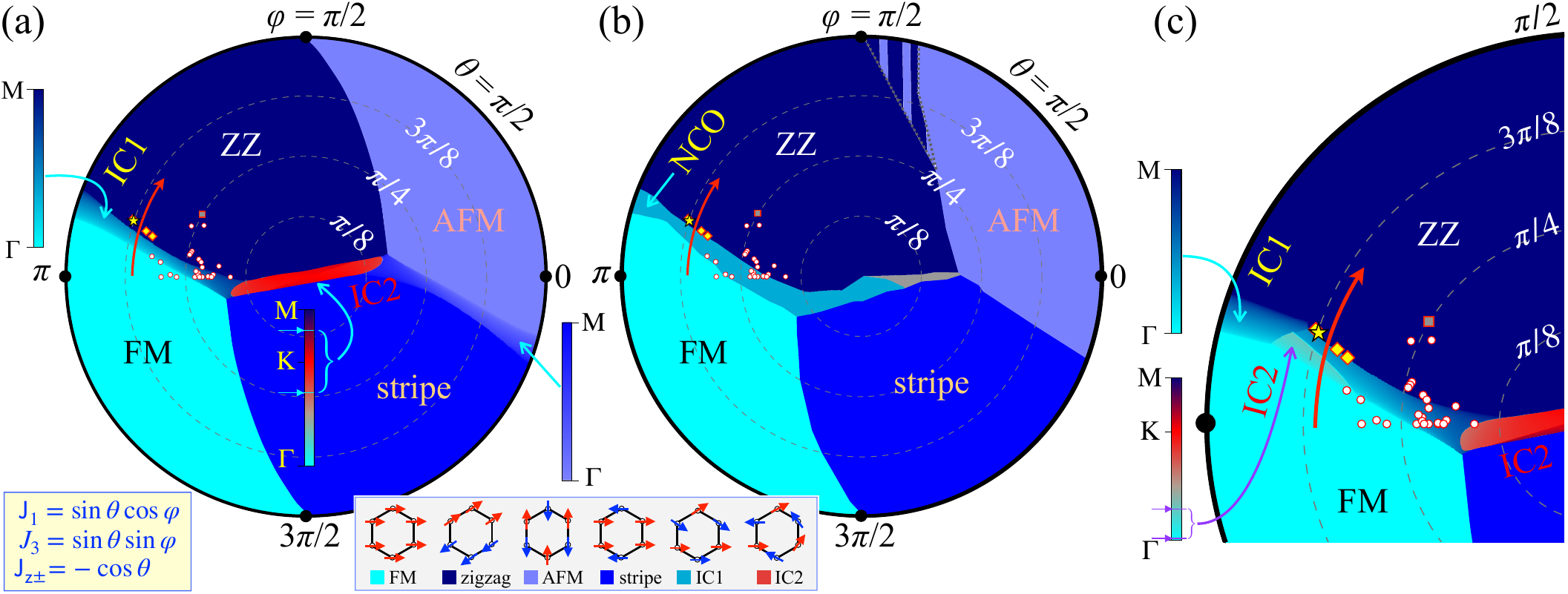}
\vskip -0.2cm
\caption{Polar phase diagrams of the model (\ref{HJpm}) for $\Delta\!=\!0$. In (a) and (b) ${\sf J_{\pm\pm}}\!=\!0$, in (c) ${\sf J_{\pm\pm}}\!=\!0.05$, in units of ${\sf J^{\rm 2}_1}+{\sf J^{\rm 2}_{z\pm}}+J_3^2\!=\!1$.  (a) and (c) are obtained by LT and (b) by ED methods, respectively. ${\sf J_1}$ and $J_3$ encode  polar and ${\sf J_{z\pm}}$ radial coordinates, see inset in (a). Other details include sketches of the phases, color-bars for the ${\bf Q}$-vectors in the IC phases,  projections of the data sets from the prior works, representative Point~$\bigstar$, and DMRG scan path from Fig.~\ref{fig_phaseDJ1J3_0}(b), see the text.}
\label{fig_phaseDpolar1}
\vskip -0.4cm
\end{figure*}

As has been noted in the past, the  LT method may fail to satisfy  its own selfconsistency constraints for some of the phases, specifically the spin length,   see Appendix~\ref{app_B}, formally finding groundstates that are  deemed unphysical. Such ``failures'' have been interpreted as the sign of the more complicated multi-${\bf Q}$ phases~\cite{multiQ,Zhu19}, candidate spin-liquid regions~\cite{Niggemann_2019}, and other states~\cite{ZaZh95}.  
In our case, LT constraints are satisfied for all  phases with the commensurate ordering vectors, the FM, ZZ, and AFM   in Figs.~\ref{fig_phaseDJ1J3_0}, \ref{fig_phaseDJ1J3_A}, and \ref{fig_phaseDJ1J3_B_LT}, and  for the stripe phase in Figs.~\ref{fig_phaseDpolar} and \ref{fig_phaseDpolar1}. But the IC phases obtained by the LT approach do violate the spin-length constraint and formally fall under the ``unphysical'' category.  

However,  the ability of the  LT method to consistently produce states that are lower in energy than the competing classical ones at the cost  of not preserving spin length can be seen as a blessing in disguise instead of being unphysical. This is because it may allow it to mimic quantum effects of the fluctuating states that are not conserving the length of the ordered moments either; a similar sentiment has also been expressed  in Ref.~\cite{KimchiLT14}. 

This conjecture will receive  rather significant  support  from the analysis of the character of the IC counter-rotating helical phases, with their periods, mutual orientations of the  ${\bf Q}$-vectors with the planes of spin rotations, and even phase shifts of the spirals, all being in  close {\it quantitative} accord  between the LT predictions and that of the  DMRG results discussed below in Sec.~\ref{Sec_IC},  yielding  additional important agreements and insights. 

\vspace{-0.3cm}
\subsubsection{Where is $\alpha$-RuCl$_3$?}
\label{Sec_cartesianPD_aRuCl3}
\vskip -0.2cm

Last, but not  least, is the study of additional constraints on the $\alpha$-RuCl$_3$ parameters  in the $J$--$J_3$ plane. 

At the quasiclassical level, the out-of-plane tilt angles   (\ref{eq_alphaKG}) that closely match  magnetic order in $\alpha$-RuCl$_3$  should be the same  throughout the ZZ  regions of Figs.~\ref{fig_phaseDJ1J3_0}(a) and \ref{fig_phaseDJ1J3_A}(a),  fixed by the choices of anisotropic parameters of Point~0 and Point~A, respectively.  Other phases that are present in the phase diagrams can only constrain the values of $|J|$ and $J_3$ from below. 

However, as we argue here, one can devise a much stronger  constraint on a combination of the two remaining isotropic parameters using  phenomenologies that have already been introduced. Although the difference of the critical fields for the transition to the paramagnetic phase $\Delta H_c$ has been utilized as a constraint for anisotropic exchanges, the values of the critical fields themselves have not been exploited yet.  As one can see in Eqs.~(\ref{Hca}) and (\ref{Hcb}), both  $H_c^{(a)}$ and $H_c^{(b)}$ fields depend on the same linear combination of $J+3J_3$, making it a natural variable~\cite{rethinking} for another strong empirical constraint that binds  $J$ and $J_3$ terms via the experimental value of one of the critical fields. 

Each of the panels in Figs.~\ref{fig_phaseDJ1J3_0} and \ref{fig_phaseDJ1J3_A} shows a straight dashed line with a constant slope of $J_3\!=\!-J/3$. They correspond to the solutions of Eq.~(\ref{Hca}) with the experimental value of $H_c^{(a)}\!=\!7$~T~\cite{Winter18,Cao16,NaglerVojta18}. Another line with the same slope is for the value of $H_c^{(a)}$ that is twice as high. The second line is introduced because the critical fields are expected to be  suppressed by quantum fluctuations, with the theoretical calculations of that effect for the KH models suggesting it potentially reaching a factor of two~\cite{Winter18,Vojta2020_NLSWT}. One can see that the lines  in Figs.~\ref{fig_phaseDJ1J3_0} and \ref{fig_phaseDJ1J3_A} carve a narrow strip of the $J$--$J_3$ space, altogether suggesting very effective additional bounds on the physical parameters of $\alpha$-RuCl$_3$. 

One may wonder why this strip is so narrow despite the doubling of the ``bare,'' unrenormalized critical field  from 7~T to 14~T. This is because a relatively small change of $J_3$ by 0.4~meV in the  $J+3J_3$ combination readily modifies $H_c$ in Eq.~(\ref{Hca}) by about 7~T. 

In the phase diagrams in  Figs.~\ref{fig_phaseDJ1J3_0} and \ref{fig_phaseDJ1J3_A}, the acceptable range of $|J|$ is also  restricted from below by the competing phases. It is also rather unlikely for $|J|$ to exceed the value of $|K|$, restricting it  from above~\cite{winter16,rethinking}.  

Altogether, the  ``complete'' sets of the prospective $\alpha$-RuCl$_3$ parameters for the model (\ref{eq_Hij}) should be chosen from the  narrow ranges that are marked by the elongated ellipses  in the $J$--$J_3$ planes in  Fig.~\ref{fig_phaseDJ1J3_0}, Fig.~\ref{fig_phaseDJ1J3_A}, and Figs.~\ref{fig_phaseDJ1J3_B_LT} and \ref{fig_phaseDJ1J3_B_ED}, for the Point~0, A, and B anisotropic parameter sets, respectively. 

One of such  choices is marked  by the star symbol in Fig.~\ref{fig_phaseDJ1J3_0}(b). It corresponds to the following complete set of the model (\ref{eq_Hij}) parameters, all in meV, 
\noindent
\begin{equation}
\begin{array}{lllllr}
     & \{\ \ K, & \Gamma,   & \Gamma',     &\ \   J, & J_3\ \}  \\ 
{\rm Point}\, \bigstar: \! & \{-7.567, \! & \! 4.276, \! & \! 2.362, \!&  \! -4.75, \! &\! 3.4 \}  \\ 
\end{array} ,
\label{eq_point_star}
\end{equation}
\noindent
which is, of course, a combination of the anisotropic parameter set of Point~0, see Sec.~\ref{Sec_rep_parameters}, and the choice of the $\{J,J_3\}$ pair from the middle of the allowed region. 

By fixing the isotropic exchanges, the parameter set for  $\alpha$-RuCl$_3$ is complete. It is not a unique choice, as both $J$ and $J_3$ can be adjusted according to the allowed range, and so can be anisotropic parameters, in a coordinated fashion, see Sec.~\ref{Sec_parameter_space}. Nevertheless, these adjustments may not be significant and are not expected to lead to drastically different physical outcomes, as the comparison between phase diagrams for  the Point~0, A, and B sets in Figs.~\ref{fig_phaseDJ1J3_0}, \ref{fig_phaseDJ1J3_A}, \ref{fig_phaseDJ1J3_B_LT}, and \ref{fig_phaseDJ1J3_B_ED} indicates. 

Moreover, the complete Point$\,\bigstar$ parameter set (\ref{eq_point_star}) can be used to test predictions and assumptions of the present study, as well as the other phenomenologies. Specifically, using Eqs.~(\ref{Hca}) and (\ref{Hcb}), this parameter set yields  the ``bare'' critical  fields $H_c^{(a)}\!=\!11.7$~T and $H_c^{(b)}\!=\!12.5$~T,  maintaining their difference at the physical 0.8~T. One of the key hypotheses of our anisotropic strategy is that this difference does not change drastically, while the fields themselves get suppressed considerably by quantum effects, see Sec.~\ref{Sec_constraints_details}. We will  provide a full vindication of both expectations  using  DMRG  in Sec.~\ref{Sec_DMRGchecks}. 

Another test is the verification by DMRG of the same out-of-plane tilt angle of spins throughout the ZZ phase for a given choice of $\{K,\Gamma,\Gamma'\}$ in Figs.~\ref{fig_phaseDJ1J3_0} and \ref{fig_phaseDJ1J3_A}, which is also expected to stay within the physically allowed range for  $\alpha$-RuCl$_3$ in the quantum limit. These selfconsistency checks of our anisotropic strategy for the quantum model will also be presented in Sec.~\ref{Sec_DMRGchecks}. 

Since we return to the  discussion of the model (\ref{eq_Hij}) in the crystallographic parametrization (\ref{HJpm}) in the next Section, it is useful to reflect on one more common thread that is exposed in the phase diagrams in Figs.~\ref{fig_phaseDJ1J3_0} and \ref{fig_phaseDJ1J3_A}. Using the upper horizontal axis for the $XXZ$ anisotropy parameter $\Delta$, one can see that the relevant region for the $\alpha$-RuCl$_3$ parameters resides solidly in the range of $-0.1\!<\!\Delta\!<\!0.3$,  in agreement with the discussions in Secs.~\ref{Sec_Jpp_Jzp_space}, \ref{Sec_comparison_Giniyat}, and \ref{Sec_compilation}. 

It is also useful to rewrite the representative Point$\,\bigstar$ parameter set (\ref{eq_point_star}) in these axes (in meV except for $\Delta$)
\noindent
\begin{equation}
\begin{array}{lllllr}
     & \{\ \ {\sf J_1}, & \! {\sf J}_{\pm \pm},   & \ \ {\sf J}_{z\pm},    \! & \! \Delta, & J_3\ \}  \\ 
{\rm Point}\, \bigstar: \! & \{-10.272,\! & \!0.623, \! &\! -4.469, \! &\!  0.124,\! &\! 3.4 \}  \\ 
\end{array},
\label{eq_point_starJzp}
\end{equation}
\noindent
to emphasize this feature. Here, the ${\sf J}_{\pm \pm}$ term is secondary as before, with ${\sf J_1}$ and ${\sf J}_{z\pm}$ dominating.

Lastly, we remark on the implications of the easy-plane,  $\Delta\!\approx\!0$  $XXZ$ character of the  $\alpha$-RuCl$_3$ parameter space. As one can easily obtain from the relations between different parametrizations in App.~\ref{app_A}, the pure $K$--$J$ model, together with its Kitaev-only points, must reside in the plane of $\Delta\!=\!1$, which places $\alpha$-RuCl$_3$ far away from that plane from the outset. This simple understanding, together with a straightforward phenomenology that allows one to detect significant $XXZ$ anisotropy in the material suspects, see Sec.~\ref{Sec_ESR_gap}, may serve as important guidance for the studies of the other Kitaev candidates.

\vspace{-0.3cm}
\subsection{${\sf J^{\it XY}_1}$--${\sf J_{z\pm}}$--$J_3$  phase diagram}
\label{Sec_polarPD}
\vskip -0.2cm

As repeatedly argued above, the crystallographic-axes parametrization (\ref{HJpm})   not only helps with a better description of $\alpha$-RuCl$_3$, it can also offer a transparent and intuitive perspective  to the studies of  anisotropic-exchange systems in general. It  allows one to take a broader conceptual view on the studies of the generalized KH and other anisotropic-exchange models, placing them in a more tangible context within    studies in frustrated magnetism. 

For  $\alpha$-RuCl$_3$, we have made it clear that its model description favors the dominant $XY$-like ($\Delta\!\approx\!0$) ferromagnetic ${\sf J_1}$ combined  with a  sizable anisotropic bond-dependent ${\sf J_{z\pm}}$,  complemented by the isotropic antiferromagnetic $J_3$, see Secs.~\ref{Sec_intro_alternative}, \ref{Sec_Jpp_Jzp_space}, \ref{Sec_comparison_Giniyat}, \ref{Sec_compilation}, and \ref{Sec_cartesianPD_aRuCl3}.  The residual $XXZ$ anisotropy $\Delta$ and  ${\sf J_{\pm\pm}}$ can be treated as secondary in their physical outcomes and studied as quantitative additions to the main model.

Here, we  analyze the phase diagram of this simplified and physically justified ${\sf J^{\it XY}_1}$--${\sf J_{z\pm}}$--$J_3$  model. Because of the  three-dimensional parameter space, and given the redundancy of the sign of the ${\sf J_{z\pm}}$ term~\cite{Zhu19}, one can explore the entirety of its parameter space with the help of the polar parametrization, shown in the inset of Fig.~\ref{fig_phaseDpolar1}(a), with the  $\{{\sf J_1}, J_3\}$ pair parameterizing polar variable and ${\sf J_{z\pm}}$ the radial one in units of ${\sf J^{\rm 2}_1}+{\sf J^{\rm 2}_{z\pm}}+J_3^2\!=\!1$. 
 
\vspace{-0.3cm}
\subsubsection{LT polar phase diagram and other details}
\label{Sec_LTpolarPD}
\vskip -0.2cm
 
Our Figure~\ref{fig_phaseDpolar1}(a) presents the  phase diagram obtained by the  LT method as in  Figs.~\ref{fig_phaseDJ1J3_0}(a) and \ref{fig_phaseDJ1J3_A}(a) above. In addition to the ZZ, FM, and IC phases, this phase diagram features  AFM and stripe phases. The former  occurs for the antiferromagnetic sign of ${\sf J_1}$  and the latter mirrors the ZZ phase. The incommensurate phase sandwiched between them, in turn, mirrors the IC1 phase separating FM and ZZ, already familiar from the Cartesian phase diagrams in Sec.~\ref{Sec_cartesianPD}. 

The incommensurate phase in the center of the circle has a ${\bf Q}$-vector that is  continuously evolving along the $\Gamma K$ line, normal to the bond of the honeycomb lattice, and then along the $KM$ BZ boundary, as is shown in the  color-coded bar. It is labeled as IC2 as it  belongs to the same phase as the IC2 phase in Figs.~\ref{fig_phaseDJ1J3_0}(a) and \ref{fig_phaseDJ1J3_A}(a), with both of them corresponding to the same type of the counter-rotated, deformed, and slightly tilted helix (spiral), as will be further discussed in Sec.~\ref{Sec_IC}.

While the ${\bf Q}$-vector in the IC2 phase evolves within the  limits marked on the color bar, similar to the finite spans for both IC1 and IC2  in  Figs.~\ref{fig_phaseDJ1J3_0}(a) and \ref{fig_phaseDJ1J3_A}(a), the IC1 phase and its AFM-stripe mirror in Fig.~\ref{fig_phaseDpolar1}(a)  span the entire range between the $\Gamma$ and $M$ points. This is due to the  ${\sf J_{z\pm}}\!=\!0$ circumference of this phase diagram, which corresponds to the  pure ${\sf J_1}$--$J_3$ model.  In this limit, the classical IC1 is a coplanar, co-rotating spiral continuously interpolating FM and ZZ~\cite{italians,J1J3us}. Away from the boundary, the IC1 phase turns into a deformed counter-rotated helix with a finite range of the ${\bf Q}$-vector that is parallel to the bond of the honeycomb lattice, see Sec.~\ref{Sec_IC}.

The inset in Figs.~\ref{fig_phaseDpolar1}(a) and \ref{fig_phaseDpolar1}(b) provides sketches of  the phases and  emphasizes the differences between the propagation vectors in the IC1 and IC2 phases.

There are small circles in Fig.~\ref{fig_phaseDpolar1}, which represent  projections of all individual $\alpha$-RuCl$_3$ parameter sets listed in Table~\ref{table3} onto the $\Delta\!=\!{\sf J_{\pm\pm}}\!=\!0$ plane of the phase diagram. While some of the parameters that are necessary to obtain the coordinates for such projections are made available in Table~\ref{table2} in App.~\ref{app_A}, we do not list  $J$ and $J_3$ values of these individual sets  in this work. 

The small square with a darker shading corresponds to the machine-learning effort of Ref.~\cite{Samarakoon22}, discussed in Secs.~\ref{Sec_ESR_gap} and \ref{Sec_compilation}. The group of three diamonds are the sets  listed in Table~\ref{table2rethinking} in App.~\ref{app_D} that were proposed in Ref.~\cite{rethinking}, which has used some of the same types of phenomenological constraints on the $\alpha$-RuCl$_3$ parameters as the ones used in the present work. These sets were deliberately not included in Table~\ref{table3} for the sake of not skewing  independent distribution of parameters from the prior literature. The projection of the representative Point~$\bigstar$ set from Fig.~\ref{fig_phaseDJ1J3_0}(b) and Sec.~\ref{Sec_cartesianPD_aRuCl3} is shown by the star symbol. Although it appears close to one of the sets from Table~\ref{table2rethinking} in App.~\ref{app_D}, the parameters in the sets  are  rather different.  

An important benefit of this comparison and the clear advantage of the bird's-eye view of the polar phase diagrams demonstrated in Fig.~\ref{fig_phaseDpolar1} is in making apparent the  commonality of the trends among the prior, although not always successful, hunts for the adequate description of  $\alpha$-RuCl$_3$  and its phenomenologies. 

In addition to  Point~$\bigstar$  from the Cartesian phase diagram in Fig.~\ref{fig_phaseDJ1J3_0}(b), one additional 
correspondence is brought into the polar field of view of Fig.~\ref{fig_phaseDpolar1} by the red line with the arrow that traverses FM, IC1, and ZZ phases at a fixed radius. It is a projection of the DMRG scan path at $J\!=\!3$ in Fig.~\ref{fig_phaseDJ1J3_0}(b), which closely corresponds to the $XXZ$ anisotropy value of $\Delta\!=\!0$, see the upper axis. As all anisotropic terms in Fig.~\ref{fig_phaseDJ1J3_0}(b) are at Point~0, by fixing $J$ (and $\Delta$) the scan changes only $J_3$. For the polar representation in  Fig.~\ref{fig_phaseDpolar1}, changing $J_3$ corresponds to varying the ratio of $J_3/{\sf J_1}$ and traveling along the  radial path. Altogether, this reference scan helps to relate the view provided by the Cartesian phase diagrams in Sec.~\ref{Sec_cartesianPD} and the polar phase diagrams in Sec.~\ref{Sec_polarPD} as offering  different slices through the same higher-dimensional parameter space. 

\vspace{-0.3cm}
\subsubsection{ED polar  phase diagram}
\label{Sec_EDpolarPD}
\vskip -0.2cm

Our Fig.~\ref{fig_phaseDpolar1}(b) offers another demonstration of the power of the numerical tour-de-force exploration of the 2D parameter space using ED in the  24-site  cluster. 

With the technical description of the approach provided  in Sec.~\ref{Sec_cartesianPD_methods} and  App.~\ref{app_C}, the phase diagram in Fig.~\ref{fig_phaseDpolar1}(b) is based on the combination of the 1D radial and polar sweeps, 1000 points each.  For the radial sweeps, the variable $\theta$ was varied from 0 to $\pi/2$ for fixed polar angles $\varphi$. For the polar sweeps,  $\varphi$  angle was swept from 0 to $2\pi$ at  fixed $\theta$.  In addition, two spiral sweeps were performed, in which both $\theta$ and $\varphi$ variables changed, making four full rotations in $\varphi$ while varying $\theta$ from 0 to $\pi/2$ and from 0 to $\pi/5$, see  App.~\ref{app_C}. Additional polar sweeps were performed for the FM-ZZ boundary with the higher discretization. To determine phase boundaries, the second derivatives of the ground-state energy along the sweeps were taken with respect to $\varphi$ and $\theta$.

As in the case of the phase diagrams in Sec.~\ref{Sec_cartesianPD_methods}, the agreement between the LT and ED phase diagrams in Figs.~\ref{fig_phaseDpolar1}(a) and \ref{fig_phaseDpolar1}(b) is {\it quantitative}, especially in the ${\sf J_1}\!<\!0$, $J_3\!>\!0$ sector relevant to $\alpha$-RuCl$_3$, where the NCO analog of the IC1 phase appears in between the FM and ZZ phases, with their boundaries in  close correspondence.  We stress here again that the two side-by-side diagrams are not only obtained in two different limits, classical and quantum, and by the two methods with different limitations, but  in the case of Fig.~\ref{fig_phaseDpolar1} they also span the entire parameter space of the model. Yet, the phases they contain and their compositions  are in a  close {\it quantitative} agreement.  

The LT phase diagram in Fig.~\ref{fig_phaseDpolar1}(a), being classical in nature, shows a complete symmetry between FM and AFM and ZZ and stripe phases, respectively. In the  quantum $S\!=\!1/2$ case in Fig.~\ref{fig_phaseDpolar1}(b), this symmetry is lost, and there is no discernible intermediate phase between the AFM and stripe phases. As we discuss below, the symmetry of the LT phase diagram in Fig.~\ref{fig_phaseDpolar1}(a) is only a property of the 
${\sf J_1}$--${\sf J_{z\pm}}$--$J_3$ model, with the finite ${\sf J_{\pm\pm}}$ also inducing asymmetries in the phase diagram, similarly to the quantum effects.

In the ED case,  the ZZ-AFM boundary, which concerns the region of small  ${\sf J_1}$, is somewhat problematic. This problem is due to the finite-size effects specific to the 24-site  cluster, which splits into three independent clusters containing only 8 sites connected by the same $J_3$ network at ${\sf J_1}\!=\!0$. Because of that, that sector of the phase diagram was clarified by DMRG. 

\vspace{-0.3cm}
\subsubsection{Finite ${\sf J_{\pm\pm}}$  phase diagram and other trends}
\label{Sec_LTpolarPD}
\vskip -0.2cm

Although not elaborated on in detail in the present study, the evolution of the phase diagram  in Fig.~\ref{fig_phaseDpolar1} with $\Delta$ away from the $\Delta\!=\!0$ limit is the following. For negative $\Delta$, the IC1  and its mirror regions expand somewhat, while the rest of the boundaries shift only mildly. At about $\Delta\!=\!-0.4$, the z-AFM and z-FM regions open at the outer edges of the ZZ and stripe sectors and continue to expand for $\Delta\!<\!-0.4$. This is in agreement with  Figs.~\ref{fig_phaseDJ1J3_0} and \ref{fig_phaseDJ1J3_A} where this range of $\Delta$  is associated with the smaller $|J|$.  For positive $\Delta$, the shift of most of the  boundaries is less  perceptible all the way to the Heisenberg limit $\Delta\!=\!1$, with the only significant effect being a nearly complete disappearance of the intermediate IC1 phase away from the outer boundary for finite ${\sf J_{z\pm}}$.

The evolution with ${\sf J_{\pm\pm}}$ is more drastic and also more relevant to the discussed physical space of $\alpha$-RuCl$_3$, featuring an asymmetry of the phase diagram mentioned above and the nucleation of another region of the IC2 phase from the FM-IC1 boundary at  ${\sf J_{\pm\pm}}\!>\!0$. Upon the further increase of ${\sf J_{\pm\pm}}$ to about 0.15, the two IC2 regions merge, while IC1 diminishes.

The effect of ${\sf J_{\pm\pm}}$ is shown in Fig.~\ref{fig_phaseDpolar1}(c) for ${\sf J_{\pm\pm}}\!=\! 0.05$ in units of ${\sf J^{\rm 2}_1}+{\sf J^{\rm 2}_{z\pm}}+J_3^2\!=\!1$, which closely corresponds to its relative value for the representative Points~0, A, and B  from the physical region for $\alpha$-RuCl$_3$.  With the full phase diagram for that choice of ${\sf J_{\pm\pm}}$ shown in App.~\ref{app_D}, where one can observe the asymmetry of the previously symmetric phases, the cutout from it with the FM-ZZ region shown in Fig.~\ref{fig_phaseDpolar1}(c) allows one to observe the emerging IC2 region within the IC1 phase. This is an important feature for the  $\alpha$-RuCl$_3$-related region of the phase diagram. The relatively small, but essential addition of ${\sf J_{\pm\pm}}$ brings the polar phase diagrams in Fig.~\ref{fig_phaseDpolar1} to a close agreement with the Cartesian ones in Figs.~\ref{fig_phaseDJ1J3_0} and \ref{fig_phaseDJ1J3_A}, which show not one, but two intermediate IC phases.

Altogether, the consideration of the phase diagram of the KH model in the crystallographic frame (\ref{HJpm}) in Sec.~\ref{Sec_polarPD} highlights the benefits of the alternative perspective on the parameter space of the anisotropic-exchange models.  The value of this perspective is also in the organic connection of the considered parameter space to the other models that are of wide interest in frustrated magnetism. 

For instance, the circumference of the phase diagrams in Fig.~\ref{fig_phaseDpolar1} corresponds to the $J_1$--$J_3$ FM-AFM model. The classical phase diagram of this model, known since the end of the 1970s~\cite{italians}, has been recently revisited for the quantum case because of the other Kitaev-candidate magnets~\cite{J1J3us,Arun23,Hickey23,Halloran23}. It was found that the IC spiral state between  FM and ZZ is completely replaced by the entirely unexpected phases, such as double-zigzag and  z-AFM N\'{e}el phase~\cite{J1J3us},  stabilized by quantum fluctuations. Not only does this consideration suggest a wider context to the studies of $\alpha$-RuCl$_3$, but it also underscores the importance and the urgency of understanding of its proximate incommensurate phases, provided next.

\section{DMRG and ED verifications}
\label{Sec_DMRGchecks}

We now look all the way back at the initial assumptions of the present study.  Our  strategy of restricting anisotropic terms of the model for $\alpha$-RuCl$_3$ by using select physical observables that  occur only because of such terms hinges on our ability to calculate these observables, which, in turn, assumes they have small or controlled quantum effects, see Secs.~\ref{Sec_intro_anisotropic} and \ref{Sec_constraints_details}.

For one of them, the ESR gap in Sec.~\ref{Sec_ESR_gap}, this logic is justified as it is deeply rooted in the more general approach of the field-induced quenching of  quantum fluctuations. However, for the other two observables that we have chosen as the physical constraints,  an {\it a posteriori} proof of the validity of such assumptions is needed.

For the out-of-plane tilt angle $\alpha$, see Sec.~\ref{Sec_tilt_angle}, we had some prior   circumstantial evidence that the effects of   quantum fluctuations in $\alpha$ are small~\cite{Chaloupka16}.  For the  observed small difference of the in-plane critical fields  $\Delta H_c$, Sec.~\ref{Sec_dHc}, the argument was that although the critical fields themselves should renormalize strongly, they should do it in sync.  This requires such a difference to be small already at the ``bare'' quasiclassical level, before accounting for quantum effects. We posed this as a falsifiable prediction of our strategy, see Sec.~\ref{Sec_dHc}. 

Below we present  a  complete  validation of both assumptions, and more. 

\vspace{-0.3cm}
\subsection{DMRG technical details}
\label{Sec_DMRG_details}

Technically, we have employed a combination of  two approaches, referred to as the DMRG ``scans'' and ``non-scans,'' using the ITensor library~\cite{itensor}. Both approaches utilized the $L_x\!\times\!L_y$-site honeycomb-lattice open cylinders of width $L_y\!=\!12$ (6 honeycomb cells). For the ``scans,'' we have used  longer cylinders of $L_x\!=\!32$ and had the  one model parameter, such as $J_3$ or a field,  varied along their  length. The representative  1D cuts with the varied $J_3$ are shown in the phase diagrams in Figs.~\ref{fig_phaseDJ1J3_0}(b) and \ref{fig_phaseDJ1J3_A}(b) by red arrows. The ``non-scans'' were done for the select parameter choices, with all parameters fixed, such as the ones marked by  diamonds and squares on the same 1D cuts in these figures. 

For the non-scans, we  used more symmetric $12\!\times\!12$ cylinders with the aspect ratio that has been demonstrated to closely approximate the 2D  thermodynamic limit~\cite{FS}. The non-scans were performed on the so-called X-cylinders (XC)~\cite{scan2}, in which one of the nearest-neighbor bond is horizontal, and on the Y-cylinders (YC), with these bonds being vertical, to study different orientations of orders. All scans were performed on the XC clusters.

This combination of approaches has been successfully employed in the past for the studies of the multi-dimensional phase diagrams of a variety of models and lattices~\cite{J1J3us,scan1,scan2,scan3,scan4,scan5,nematic-our}.  Scans give direct snapshots of the phases along their 1D cuts~\cite{scan1,scan2}, help to identify  phase boundaries~\cite{scan3,scan4,scan5},  distinguish first- and second-order transitions~\cite{nematic-our},  and  uncover hidden phases~\cite{J1J3us}. The non-scans allow one to study the given parameter set in finer detail.

We performed a sufficient number of   DMRG sweeps to reach a maximum bond dimension of $m\!\sim\!1600$ and to ensure good convergence with a truncation error of $\mathcal{O}(10^{-5})$. Generally, there is no spin-rotational symmetry to utilize in an anisotropic-exchange model (\ref{eq_Hij}), with the DMRG ground states typically  breaking any remaining lattice or emergent spin symmetries. We find the local magnetic order in these states changes little with increasing bond dimension, a signature of mimicking the thermodynamic limit in 2D ~\cite{tt'j}, enabling us to measure the local ordered moment $\langle {\bf S}_i\rangle$ directly.

With the selfconsistency checks of our anisotropic strategy discussed below, the nature of the incommensurate phases is uncovered in Sec.~\ref{Sec_IC}.

\begin{figure*}[t]
\includegraphics[width=.99\linewidth]{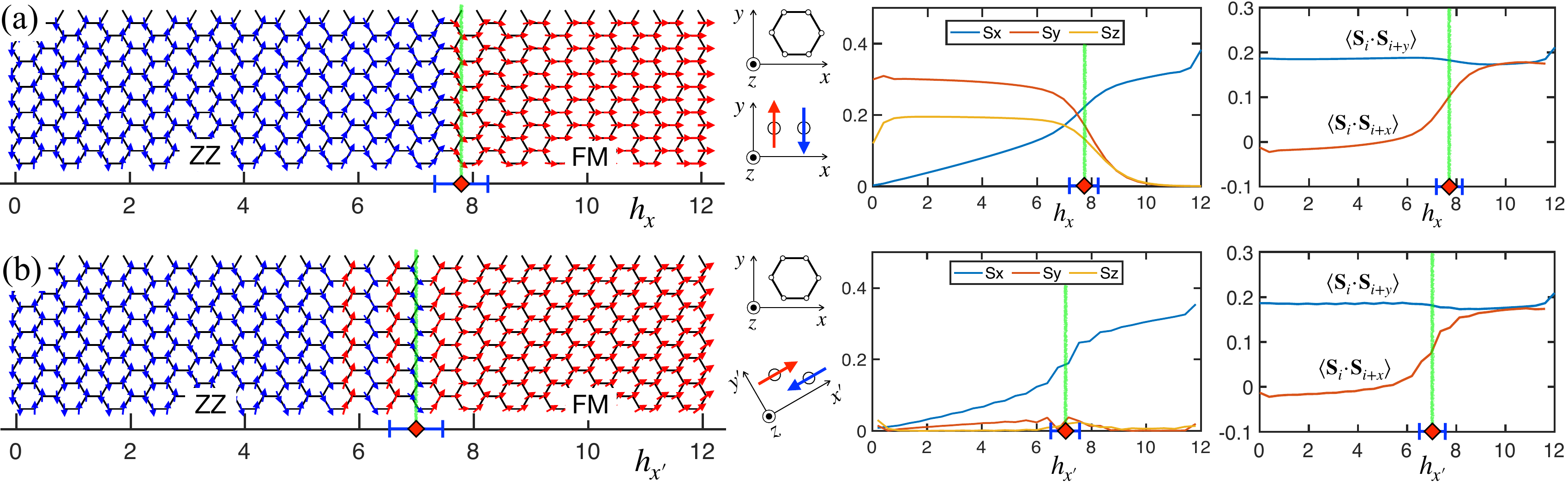} 
\vskip -0.2cm
\caption{The $32\!\times\!12$ XC long-cylinder scans for the Point~$\bigstar$ parameter set (\ref{eq_point_star}) vs field (a) parallel to the bond, $b(x)$-direction, and (b) perpendicular to the bond, $a(x')$-direction. The arrows are the ordered moments' projections onto the $xy$ plane. Arrows within 45$\degree$ to the field are colored in red and otherwise in blue.  The second and third panels show $\langle  S^\alpha_i\rangle$  and  nearest-neighbor correlators  $\langle {\bf S}_i \!\cdot\! {\bf S}_{i+x(y)} \rangle$, respectively, over the circumference of the cylinder. In (b) axes for spins are rotated to align $x'$ axis with the field, see text. Transitions to the polarized phase and  error bars are indicated.}
\label{fig_DMRG_Hc_star}
\vskip -0.4cm
\end{figure*}

\vspace{-0.3cm}
\subsection{Selfconsistency checks}
\label{Sec_DMRGchecks_ZZ_dH}

\subsubsection{Tilt angle}
\label{Sec_DMRGchecks_alpha}
\vskip -0.2cm

According to the classical energy minimization result in Eq.~(\ref{eq_alphaKG}), the  tilt angle of the spins in the ZZ phase for the parameters relevant to $\alpha$-RuCl$_3$ is significant, with the ``bare'' $30\degree$-to-$37\degree$ physical range proposed in Sec.~\ref{Sec_tilt_angle}. It is also  independent  on the isotropic exchanges $J$ and $J_3$.

For the ZZ phases observed in the DMRG scans that are discussed in the next Section, the tilt angles are indeed significant. We have also performed their more precise evaluations from the numerical values of the spin projections using DMRG non-scans  for  Point~0 and Point~A sets of parameters. They give the renormalized  angles $33.7\degree$ and  $29.4\degree$ against their bare values of $35\degree$ and  $32\degree$, respectively, see  Sec.~\ref{Sec_rep_parameters}. Thus, in agreement with our expectations, the quantum renormalization effect in  $\alpha$ is small, leaving it safely within the physical range and fully justifying the use of its bare expression in Eq.~(\ref{eq_alphaKG}) for restricting model parameters.

\vspace{-0.3cm}
\subsubsection{Critical fields}
\label{Sec_DMRGchecks_dH}

The most challenging assumption is  the use of $\Delta H_c$ as a phenomenological constraint. To verify this, we chose the representative set of the Point~$\bigstar$, see Sec.~\ref{Sec_cartesianPD_aRuCl3} and Eq.~(\ref{eq_point_star}), which is Point~0 of the anisotropic set of $\{K,\Gamma,\Gamma'\}$  with the isotropic pair $\{J,J_3\}$ fixed in the plausible region for $\alpha$-RuCl$_3$, as is described in Sec.~\ref{Sec_cartesianPD_aRuCl3}. One can use Eqs.~(\ref{Hca}) and (\ref{Hcb}) for the Point~$\bigstar$ set to obtain the bare values of $H_{c,0}^{(b)}\!=\!12.51$~T and $H_{c,0}^{(a)}\!=\!11.71$~T, with their bare difference  $\Delta H_{c,0}\!=\!0.8$~T corresponding to how the selection of the Point~0 set was made, see Sec.~\ref{Sec_rep_parameters}. Recall that the experimental values are $\Delta H_c^{\rm exp}\!\approx\!0.8$~T,  $H_{c,\rm exp}^{(b)}\!\approx\!7.8$~T, and $H_{c,\rm exp}^{(a)}\!\approx\!7.0$~T, respectively.

We performed DMRG scans vs field in the two principal in-plane directions, $a$ and $b$, perpendicular and parallel to the bond, respectively, for the Point~$\bigstar$ parameter set. The results are presented in Fig.~\ref{fig_DMRG_Hc_star}. 

In Fig.~\ref{fig_DMRG_Hc_star}(a), the first panel shows the $32\!\times\!12$ XC long-cylinder scan vs field parallel to the bond, $b(x)$-direction. The arrows represent the local ordered moments $\langle {\bf S}_i \rangle$ projected onto the $xy$ plane, with arrows colored red for the spins within 45$\degree$ to the field and blue otherwise. The honeycomb lattice is in the $xy$ plane. The second and third panels show the evolution of the three components of the on-site ordered moment, $\langle S_i^\alpha \rangle$, and the nearest-neighbor correlators, $\langle {\bf S}_i \!\cdot\! {\bf S}_{i+x} \rangle$ and $\langle {\bf S}_i \!\cdot\! {\bf S}_{i+y} \rangle$, averaged over the circumference of the cylinder, vs field $h_x$ (in Tesla). 

Fig.~\ref{fig_DMRG_Hc_star}(b) shows the same for the field  perpendicular to the bond, $a(x')$-direction. To compare the results for the two field directions fairly, we kept the same XC cylinder for $H^{(a)}$, but used a tilted direction of the field to have it perpendicular to the bond, with the spin axes tilted accordingly. Because of the tilt, the averaging of the spin components $\langle S_i^\alpha \rangle$ and correlators $\langle {\bf S}_i \!\cdot\! {\bf S}_{i+x} \rangle$ and $\langle {\bf S}_i \!\cdot\! {\bf S}_{i+y} \rangle$ was done over the two nearest vertical zigzag columns to minimize the oscillatory trends in these quantities.

The data in the second and third panels in Figs.~\ref{fig_DMRG_Hc_star}(a) and \ref{fig_DMRG_Hc_star}(b) together with the scans themselves show clear transitions from the ZZ to polarized FM states, with their positions  and  error bars determined from the inflection points in the ordered moment curves and the widths of the transition regions, respectively. The DMRG results do indeed confirm that the downward renormalization of the critical fields is quite significant in both directions, reaching some 40\% of their bare values, in agreement with the expectations laid out above. 

Notably, the critical fields are renormalized down from their classical values  by about the same factor, maintaining close proximity to each other and providing strong support to the suggested constraint. Their resultant values are also very close to the experimental ones, $H_{c}^{(b)}\!\approx\!7.8(5)$~T and $H_{c}^{(a)}\!\approx\!7.0(5)$~T, effectively maintaining the original value of the critical field difference of $\Delta H_c\!\approx\!0.8(5)$~T.

While such a close agreement with the experimental data for the chosen representative set of parameters may be somewhat fortuitous, the demonstrated confirmation of the original logic of our approach is rather exceptional, providing a very strong support to the validity of our strategy, approach, and their results.

\vspace{-0.3cm}
\subsubsection{ESR gap}
\label{Sec_DMRGchecks_ESR}
\vskip -0.2cm

For completeness, we also provide a verification of the  ESR gap criterion for the Point~$\bigstar$ set using an ED calculation of the excitation spectrum on a 24-site cluster. 

\begin{figure}[b]
\vskip -0.2cm
\includegraphics[width=.99\linewidth]{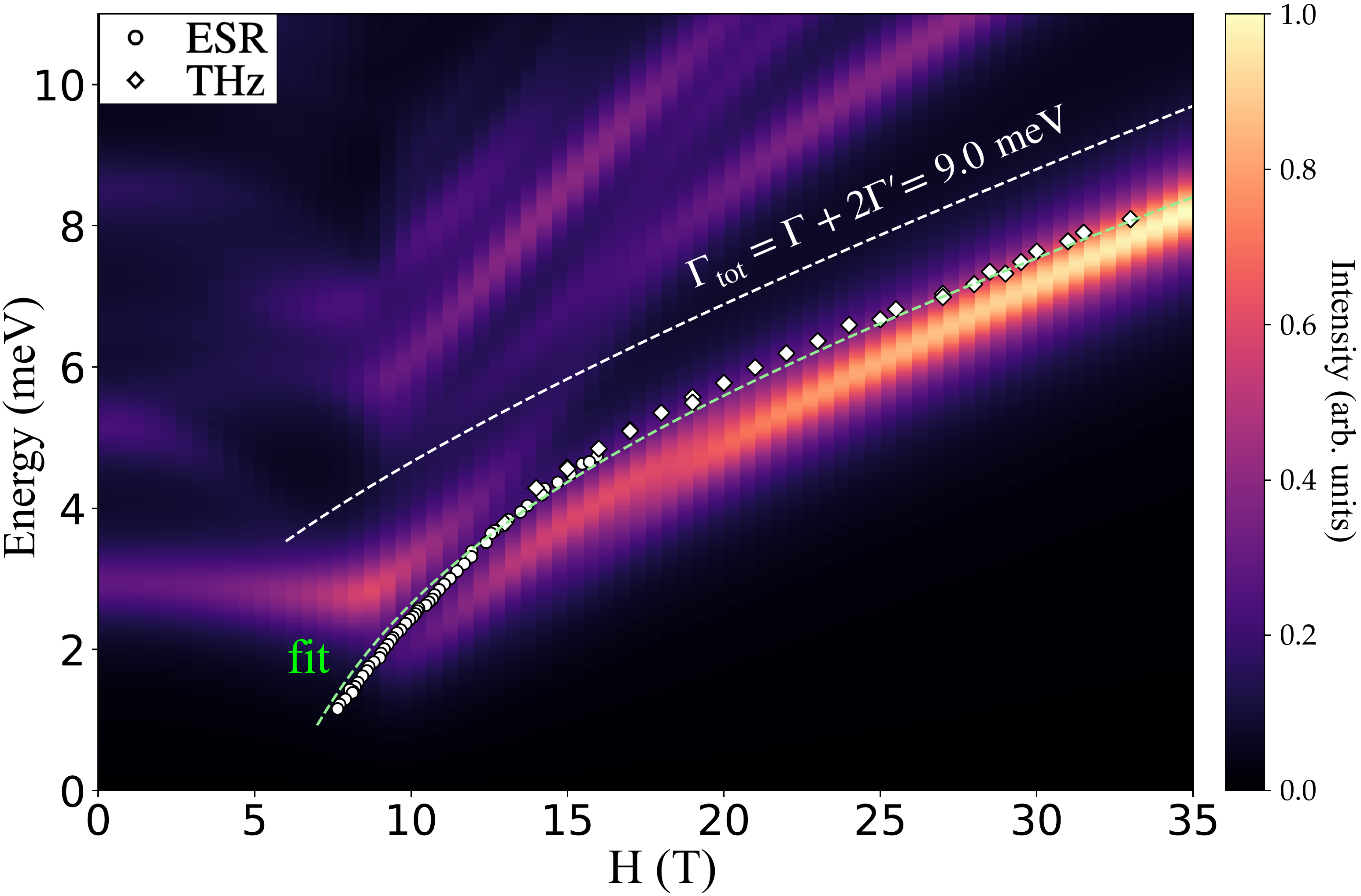} 
\vskip -0.2cm
\caption{The intensity map of ${\cal S}^{bb}(0,\omega)$  versus field in the $a$-direction obtained using a 24-site cluster ED for the Point~$\bigstar$ set, experimental ESR~\cite{Zvyagin17} and THz~\cite{kaib19} data  for the lowest-energy spin excitation energy from Fig.~\ref{fig_ESRfit}, and LSWT results from Eq.~(\ref{Ek0}) for $\Gamma_{\rm tot}\!=\!9.0$~meV for the Point~$\bigstar$ set.}
\label{fig_ED_ESR}
\end{figure}

In Figure~\ref{fig_ED_ESR}, the intensity map of the ${\bf q}\!=\!0$ dynamical spin-spin correlation function, ${\cal S}^{bb}(0,\omega)$,   obtained by ED is shown versus the field in the $a$-direction.  Also shown are the experimental data presented in Fig.~\ref{fig_ESRfit} and  the ``bare'' LSWT magnon energy $E_0$  from Eq.~(\ref{Ek0}) for $\Gamma_{\rm tot}\!=\!\Gamma+2\Gamma'\!=\!9.0$~meV, which corresponds to the Point~0 (and Point~$\bigstar$) set. The $g$-factor in ED is the same as in Fig.~\ref{fig_ESRfit}. 

The ED calculations were carried out using the Lanczos algorithm~\cite{lanczos1950iteration}, employing the continued fraction method~\cite{dagotto1994correlated} to obtain the dynamical correlation functions.   A Krylov dimension of $150$ and a Lorentzian broadening of $0.5$~meV were used; see also App.~\ref{app_C3}. 

The downward-renormalization effects of   quantum fluctuations on the ESR gap, emphasized in Sec.~\ref{Sec_ESR_gap} and Fig.~\ref{fig_ESRfit} as necessary arguments for the large value of $\Gamma+2\Gamma'$, are clear from the difference between its quasiclassical and ED values. Although not providing an ideal fit to the experimental data, the lowest-energy excitation in the ED spectrum provides a  quantitative agreement that is much closer than that of the other parameter sets for which similar calculations have been performed~\cite{Winter18,kaib19}. 

All three successful comparisons serve as a direct  {\it a posteriori} validation of our strategy for constraining the
anisotropic terms of the effective model.

\subsection{Magnetization curve}
\label{Sec_DMRGchecks_MH}

\begin{figure}[t]
\includegraphics[width=.99\linewidth]{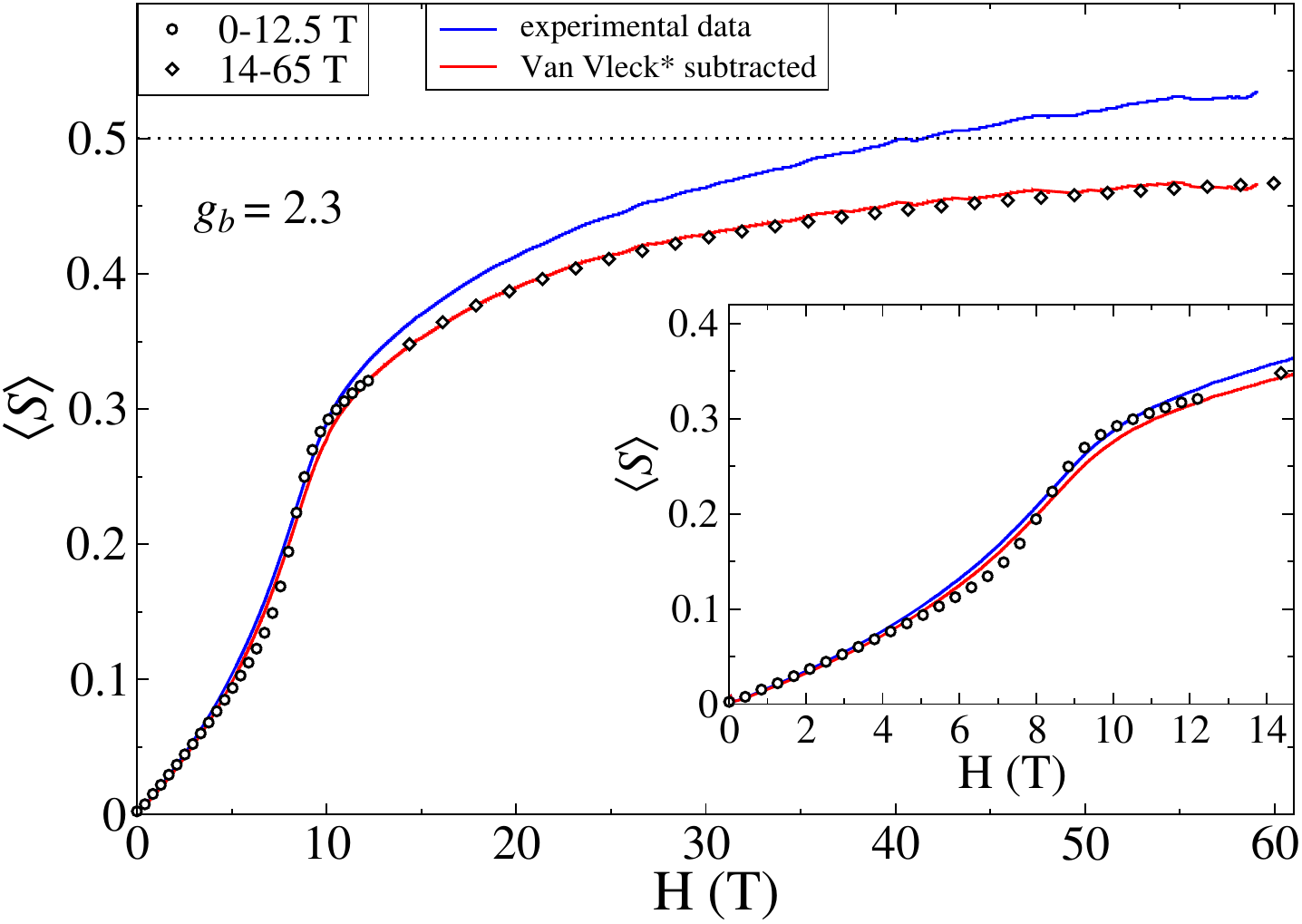} 
\vskip -0.2cm
\caption{Upper curve is the ordered moment extracted from the $M(H)$ results for $\alpha$-RuCl$_3$ in the in-plane field, Ref.~\cite{Coldea15}, symbols are DMRG results, and the lower curve is the experimental data with the Van~Vleck contribution subtracted, see text. Inset: same for the  smaller range of fields.}
\label{fig_MH}
\vskip -0.4cm
\end{figure}

As a corollary of this study, our Figure~\ref{fig_MH}   shows a comparison of the DMRG and experimental results for the ordered moment along the field  vs the in-plane field. DMRG is  for the Point~$\bigstar$ parameter set (\ref{eq_point_star}) and the low-temperature magnetization results are from Ref.~\cite{Coldea15}, which provided $M(H)$ data for $\alpha$-RuCl$_3$ for the fields up to 60~T. 

The DMRG results, shown by the symbols, are from the two $32\!\times\!12$ XC long-cylinder scans for the field along the $b$-direction. The first one is from 0~T to 12.5~T (circles), also shown in  the second panel of Fig.~\ref{fig_DMRG_Hc_star}(a), and the second scan is from 14~T to 65~T (diamonds). The upper curve is the original experimental data from Ref.~\cite{Coldea15}. 

As one can see, even at the highest field the ordered moment in DMRG is not fully saturated and retains some residual field dependence. Thus, to subtract the Van~Vleck contribution from the experimental curve more consistently, we extract the linear slope in the high-field range of $\agt\!40$~T from {\it both} the experimental and DMRG data. Then the linear contribution with the {\it difference} of these slopes is subtracted from the upper curve to obtain the lower one for all $H$. In this study, we also used a slightly smaller $g_b\!=\!2.3$ to match the value quoted in Ref.~\cite{Coldea15}. 

Needless to say, the agreement demonstrated in Fig.~\ref{fig_MH} is rather close. The DMRG results also naturally reproduce the low value of the saturated moment in the nominally polarized phase just above $H_c^{(b)}$, the slow approach of $M$ to the saturation, and its  unusual behavior showing an inflection point in the lower fields, see inset. This is not to claim that the  Point~$\bigstar$ parameter choice in Eq.~(\ref{eq_point_star}) is unique in describing $\alpha$-RuCl$_3$, but it certainly makes the case that the region from where this representative set was extracted has a lot to do with this materials' phenomenology.

\begin{figure*}[t]
\includegraphics[width=\linewidth]{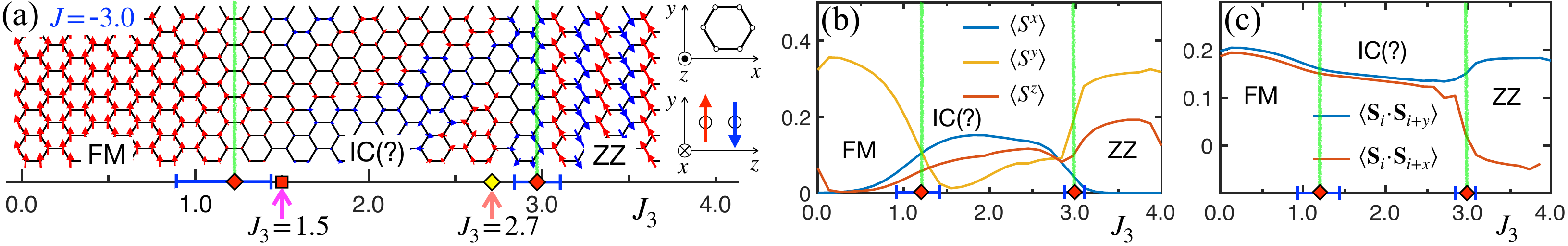} 
\caption{(a) The $32\!\times\!12$ XC long-cylinder scan with $J_3$  from 0 to 4.0~meV for the Point~0 set and $J\!=\!-3.0$~meV. The blue and red arrows arrows are the ordered moments' projections onto the $yz$ plane with $\langle S^y_i \rangle\!<\!0$ and $\langle S^y_i \rangle\!>\!0$, respectively. IC(?) marks the phase that is not identified by the scan. The square and diamond with the arrows at $J_3\!=\!1.5$~meV and  $J_3\!=\!2.7$~meV correspond to the same symbols in the phase diagram in Fig.~\ref{fig_phaseDJ1J3_0}(b) along the red line. The two non-scans  at  these values of $J_3$ are studied below, see   Figs.~\ref{fig_DMRG_IC0_1} and \ref{fig_DMRG_IC0_2}. (b) and (c) $\langle  S^\alpha_i\rangle$ components and nearest-neighbor correlators $\langle {\bf S}_i \!\cdot\! {\bf S}_{i+x(y)} \rangle$, respectively, averaged over the circumference of the cylinder,  vs $J_3$. Transitions between  phases and  error bars are indicated, see text.}
\label{fig_DMRG_IC0scan}
\end{figure*}

\section{Nature of incommensurate phases}
\label{Sec_IC}

As was made abundantly clear in the preceding pages,  the {\it only}  phases  that occur in the vicinity of the ZZ region of the advocated $\alpha$-RuCl$_3$ parameter space are the FM and  incommensurate  IC1 and IC2 phases. Although the FM state has been noted as  potentially relevant to the physics of $\alpha$-RuCl$_3$ in the past~\cite{Winter18,winter17,Keimer20}, the IC phases have not been closely scrutinized. More generally, the phase diagrams of a wide range of honeycomb-lattice Kitaev materials and models exhibit a  sequence of   ZZ, FM, and intermediate  phases~\cite{Halloran23,Winter_Co_2022,WinterReview,RauGp}, the latter of which are the singularly most understudied states. 

Given the close proximity of the IC phases to where $\alpha$-RuCl$_3$ may reside, they are in a dire need of understanding, and exploring their nature is  imperative. This exploration is conducted using LT and DMRG calculations, as presented below.  This Section also allows us to demonstrate the synchrony of our different methods and to showcase their power.
 
\subsection{IC1 and IC2: Deformed counter-rotating helices}
\label{Sec_helices}

\begin{figure*}[t]
\includegraphics[width=\linewidth]{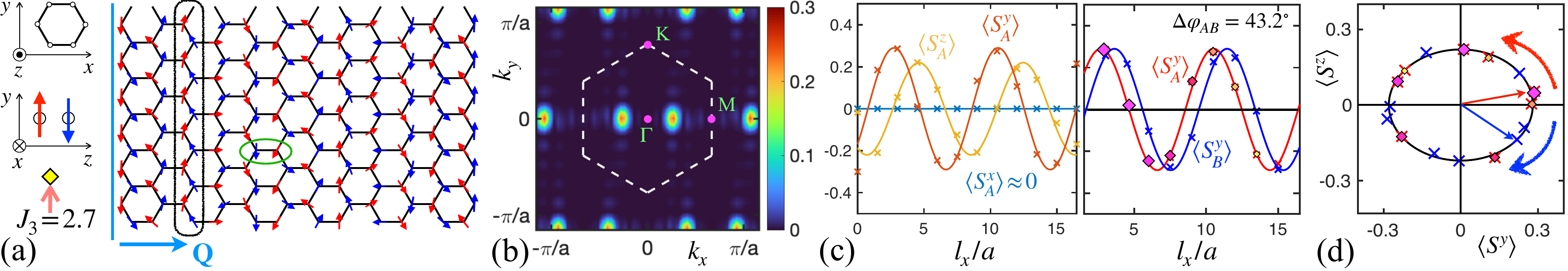} 
\caption{(a) The $12\!\times\!12$ XC-cylinder non-scan for the Point~0, $J\!=\!-3.0$~meV,  and   $J_3\!=\!2.7$~meV. Ordered moments are colored according to the A (red) and B (blue) sublattices. The unit cell (small ellipse), propagation vector ${\bf Q}$ of the counter-rotating spirals, and the vertical zigzag column (rounded rectangle) normal to ${\bf Q}$  are indicated. (b) $|\langle S^y_{B,{\bf q}}\rangle|^2$ shows the periodic single-${\bf Q}$ character of the state in (a).  First panel in (c):  $\langle S_A^{x(y,z)}\rangle$ spin components in the A sublattice along the cylinder. Spins rotate in the $yz$ plane, normal to ${\bf Q}\!\parallel\!x$. Second panel in (c): $\langle S_A^{y}\rangle$ and $\langle S_B^{y}\rangle$ vs $x$. There is a phase offset $\Delta\varphi_{AB}$ between the A and B spirals. Symbols are the data, solid lines are the spiral ansatz for the deformed counter-rotating spin helices in Eq.~(\ref{ansatz}). Colored diamonds of decreasing size mark the data for sublattice A vs $x$ to correlate with (d).  (d)  Cross-sections of the A and B helices, viewed along the $x$-axis.  Straight arrows are ordered moments from the rounded black rectangle in (a)  averaged over the vertical columns.  The  curved arrows are the directions of the spins' rotation along  ${\bf Q}$. Following the diamonds as they decrease in size from that initial column, one can see the counterclockwise rotation about the spiral for the A sublattice.}
\label{fig_DMRG_IC0_1}
\end{figure*}

Our main representative results are shown in Figs.~\ref{fig_DMRG_IC0scan}, \ref{fig_DMRG_IC0_1}, and \ref{fig_DMRG_IC0_2}. 

In Fig.~\ref{fig_DMRG_IC0scan}(a), we show the $32\!\times\!12$ XC long-cylinder DMRG scan for the Point~0 set of $\{K,\Gamma,\Gamma'\}$ with $J_3$ varying  from 0 to 4.0~meV for  fixed $J\!=\!-3.0$~meV, the path corresponding to the solid red line in the Cartesian phase diagram in Fig.~\ref{fig_phaseDJ1J3_0}(b). The arrows represent the local ordered moments $\langle {\bf S}_i \rangle$ projected onto the $yz$ plane. The honeycomb lattice is in the $xy$ plane. Spins with positive and negative $\langle S^y_i \rangle$ are shown in red and blue, respectively. 

This 1D cut through the phase diagram  as a function of $J_3$ provides a direct visualization of the FM phase, an  intermediate phase marked as IC(?) although it is not identified within the scan, and the ZZ phase.  Fig.~\ref{fig_DMRG_IC0scan}(b) shows the evolution of the three components of the on-site ordered moment, $\langle S_i^\alpha \rangle$, averaged over the vertical direction (circumference of the cylinder), vs $J_3$. Fig.~\ref{fig_DMRG_IC0scan}(c) shows the same for the nearest-neighbor correlators $\langle {\bf S}_i \!\cdot\! {\bf S}_{i+x} \rangle$ and $\langle {\bf S}_i \!\cdot\! {\bf S}_{i+y} \rangle$ averaged the same way. 

The transition points and their error bars are determined from the inflection points in the ordered moment curves and the widths of the transition regions, respectively. One can see that the transition region is  narrower and  sharper on the IC-ZZ side, consistent with both ED and LT approaches indicating the first-order character of that transition. This transition point at $J_3\!\approx\!3$~meV is also in a very close agreement with the ED results of Fig.~\ref{fig_phaseDJ1J3_0}(b) discussed in  Sec.~\ref{Sec_cartesianPD_methods}. The FM-IC transition is ``softer'' by all  indicators shown in Fig.~\ref{fig_DMRG_IC0scan}, again in agreement with both ED and LT that suggested a nearly second-order type of it. The DMRG scan also shows the FM-IC transition at $J_3\!\approx\!1.2$~meV, a somewhat lower value  than the one found by ED in  Fig.~\ref{fig_phaseDJ1J3_0}(b). 

From the ordered moment curves in Fig.~\ref{fig_DMRG_IC0scan}(b), one may also find an indication of a transition within the IC region at about $J_3\!\approx\!2.0$~meV,  suggesting two different IC phases  in the scan, with this suggestion  certainly requiring a less speedy look. Such a look is provided below in Figs.~\ref{fig_DMRG_IC0_1} and \ref{fig_DMRG_IC0_2} by the two non-scans within the IC phases at   $J_3\!=\!2.7$~meV  and $J_3\!=\!1.5$~meV, respectively. They are marked by the square and the diamond (with arrows) in Fig.~\ref{fig_DMRG_IC0scan}(a),  corresponding to the  symbols used for these points in the phase diagram in Fig.~\ref{fig_phaseDJ1J3_0}(b).

\subsubsection{IC1}
\label{Sec_helices_IC1}

Starting from $J_3\!=\!2.7$~meV, which is closer to the ZZ phase, Fig.~\ref{fig_DMRG_IC0_1}(a) shows the  $12\!\times\!12$ XC cylinder with all parameters fixed as discussed. To facilitate the visual perception of its real-space spin configuration, the local ordered moments are colored in red and blue according to their sublattices of the honeycomb lattice, A and B, respectively, with the unit cell marked by the green oval. Spins are shown in the $yz$ plane and the honeycomb lattice is in the $xy$ plane, as before.  

Focusing on the spins in the vertical zigzag columns  in Fig.~\ref{fig_DMRG_IC0_1}(a), such as the one marked by the rounded rectangle, and following their evolution along the cylinder length, it is apparent that spins in the two sublattices form  counter-rotating spirals, with  red spins rotating clockwise and blue spins counterclockwise. 

Fig.~\ref{fig_DMRG_IC0_1}(b)  complements these real-space observations. It shows a  proxy of the static structure factor, $|\langle S^y_{B,{\bf q}}\rangle|^2\!=\!|\sum_{i\in B} S_i^y e^{i{\bf q}{\bf r}_i}|^2$,  the square of the norm of the Fourier transform of one of the spin components from one sublattice, highlighting its periodicity. The peaks correspond to the propagation vector $\pm{\bf Q}$ of the spiral, with the dashed hexagon showing the Brillouin zone of the honeycomb lattice. Since the ${\bf Q}$ vector is directed along the $\Gamma$M line, parallel to the horizontal nearest-neighbor bonds, the spin state realized in the DMRG cluster in Fig.~\ref{fig_DMRG_IC0_1}(a) identifies closely with the  IC1 state of the LT approach in Fig.~\ref{fig_phaseDJ1J3_0}(a), discussed in Sec.~\ref{Sec_cartesianPD_methods}.  

Figs.~\ref{fig_DMRG_IC0_1}(c) and  \ref{fig_DMRG_IC0_1}(d) quantify this IC1 state of the two-sublattice counter-rotating spirals in more detail. To be precise, these spirals will be referred to as the deformed spin helices. The first panel of Fig.~\ref{fig_DMRG_IC0_1}(c) explicates that the helices rotate entirely in the $yz$ plane, that is, in a screw-like fashion, normal to the propagation vector ${\bf Q}||x$ of the helix.  For clarity, only the results for the A-sublattice (red spins) are shown, with the ones for the B-sublattice being very similar. 

The second panel of Fig.~\ref{fig_DMRG_IC0_1}(c)  demonstrates the phase offset between the  A and B helices. Fig.~\ref{fig_DMRG_IC0_1}(d) provides the parametric plot of the $\langle S^z_i\rangle$ vs  $\langle S^y_i\rangle$ in both sublattices, which is also  the cross-sectional view of the helices along the $x$-axis. The spins on the edges are excluded,  straight arrows show the ordered moments from the rounded black rectangle in Fig.~\ref{fig_DMRG_IC0_1}(a),  and  curved arrows show the direction of the rotation of the helices along the propagation vector ${\bf Q}$.

In Figs.~\ref{fig_DMRG_IC0_1}(c) and \ref{fig_DMRG_IC0_1}(d), the symbols are the DMRG data for  the ordered moments averaged over the vertical zigzag columns and plotted vs the $x$-coordinate of the cylinder in units of the interatomic distance $a$.  The solid curves are given by the spiral ansatz with ${\bf Q}\!\parallel\!\hat{\bf x}$
\begin{align}
{\bf \langle S\rangle}_{i,\gamma}=\hat{\bf y}\langle S^y\rangle\cos \theta_{i,\gamma} \pm\hat{\bf z}\langle S^z\rangle\sin \theta_{i,\gamma},
\label{ansatz}
\end{align}
where $\hat{\bf y}$ and $\hat{\bf z}$ are the unit vectors for the $y$ and $z$ axes, $\pm$ is for $\gamma\!=\!A(B)$, and the phases are $\theta_{i,\gamma}\!=\!{\bf Qr}_i+\varphi_\gamma$.

The in-plane and the out-of-plane ordered moments  in  Figs.~\ref{fig_DMRG_IC0_1}(c) and \ref{fig_DMRG_IC0_1}(d) are $\langle S^y\rangle\!=\!0.28$ and $\langle S^z\rangle\!=\!0.22$, and the phase offset between the A and B helices is $\Delta\varphi_{AB}\!=\!43.2\degree$. One can read the pitch of the helix from  Fig.~\ref{fig_DMRG_IC0_1}(c) as $\ell\!\approx\!7.85a$, which corresponds to the incommensurate $|{\bf Q}|\!\approx\!0.19$~r.l.u. [reciprocal lattice vector is ${\bf G}\!=\!(4\pi/3a,0)$]. 

Before moving on to the next non-scan, we note, for the record, that the DMRG non-scan in the XC cylinder in Fig.~\ref{fig_DMRG_IC0_1}(a) realizes an incommensurate spiral state. 

\subsubsection{IC2}
\label{Sec_helices_IC2}

\begin{figure*}[t]
\includegraphics[width=\linewidth]{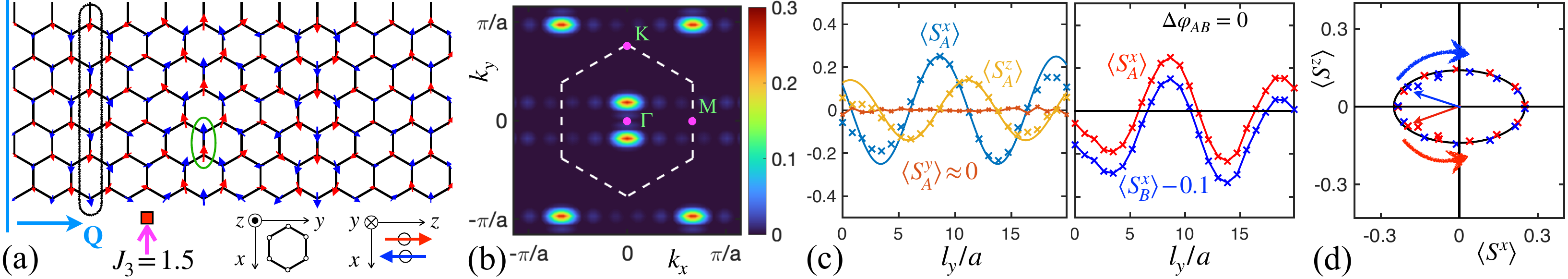} 
\caption{Same as Fig.~\ref{fig_DMRG_IC0_1} for the $6\!\times\!24$ YC-cylinder at $J_3\!=\!1.5$~meV and $S^y\!\Rightarrow\!S^x$. In the second panel of (c), the  B-sublattice data are shifted to emphasize zero phase offset.}
\label{fig_DMRG_IC0_2}
\end{figure*}

For $J_3\!=\!1.5$~meV, the point in the proximity of the FM-IC phase boundary, see Fig.~\ref{fig_DMRG_IC0scan}(a), a different state is realized, for which we found the ``rotated'' YC orientation of the cylinders to be  optimal. In Fig.~\ref{fig_DMRG_IC0_2}(a), the  $6\!\times\!24$ YC cylinder is shown, with the same color-coding for the spins according to their sublattices as in Fig.~\ref{fig_DMRG_IC0_1}(a), but  spin are now shown in the $zx$ plane, while the lattice axes are maintained as $xy$ for consistency.  

With the  evidence presented in Figs.~\ref{fig_DMRG_IC0_2}(a), \ref{fig_DMRG_IC0_2}(b), \ref{fig_DMRG_IC0_2}(c) and \ref{fig_DMRG_IC0_2}(d), following the same exposition as in Fig.~\ref{fig_DMRG_IC0_1}, it is easy to see that the state presented here  identifies  with the  IC2 state of the LT approach in Fig.~\ref{fig_phaseDJ1J3_0}(a), also discussed in Sec.~\ref{Sec_cartesianPD_methods}.  It is also composed of the two deformed counter-rotating helices in the two sublattices, with the spin plane normal to the propagation vector of the helices, but now with  the ${\bf Q}$ vector  directed perpendicular to the nearest-neighbor bonds of the honeycomb lattice, ${\bf Q}\!\parallel\!\hat{\bf y}$, which is parallel to $\Gamma$K line
\begin{align}
{\bf \langle S\rangle}_{i,\gamma}=\hat{\bf x}\langle S^x\rangle\cos \theta_{i,\gamma} \pm\hat{\bf z}\langle S^z\rangle\sin 
\theta_{i,\gamma},
\label{ansatz1}
\end{align}
Apart from the different direction of the ordering vector in the IC2 phase compared to that of the IC1, the qualitative difference is also in the lack of the phase offset between the A and B helices, meaning that in the spiral ansatz in Eq.~(\ref{ansatz1}), $\Delta\varphi_{AB}$ is zero, as is demonstrated in the second panel of Fig.~\ref{fig_DMRG_IC0_2}(c). 

Quantitatively, the in-plane and the out-of-plane ordered moments in the spiral ansatz (\ref{ansatz1}) for the solid lines in Fig.~\ref{fig_DMRG_IC0_2}(d)  are $\langle S^x\rangle\!=\!0.25$ and $\langle S^z\rangle\!=\!0.14$, exhibiting a stronger  deformation of the helices than in the IC1 state of Fig.~\ref{fig_DMRG_IC0_1}(d). The pitch of the helix in Fig.~\ref{fig_DMRG_IC0_2}(c) is also longer, $\ell\!\approx\!10.39a$, corresponding to  $|{\bf Q}|\!\approx\!0.144$~r.l.u.,  translating to the the nearly commensurate periodicity of six hexagons along the YC cylinder, as one can observe  in Fig.~\ref{fig_DMRG_IC0_2}(a).

\subsubsection{Comparison with LT and classical spiral ansatz}
\label{Sec_helices_compLT}

The IC phases were found as potentially proximate to the $\alpha$-RuCl$_3$ parameter space using the LT approach and verified with ED and DMRG.

In order to show the almost unnatural closeness of the agreement of the LT and DMRG results, we  performed additional checks of the IC states within the LT approach for the representative choices of parameters for the Point~0 selection of the anisotropic terms and two choices of the $\{J,J_3\}$ pair: $J\!=\!-3.5$~meV and $J_3\!=\!2.4$~meV for the IC1 phase and $J\!=\!-2.0$~meV and $J_3\!=\!1.8$~meV for the IC2 phase, shown by the diamond and the square, respectively, in the Cartesian LT phase diagram in Fig.~\ref{fig_phaseDJ1J3_0}(a).  

The IC1 representative choice in the LT calculations yielded the phase offset between  the A and B helices as $\Delta\varphi_{AB}\!=\!41\degree$ and the pitch of the helix as $\ell\!\approx\!7.805a$, both nearly coincidental with the corresponding numbers in the DMRG results in Sec.~\ref{Sec_helices_IC1} above.  For the IC2 choice, the LT gives zero phase offset $\Delta\varphi_{AB}$, in a complete accord with the same DMRG answer in Sec.~\ref{Sec_helices_IC2}. The pitch of the helix is somewhat shorter than the DMRG counterpart, $\ell\!\approx\!8.03a$, but not by much. 

The purely classical spiral solution suggests no dependence of the spiral's pitch on the isotropic exchanges for the fixed anisotropic ones, see App.~\ref{app_E}. This is in a broad agreement with all non-scans for the IC states that we have studied by DMRG, all showing similar values of their $|{\bf Q}|$. The LT results are less constraining, but the windows for ${\bf Q}$ for both IC1 and IC2 are narrow. 

Here is the summary of the rather amazing agreement of the LT and DMRG results on the IC phases. 
\begin{itemize}
\vspace{-0.15cm}
\item[$\circ$]  Two distinct states are realized in both approaches, IC1 and IC2, with the ordering vectors ${\bf Q}$   directed along $\Gamma$M and $\Gamma$K of the honeycomb-lattice BZ, respectively,  parallel and  perpendicular to the nearest-neighbor bonds, with IC1 bordering the ZZ phase and IC2 the FM phase.   
\vspace{-0.25cm}
\item[$\circ$] Despite the different orientation of the ordering vectors, there is no sign of a significantly sharp transition between these two ICs in either of the methods, suggesting a weak first-order transition in the thermodynamic limit. They are nearly degenerate in LT and classical calculations and DMRG scans shows only minor signs of a transition. For some of the parameters, DMRG non-scans in XC vs YC cylinders  showed different ${\bf Q}$ orientations and dependence on the initial state also hinted at a close degeneracy of the different IC states. 
\vspace{-0.25cm}
\item[$\circ$] All approaches show an impressive consistency regarding the first-order like transition at the ZZ-IC1  boundary and the  second-order or weakly first-order transition for the FM-IC2  boundary.
\vspace{-0.25cm}
\item[$\circ$] Both LT and DMRG identify IC states as the screw-like spirals, or helices, with the plane of  spin rotation orthogonal to the  propagation vector ${\bf Q}$ (nearly orthogonal for IC2 in the LT case).
\vspace{-0.25cm}
\item[$\circ$] Both LT and DMRG identify IC  states as the counter-rotating helices of spins in the two sublattices of the honeycomb structure.
\vspace{-0.25cm}
\item[$\circ$] The phase offset between the IC2 helices is zero in both LT and DMRG, and it is finite for  IC1, with values that are close between the two techniques.
\vspace{-0.25cm}
\item[$\circ$] The pitch of the helices is independent of $J$ and $J_3$, i.e., fully defined by the $\{K,\Gamma,\Gamma'\}$ choice in the classical calculations of App.~\ref{app_E}, which is in a broad agreement with all non-scans probed by DMRG. 
\end{itemize}
\vspace{-0.15cm}

This rather close accord between the classical LT and purely quantum DMRG techniques reinforces the sentiment expressed earlier in Sec.~\ref{Sec_cartesianPD_methods} on  the ability of the  LT method to mimic quantum effects of the fluctuating states that are not conserving the length of the ordered moments~\cite{KimchiLT14}, as it allows one to ``squish'' the helices in the IC phases, thus lowering the energy and letting LT to guess the states correctly.

We would be remiss if we did not mention the earlier discovery and  detailed analysis of the counter-rotating spirals in the lithium-iridate compounds, described by a similar model on related honeycomb-based structures and rationalized as a sign of their significant bond-dependent Kitaev-like exchanges, see Refs.~\cite{ColdeaPRL14,ColdeaPRB14,Kimchi1,Kimchi2,Kimchi3}.

\begin{figure*}[t]
\includegraphics[width=.99\linewidth]{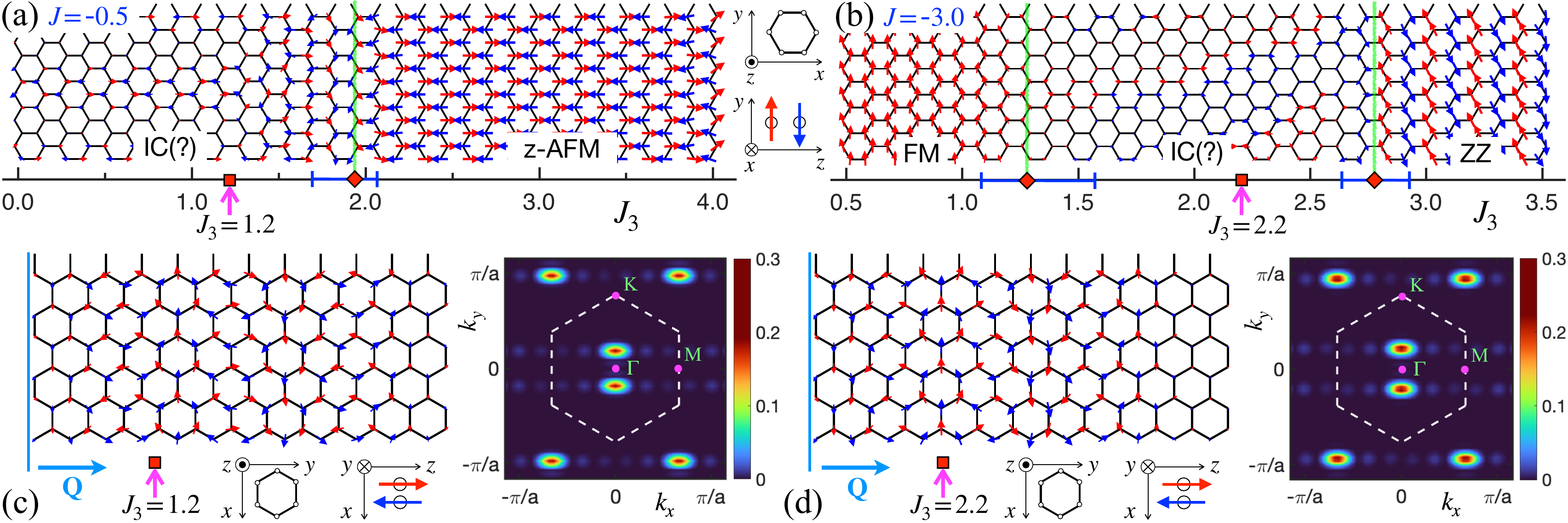} 
\caption{(a) and (b) The $32\!\times\!12$ XC long-cylinder scans for Point A and $J\!=\!-0.5$~meV and $J\!=\!-3.0$~meV, respectively, same as in Fig.~\ref{fig_DMRG_IC0scan}(a),  $J_3$'s for the non-scans in (c) and (d) are indicated. (c) and (d) The $6\!\times\!24$ YC-cylinder non-scans and $|\langle S^x_{B,{\bf q}}\rangle|^2$ for $J_3\!=\!1.2$~meV and $J_3\!=\!2.2$~meV, respectively, same as  in Figs.~\ref{fig_DMRG_IC0_2}(a) and \ref{fig_DMRG_IC0_2}(b).}
\label{fig_DMRG_ICA}
\end{figure*}

\subsubsection{Point A, more phases, more helices}
\label{Sec_helices_pointA}

Here we provide some additional DMRG analysis of the phase diagram for the Point~A anisotropic parameter set in order to reinforce the earlier findings, expose additional phases, and verify the ubiquity of the IC states.  

Our Figures~\ref{fig_DMRG_ICA}(a) and \ref{fig_DMRG_ICA}(b) provide the 1D scans through the phase diagram for the Point~A in Fig.~\ref{fig_phaseDJ1J3_A}(b) along the two vertical red lines, at $J\!=\!-0.5$~meV and $J\!=\!-3.0$~meV, respectively, vs $J_3$. Notations, color-coding, and analysis are the same as for Fig.~\ref{fig_DMRG_IC0scan}(a) above.

One can see that in both cases, the agreement for the  upper transition boundary with the ED results in  Fig.~\ref{fig_phaseDJ1J3_A}(b) is very close, while the lower boundary with the FM phase is shifted down significantly compared to ED, the trend already noted in Sec.~\ref{Sec_helices} above. This is also in agreement with the ``softer'' character of that transition. In Fig.~\ref{fig_DMRG_ICA}(a) the FM phase is simply absent, and the upper boundary is with the z-AFM N\'{e}el phase that has spins pointing along the $z$ axis, promoted by  the antiferromagnetic out-of-plane nearest-neighbor exchange in the crystallographic parametrization of the model (\ref{HJpm}) and the $J_3$ term, as discussed in Sec.~\ref{Sec_cartesianPD_methods}. 
 
In Figs.~\ref{fig_DMRG_ICA}(c) and \ref{fig_DMRG_ICA}(d), the IC states, corresponding to the choices of  $\{J,J_3\}\!=\!\{-0.5,1.2\}$~meV and $\{J,J_3\}\!=\!\{-3.0,2.2\}$~meV, respectively, and marked by the arrows in Figs.~\ref{fig_DMRG_ICA}(a) and \ref{fig_DMRG_ICA}(b), are exposed in the same  fashion as in Figs.~\ref{fig_DMRG_IC0_1}(a), \ref{fig_DMRG_IC0_1}(b) and \ref{fig_DMRG_IC0_2}(a), \ref{fig_DMRG_IC0_2}(b) above. 

In Figs.~\ref{fig_DMRG_ICA}(c) and \ref{fig_DMRG_ICA}(d), we use the $24\!\times\!6$ YC cylinders, with the spins shown in the $zx$ plane and the square of the norm of the Fourier transform of $|\langle S^x_{B,{\bf q}}\rangle|^2$ shown in the second panels. By following the same analysis  as in Figs.~\ref{fig_DMRG_IC0_1}(c) and \ref{fig_DMRG_IC0_2}(c), both states are identified as the representatives of the IC2 phase. As for the IC2 state in Fig.~\ref{fig_DMRG_IC0_2},  they are described by the two-sublattice, deformed counter-rotating helices with the spin plane normal to the propagation vector of the helices and with  the ${\bf Q}$ vector  directed along the $\Gamma$K line. 

One notable feature of all  DMRG non-scans of the IC states analyzed in our study is the close lengths of their ${\bf Q}$ vectors, suggesting the pitches of their helices to be roughly independent of the isotropic exchanges $J$ and $J_3$, in agreement with the classical consideration of App.~\ref{app_E}.

\subsubsection{IC summary}
\label{Sec_helices_summary}

The puzzling incommensurate phases in the phase diagram of $\alpha$-RuCl$_3$ are  enigmas no more.

The IC1 and IC2 phases were identified in Sec.~\ref{Sec_PhDs} as proximate to the   parameter space relevant to $\alpha$-RuCl$_3$, see Figs.~\ref{fig_phaseDJ1J3_0}, \ref{fig_phaseDJ1J3_A}, and \ref{fig_phaseDpolar1}. They were  discussed as vitally important to the understanding of their possible  effects onto the $\alpha$-RuCl$_3$ properties. They are now fully exposed and understood as  constituting two types of the counter-rotated helices with different directions of their propagation vectors as is thoroughly demonstrated above. 

The IC1 phase, bordering the immediate region of the ZZ phase  pertinent to $\alpha$-RuCl$_3$, is separated from it by a first-order transition, suggesting limited direct effects of the IC phases on the phenomenology of this material. 

If anything, this consideration also dispels the unfounded, but persistent misconception that the standard DMRG approach is incapable of identifying spiral phases, especially incommensurate ones, and would always select a different ground-state instead. Figures~\ref{fig_DMRG_IC0scan}, \ref{fig_DMRG_IC0_1}, \ref{fig_DMRG_IC0_2}, and \ref{fig_DMRG_ICA} and the detailed analysis provided above shatter that belief.

\section{Outlook}
\label{Sec_Outlook}

With the progress in finding a definitive set of parameters for an effective model of $\alpha$-RuCl$_3$ described above, is this truly the end of the $\alpha$-RuCl$_3$ parameters' drama? Yes and no. 

It is an unequivocal  ``'yes'' in the sense of much better clarity on the overall  strategy for finding such parameters, on where  the parameters of the effective model lie, and what phases are proximate to that region. 

It is a more cautious ``no,'' or ``maybe,'' for the following reasons. 

First, there is a technical issue of evaluating a wide range of  physical observables for $\alpha$-RuCl$_3$ using  parameters from the  physical regions proposed in this work.  Preferably, such calculations should be performed using unbiased numerical methods, with the goal of possibly identifying a more precise set of these parameters. A more complex reason is that the description of some experiments may require further ``dressing'' of the $\alpha$-RuCl$_3$ effective model with additional exchanges. 

There are subleading, but potentially important, 3D couplings between the honeycomb planes of $\alpha$-RuCl$_3$, with  estimates  in the range of 0.5~meV compared to 5--10~meV for the parameters of the 2D plane model considered in this work~\cite{Balz19}.  However, they cannot be simply added on top of the  2D parameters discussed here, as the parameters for that modified 3D model need to be re-evaluated to meet the same phenomenological constraints. This may or may not be a trivial rescaling of some subset of existing  2D terms, as  discussed above, depending on how  isotropic these 3D couplings are~\cite{Janssen20}. 

There is also a possibility of lower symmetry within the $\alpha$-RuCl$_3$ honeycomb planes, reducing it to $C_2$ and making exchanges on two nearest-neighbor bonds differ from those on the third~\cite{WinterReview}. Certainly, this would modify and complicate the implementation of our proposed constraints. Whether such a change can or cannot be accounted for without modifying the effective five-parameter model discussed in this work has not been investigated. The answer may also depend on which specific experiments are to be described. 

A  major direction that is not pursued in this work and left for future studies is the quantitative description of inelastic neutron scattering (INS) results for $\alpha$-RuCl$_3$ using parameters from the  physical ranges proposed here. INS measurements have clearly indicated non-negligible 3D couplings in $\alpha$-RuCl$_3$~\cite{Balz19}, so the model needs to be modified accordingly before pursuing such a study.

Another specific issue with such studies is the use of the linear spin-wave theory (LSWT) for anisotropic-exchange models in general and for the suggested ranges of the effective  $\alpha$-RuCl$_3$  model in particular, both in the zigzag state and in the nominally polarized paramagnetic one. The problem is strong quantum effects in the spectrum. For instance, one can verify for any parameter sets proposed in this work, or from their broad vicinity, that the predicted zero-field LSWT single-magnon spectrum will have an unreasonably large width and a vanishingly small gap for the lowest excitation at the accidental M-points, not affiliated with the ordering vector. 

For the  problem of large width, one can technically mitigate this effect by considering strong mixing with two-magnon continua. This should also eliminate much of the coherent spectrum at higher energies~\cite{winter17, rethinking,Smit20}, bringing the results for that part of the spectrum  into broad accord with experimental observations. For the vanishing gap,  the observed lowest-energy experimental mode is at the same ${\bf q}$-vector, but the gap is not zero. The technical solution to the problem requires a rather involved  calculation that includes fluctuation-induced corrections selfconsistently~\cite{Rau_Ir23}, making any na\"ive comparison of the LSWT spectra with the experimental ones in the fluctuating regime somewhat meaningless.

Another approach is to model the INS spectrum using numerical methods. In App.~\ref{app_C3}, we present  results for one such calculation of the dynamic spin correlation function that can be used as a proxy for the INS dynamical structure factor. The calculations are for zero field and for the parameter set suggested in this work. While  a more detailed analysis is needed to  compare these results with the available INS data, certain improvements relative to similar comparisons can be pointed out, such as a better quantitative matching of the lowest-energy excitations, in accord with experimental expectations.

One may also envision a re-investigation of the theoretical calculations of field-induced thermal transport phenomena  in $\alpha$-RuCl$_3$  using the  parameter  sets  proposed in this work~\cite{Nevidomskyy23,moore18}. However, given the emergent consensus on the substantial contribution of phonons to such effects, this research direction is more tenuous.  

Thus, the saga of $\alpha$-RuCl$_3$ continues.

\vspace{-0.3cm}
\section{Conclusions}
\label{Sec_conclusions}
\vskip -0.2cm

In this paper, we have offered a historical overview of the searches for the best effective model of $\alpha$-RuCl$_3$, which are complicated by the material's strongly fluctuating ground and field-induced states and the complex structure of its low-energy description. We have outlined an ``anisotropic strategy'' for constraining the most important spin-orbit-induced anisotropic-exchange model  parameters using existing phenomenology and  demonstrated its success for $\alpha$-RuCl$_3$ using a combination of quasiclassical analysis and unbiased numerical approaches. The same strategy can be applied to other anisotropic-exchange materials with complex models.

The resulting constrained parameter space allowed us to focus on a much narrower  region of the multi-dimensional phase diagram of the $\alpha$-RuCl$_3$ model.  The selected  representative choices of  anisotropic exchanges enabled a detailed study of the remaining  dimensions of its parameter space. 

Not only has this approach systematically led us to  the definitive parameter region of $\alpha$-RuCl$_3$ in the phase diagram, but it has also facilitated an otherwise prohibitively costly numerical exploration of its  parameter space. This exploration, with both semi-classical and quantum methods, has demonstrated  close agreement on the structure, properties, and hierarchy of the phases, demystifying relevant proximate phases of $\alpha$-RuCl$_3$. Specifically, the enigmatic nature of the incommensurate phases has been resolved in the present study, identifying them as  counter-rotating helical states.

Moreover, a verification of our proposed general anisotropic strategy and our specific choices of  phenomenological constraints has been performed using DMRG and ED, confirming the  selfconsistency of our assumptions and results, showing a notable accord between experiments and theory, and further justifying  the validity of our strategy and proposed parameters.  

In addition to a systematic analysis of  prior attempts at determining $\alpha$-RuCl$_3$ parameters and bringing closer together several approaches to the derivation of anisotropic-exchange models, our work has suggested an intuitive description of its model via a different crystallographic parametrization of the exchange matrix, offering a unifying view of the earlier assessments of its parameters and yielding important physical insights. This parametrization has also provided a connection to a broader class of  relevant paradigmatic  models in frustrated magnetism, thus expanding the context of  studies of $\alpha$-RuCl$_3$.

\begin{acknowledgments}

We would like to thank Matthias Vojta and Cristian Batista for fruitful discussions  and Radu Coldea and Steven Nagler for    sharing their experimental results. We are indebted to David Kaib for numerous insights and educational efforts. We are grateful to Giniyat Khaliullin for sharing the data and inspirational conversations. 
A.~L.~C. thanks A.~A.~Chernyshev for precious editorial inputs and appreciation of cultural references.

This work was primarily supported by the U.S. Department of Energy, Office of Science, Basic Energy Sciences under Award No. DE-SC0021221 (A.~L.~C.). 
The work by  S.~J. and S.~R.~W. was supported by the National Science Foundation under DMR-2412638. 
S.~J. is also supported by the Department of Energy (DOE), Office of Sciences, Basic Energy Sciences, Materials Sciences and Engineering Division, under Contract No. DEAC02-76SF00515.
M.~M. and R.~V. gratefully acknowledge support by the Deutsche Forschungsgemeinschaft (DFG, German Research Foundation) for funding through TRR 288--422213477 (project A05) and CRC 1487--443703006 (project A01).

We would like to thank Aspen Center for Physics (A.~L.~C.) and the Kavli Institute for Theoretical Physics (KITP, A.~L.~C. and R.~V.) where different stages of this work were advanced. The Aspen Center for Physics is supported by National Science Foundation Grant No. PHY-2210452 and KITP is supported  by the National Science Foundation under  Grant No. NSF PHY-2309135.

\end{acknowledgments}

\appendix

\section{Details}
\label{app_A}

The transformation matrix $\hat{\mathbf{R}}_c$ from the cubic to crystallographic reference frame in Fig.~\ref{fig_axes}(b) is 
\begin{align}
\hat{\mathbf{R}}_c=\left(
\begin{array}{ccc}
 -\frac{1}{\sqrt{2}} & \frac{1}{\sqrt{2}} & 0 \\
 -\frac{1}{\sqrt{6}} & -\frac{1}{\sqrt{6}} &  \sqrt{\frac{2}{3}}  \\
 \frac{1}{\sqrt{3}} & \frac{1}{\sqrt{3}} & \frac{1}{\sqrt{3}} \\
\end{array} 
\right),
\label{eq_cubic_transform}
\end{align}

\begin{table*}[t]
  \begin{tabular}{ | l | c || c | c | c | c | c | c |}
    \hline
\text{Reference} 
& \text{Method} 
& $K$ & $\Gamma$ & $\Gamma'$  & ${\sf J_{\pm \pm}}$ & ${\sf J_{z\pm}}$ & ${\sf J}_1(1-\Delta)$ \\ \hline\hline
\multirow{3}{2.5cm}{Kim et al. \cite{kee16}} 
& DFT+$t/U$, $P3$  & {-6.55} & {5.25} & -0.95  & -0.98 & {\bf -6.01} & -3.35 \\ \cline{2-8}
& DFT+SOC+$t/U$   & {-8.21} & {4.16} & -0.93   & -0.33 & -6.26 & -2.3   \\ \cline{2-8}
& same+fixed lattice & -3.55  &  7.08    & -0.54  &-1.95 & {\bf -5.27} & -6.01 \\ \cline{2-8} \hline   
Winter et al. \cite{winter16}
& DFT+ED, $C2$         & {-6.67} & { 6.6}  & -0.87 & -1.38 & -6.66 & -4.87 \\ \cline{2-8} \hline
&  DFT+$t/U$, $U\!=\!2.5$eV  & -14.43  & 6.43 &    & {\bf 0.26} & -9.84 &-6.43 \\ \cline{2-8}
Hou et al. \cite{gong17} 
& same, $U\!=\!3.0$eV & -12.23 & 4.83 &          & {\bf 0.43} & -8.05 & -4.83  \\ \cline{2-8}
& same, $U\!=\!3.5$eV & -10.67 & 3.8   &          & {\bf 0.51} & -6.82 & -3.80  \\ \hline
\multirow{2}{2.5cm}{Wang et al. \cite{li17}} 
& DFT+$t/U$, $P3$    & {-10.9}  &  6.1    &           & -0.22 & -8.01 & -6.1 \\ \cline{2-8}
& same, $C2$              & {-5.5}    & {7.6}   &           & -1.62 & -6.18 & {\bf -7.6}  \\ \hline
Eichstaedt\! et\! al. \cite{berlijn19} 
& DFT+Wannier+$t/U$ & -14.3  & 9.8      & -2.23 & -1.63 & -12.41 & -5.33 \\ \hline
Ran et al. \cite{wen17} 
& LSWT, INS fit           & {-6.8}    &   9.5   &            &-2.03 & -7.68 & {\bf -9.5} \\ \hline
Winter et al. \cite{winter17} 
& \textit{Ab initio}+INS fit & -5.0  & 2.5  &           & 0.0    & {\bf -3.54}  & -2.5    \\ \hline
Suzuki et al. \cite{suga18} 
& ED, $C_p$ fit            & -24.41     & 5.25 & -0.95 & 2.0   & -14.43 & -3.35  \\ \hline
Cookmeyer\! et\! al. \cite{moore18} 
& thermal Hall fit        & {-5.0}     &     2.5 &           & 0.0   & {\bf -3.54}   & -2.5  \\ \hline
 Wu et al. \cite{orenstein18} 
& LSWT, THz fit           & -2.8        &     2.4 &           & -0.33 & {\bf -2.45}  & -2.4   \\ \hline
Ozel et al. \cite{gedik19}
& same                         & -3.5        &  2.35  &           & -0.2   & {\bf -2.76}   & -2.35   \\ \hline
Sahasrabudhe et al. \cite{kaib19} 
& ED, Raman fit           & {-10.0}  & 3.75   &            & {\bf 0.42}  &  -6.48   & -3.75 \\ \hline
\multirow{2}{2.5cm}{Sears et al. \cite{kim19}}
& \multirow{2}{2.9cm}{Magnetization fit} & -10.0 & 10.6 & -0.9 & -2.17 & -10.14 & {\bf -8.8}  \\  \cline{3-8}
&                                     & {-10.0}  & 8.8    &             & -1.27 & -8.86   & {\bf -8.80}  \\ \hline
Laurell et al. \cite{okamoto19} 
& ED, $C_p$ fit            & {-15.1}   & 10.1  & -0.12   & -0.89 & -11.94 & {\bf -9.86}  \\ \hline
Suzuki  et al. \cite{Keimer20} 
& RIXS                           & -5.0        & 2.5     & +0.1      & 0.03   &  {\bf -3.49}  & -2.7 \\ \hline
Kaib et al. \cite{Kaib20}
& GGA+U                     & -10.12    & 9.35   & -0.73   & -1.67 & -9.52   & {\bf -7.89}  \\ \hline
Andrade et al. \cite{Andrade20}
& $\chi$                       & -6.6        & 6.6      &             & -1.1   &  -6.22  &  -6.6 \\ \hline
Janssen et al. \cite{Janssen20}
& LSWT+3D                 & -10.0      & 5.0      &             &  0.0   &  -7.07  &  -5.0  \\ \hline
Li et al. \cite{Li21}
& $C_m$, $\chi$        &-25.0       & 7.5      & -0.5     & 1.5    & -15.56 & -6.5  \\ \hline
Ran et al. \cite{Ran22}
& polarized INS           & -7.2        & 5.6      &             & -0.67 & {\bf -6.03}   & -5.6  \\ \hline
Samarakoon et al. \cite{Samarakoon22} \ 
& \ Machine learning, INS \ & -5.3  &  0.15   &        & {\bf 0.83}  & {\bf -2.57}   & -0.15  \\ \hline
Liu et al. \cite{Giniyat22}
& downfolding            & -5.0        &  2.8     & +0.7      &  {\bf 0.13} & {\bf -3.35}   & -4.2 \\ \hline
    \hline
\multirow{4}{2.5cm}{ \ \ \ \ \  \bf This work}  
&  {\bf realistic range}  
&\ {\bf [-10.0,-4.4]}\  &\  {\bf [3.2,5.0]} \   &\  {\bf [1.8,2.85]}\  & \ {\bf [0.16,1.0]} \ &  \ {\bf [-5.9,-2.6]} \ & {\bf [-10.0,-7.5]} \\ \cline{2-8}
&  {\bf point 0}   & {\bf -7.57}  & {\bf 4.28} & {\bf 2.36}   & {\bf 0.62}  & {\bf -4.47} &  {\bf -9.0}  \\ \cline{2-8}
& {\bf point A}    & {\bf -5.43} & {\bf 3.65}  & {\bf 2.18}   & {\bf 0.414}  & {\bf -3.25}  & {\bf -8.0}    \\ \cline{2-8}
&   {\bf point B}  & {\bf  -8.73}  & {\bf 4.71}   & {\bf 2.39} & {\bf 0.68}  & {\bf -5.21}  & {\bf -9.5}  \\ \hline
    \hline    
\end{tabular}
\caption{Same as Table~\ref{table3} with the $\{ {\sf J_1}(1-\Delta),{\sf J}_{\pm \pm},{\sf J}_{z\pm}\}$ parameters of the model in the crystallographic parametrization (\ref{HJpm})  converted from the generalized KH model  (\ref{H_JKGGp}) using Eq.~(\ref{eq_JpmKG}), see Secs~\ref{Sec_intro_alternative} and \ref{Sec_parameter_space}.}
\label{table2}
\vskip -0.2cm
\end{table*}

The form of the Hamiltonian \eqref{eq_Hij} in the crystallographic $\{x,y,z\}$ axes 
of the honeycomb plane is given by Eq.~(\ref{HJpm}). Its parameters are related to that of the generalized KH model in 
the cubic axes \eqref{eq_Hij} via
\begin{align}
{\sf J_1}&=J+\frac{1}{3} \big( K-\Gamma-2\Gamma'\big),\nonumber\\
\big(1-\Delta\big) {\sf J_1}&=- \big( \Gamma+2\Gamma'\big),\nonumber\\
\label{eq_JpmKG}
{\sf J_{\pm\pm}}&=-\frac{1}{6} \big( K+2\Gamma-2\Gamma'\big),\\
{\sf J_{z\pm}}&=\frac{\sqrt{2}}{3} \big( K-\Gamma+\Gamma'\big).\nonumber
\end{align}
The inverse relation is
\begin{align}
J&=\frac{1}{3}\left( 2{\sf J_1}+\Delta {\sf J_1}+2{\sf J_{\pm\pm}}-\sqrt{2} {\sf J_{z\pm}}\right),\nonumber\\
\label{eq_jkg_transform}
K&=-2{\sf J_{\pm\pm}}+\sqrt{2}{\sf J_{z\pm}},\\
\Gamma&=\frac{1}{3} \left( -{\sf J_1}+\Delta {\sf J_1}-4{\sf J_{\pm\pm}}-\sqrt{2} {\sf J_{z\pm}}\right),\nonumber\\
\Gamma'&=\frac{1}{6} \left( -2{\sf J_1}+2\Delta {\sf J_1}+4{\sf J_{\pm\pm}}+\sqrt{2} {\sf J_{z\pm}}\right).\nonumber
\end{align}
Using Eq.~(\ref{eq_JpmKG}), one can convert  the comprehensive compilation of the  $\{ K,\Gamma,\Gamma'\}$ parameter sets of the generalized KH model  (\ref{H_JKGGp})  in Table~\ref{table3},  which were previously proposed for $\alpha$-RuCl$_3$, to  the $\{ {\sf J_1}(1-\Delta),{\sf J}_{\pm \pm},{\sf J}_{z\pm}\}$ parameters of the model in the crystallographic parametrization (\ref{HJpm}). The result of such a conversion is presented in our  Table~\ref{table2}, see also Secs~\ref{Sec_intro_alternative} and \ref{Sec_parameter_space}.

\section{LT Details}
\label{app_B}

The phase diagrams in Fig.~\ref{fig_phaseDJ1J3_B_LT}, Fig.~\ref{fig_phaseDpolar}, Fig.~\ref{fig_phaseDJ1J3_0}(a), Fig.~\ref{fig_phaseDJ1J3_A}(a), Figs.~\ref{fig_phaseDpolar1}(a) and (c), and  Fig.~\ref{fig_phaseDpolarJpp} are obtained using the Luttinger-Tisza (LT) method~\cite{lt_original,Lyons_Kaplan_1960,Friedman_1974,Litvin_1974}. Here we briefly outline its basics. 

The goal of the LT method is to find the spin arrangement of the  classical spins $\mathbf{S}_{i}$, interacting via binary interactions, that minimizes their  energy under the following ``strong'' condition on the spin length 
\begin{align}
	\big(\mathbf{S}_{i}\big)^2 = S^2_i, 
	\label{eq:strongconditionsoriginalLT}
\end{align}
where $S_i$ is the spin length on the site $i$. Since, in practice, that would require a macroscopic number of Lagrange multipliers, Luttinger and Tisza introduced the so-called ``weak'' condition
\begin{align}
	\sum_{i}\big(\mathbf{S}_{i}\big)^2 = \sum_{i} S^2_i ,
	\label{eq:weakconditionoriginalLT}
\end{align}
replacing the strong local condition on the spin magnitude by the global constraint on  the average spin length. 

The weak condition is a necessary, but not sufficient condition for the spin arrangement to satisfy the strong condition. If a solution obtained by enforcing the weak condition also fulfills the strong condition, the original problem is solved. Modifications of the  LT method using alternative weak conditions have also been proposed, see Refs.~\cite{Freiser61,Lyons_Kaplan_1960,Yosida68}. 

The numerical method typically carried out today under the name of Luttinger-Tisza is simplified and involves diagonalization of the Fourier transform of the exchange matrix~\cite{Li_2016,kaplanLT2007,
multiQ,Wang17,Zhu19,Niggemann_2019,K1K2,Maksimov22} and finds the lower limit of the energy  by applying the weak condition.  In the  LT approach, the lowest eigenenergy of the exchange matrix  in the momentum space and the corresponding ordering vector are identified by a scan through the reciprocal space. The associated spin arrangement is found by the  Fourier transform back into the real space. The implementation of the LT method is computationally   straightforward. 

\subsection{LT formalism}
\label{app_B1}

The most general lattice Hamiltonian with binary interactions of the classical spins  is given by
\begin{align}
\label{cl_model1}
{\cal H}=\sum_{\langle ij\rangle} \mathbf{S}_i^{\rm T} \hat{\rm \bf J}_{ij} \mathbf{S}_j\,,
\end{align}
where the lattice indices $i$ and $j$ run over {\it all} sites of the lattice, $\langle ij\rangle$ denotes the corresponding bonds, and the $3\times3$ exchange matrix depends only on $\mathbf{r}_{i}-\mathbf{r}_{j}$ because of the translational invariance.  To capitalize on the latter, it is convenient to rewrite the coordinates of the spins as $\mathbf{r}_{i}\!=\!\mathbf{R}_\ell+{\bm \rho}_\alpha$, where  $\mathbf{R}_\ell$ is the  unit cell coordinate and ${\bm \rho}_\alpha$ is the the coordinate
of the spin $\alpha$ sublattice within the unit cell. Then, the model (\ref{cl_model1})  can be rewritten as
\begin{align}
\label{cl_model2}
{\cal H}=\frac12\sum_{\ell,\ell'}\sum_{\alpha,\beta}\mathbf{S}_{\ell,\alpha}^{\rm T} \hat{\rm \bf J}^{\alpha\beta}_{\ell\ell'} \mathbf{S}_{\ell',\beta}\,,
\end{align}
separating summation over $i=\{\ell,\alpha\}$ into the ones over the unit cells and sublattices.

The Fourier transform of the 3D vectors ${\bf S}_{\ell,\alpha}$
\begin{eqnarray}
{\bf S}_{\ell,\alpha}=
\frac{1}{\sqrt{N}} \sum_{{\bf q}} {\bf S}_{\alpha {\bf q}} \, e^{i{\bf q}({\bf R}_\ell+{\bm \rho}_\alpha)}\, ,
\label{eq_fourier}
\end{eqnarray}
where $N$ is the number of  unit cells, with ${\bf R}_\ell$ and  ${\bm \rho}_\alpha$ defined above, allows one to rewrite the model in (\ref{cl_model2}) in the reciprocal space, using ${\bf S}_{\alpha-\mathbf{q}}\!=\!{\bf S}^*_{\alpha\mathbf{q}}$, as
\begin{align}
\label{cl_model_q1}
{\cal H}=\sum_\mathbf{q}\sum_{\alpha\beta} \mathbf{S}_{\alpha\mathbf{q}}^{\dagger} \hat{\rm \bf J}_{\alpha\beta}(\mathbf{q}) \mathbf{S}_{\beta\mathbf{q}}\,,
\end{align}
with  the exchange matrix $\hat{\rm \bf J}_{\alpha\beta}(\mathbf{q})$ in the momentum space  
\begin{align}
\label{J_q}
\hat{\rm \bf J}_{\alpha\beta}(\mathbf{q})=\frac{1}{2}\sum_{\Delta{\bf R}_{\ell\ell'}} \hat{\rm \bf J}^{\alpha\beta}_{\ell\ell'} \ e^{i\mathbf{q} (\Delta{\bf R}_{\ell\ell'}+{\bm \rho}_\beta-{\bm \rho}_\alpha)},
\end{align}
where $\Delta{\bf R}_{\ell\ell'}\!=\!\mathbf{R}_{\ell'}-\mathbf{R}_\ell$.

The form in (\ref{cl_model_q1}) allows for a more compact writing using the ``combined'' vector of the spins in all sublattices ${\widetilde{\bf S}}^\dag({\bf q})=\big({\bf S}^\dag_{1\mathbf{q}},{\bf S}^\dag_{2\mathbf{q}},\dots,{\bf S}^\dag_{\alpha\mathbf{q}},\dots,{\bf S}^\dag_{N_s\mathbf{q}} \big)$, where $N_s$ is the number of sublattices, resulting in 
\begin{align}
\label{cl_model_q2}
{\cal H}=\sum_\mathbf{q}	\mathbf{\widetilde{S}}^{\dagger}(\mathbf{q}) {\tilde{\rm \bf J}}(\mathbf{q}) \mathbf{\widetilde{S}}(\mathbf{q}),
\end{align}
with the $3N_s\times3N_s$ matrix ${\tilde{\rm \bf J}}(\mathbf{q})$ built from the  $3\times3$ blocks of $\hat{\rm \bf J}_{\alpha\beta}(\mathbf{q})$ matrices.

Since the exchange matrix ${\tilde{\rm \bf J}}(\mathbf{q})$ is  hermitian, there exist unitary matrices $U(\mathbf{q})$  that diagonalize it
\begin{align}
	U(\mathbf{q}) {\tilde{\rm \bf J}}(\mathbf{q}) U^{\dagger}(\mathbf{q}) = \hat{\bm\lambda}(\mathbf{q}),
\end{align}
where $\hat{\bm\lambda}(\mathbf{q})$ is a diagonal $3N_s\times3N_s$ matrix for all $\mathbf{q}$. 

Straightforwardly, the Hamiltonian (\ref{cl_model_q2})  is diagonal  in the basis of the ``rotated'' spin vectors,
${\bf \bar{S}}({\bf q})\!=\!U {\bf \widetilde{S}}({\bf q})$,
\begin{align}
\label{cl_model_q3}
{\cal H}=\sum_\mathbf{q}	{\bf \bar{S}}^{\dagger}(\mathbf{q}) \hat{\bm\lambda}(\mathbf{q}){\bf \bar{S}}({\bf q})=
\sum_{{\bf q},\alpha}\lambda_{\alpha{\bf q}}\big|{\bf \bar{S}_{\alpha{\bf q}}}\big|^2,
\end{align}
with $\lambda_{\alpha{\bf q}}$ being the eigenvalue of the exchange matrix ${\tilde{\rm \bf J}}(\mathbf{q})$ in the $\alpha$-sublattice sector for a given ${\bf q}$. 

If there is a momentum ${\bf Q}$ at which these eigenvalues achieve the minimal $\lambda_{\bf Q}$, such that  $\lambda_{\alpha{\bf q}}\!\geq\!\lambda_{\bf Q}$ for all ${\bf q}$ and $\alpha$, it means that  the spin configuration minimizing the classical energy has been identified, because 
\begin{align}
{\cal H} \geq \lambda_{\bf Q}\sum_{{\bf q},\alpha}\big|{\bf S_{\alpha{\bf q}}}\big|^2 = \lambda_{\bf Q} NN_sS^2\,, 
\end{align}
where the assumption on spin length being the same for all  sublattices, $S_i=S$, has been made for simplicity.

The connection to the LT method is in the last equation, as it corresponds to applying the weak condition in Eq.~\eqref{eq:weakconditionoriginalLT}, 
\begin{align}
	\sum_{{\bf q},\alpha}\big|{\bf S_{\alpha{\bf q}}}\big|^2  = \sum_{\ell,\alpha} \big( \mathbf{S}_{\ell,\alpha}\big)^2=  \sum_{i} \big( \mathbf{S}_{i}\big)^2=N N_s S^2.
\end{align}
One must note that the same result can be obtained  by  minimizing the classical energy in Eq.~(\ref{cl_model_q2})  using the Lagrange multiplier for the weak constraint in (\ref{eq:weakconditionoriginalLT}), the procedure that yields the  eigenvalue problem directly for $\lambda_{\bf Q}$~\cite{K1K2}, much in the spirit of the original LT approach.

The spin eigenstates corresponding to  the minimal eigenvalues $\lambda_{\bf Q}$ found within the LT approach usually are the so-called  single-$\mathbf{Q}$ structures, for which  $\mathbf{S} (\pm \mathbf{Q})$ must be kept to determine the type of the spin arrangement, although the generalizations to the multi-${\bf Q}$ states are also possible, see, e.g., Ref.~\cite{multiQ}. In many cases,  the commensurate ordering vectors  $\mathbf{Q}$ already give  a clear idea of the type of spin ordering. 

If the  LT eigenstate also satisfies the strong constraint~\eqref{eq:strongconditionsoriginalLT}, then LT method gives the correct classical ground state. In our case, the strong constraint is  satisfied for the FM, ZZ, AFM, and stripy phases. Otherwise, it formally  breaks down, see, however,  the discussions in Secs.~\ref{Sec_helices_compLT} and \ref{Sec_cartesianPD_methods}.
 
Below, we list  explicit expressions for the  elements of the  exchange matrices that were used to create the LT phase diagrams in Fig.~\ref{fig_phaseDJ1J3_B_LT}, Fig.~\ref{fig_phaseDpolar}, Fig.~\ref{fig_phaseDJ1J3_0}(a), Fig.~\ref{fig_phaseDJ1J3_A}(a), Figs.~\ref{fig_phaseDpolar1}(a) and (c), and  Fig.~\ref{fig_phaseDpolarJpp}.
 
\begin{figure*}[t]
\includegraphics[width=1\linewidth]{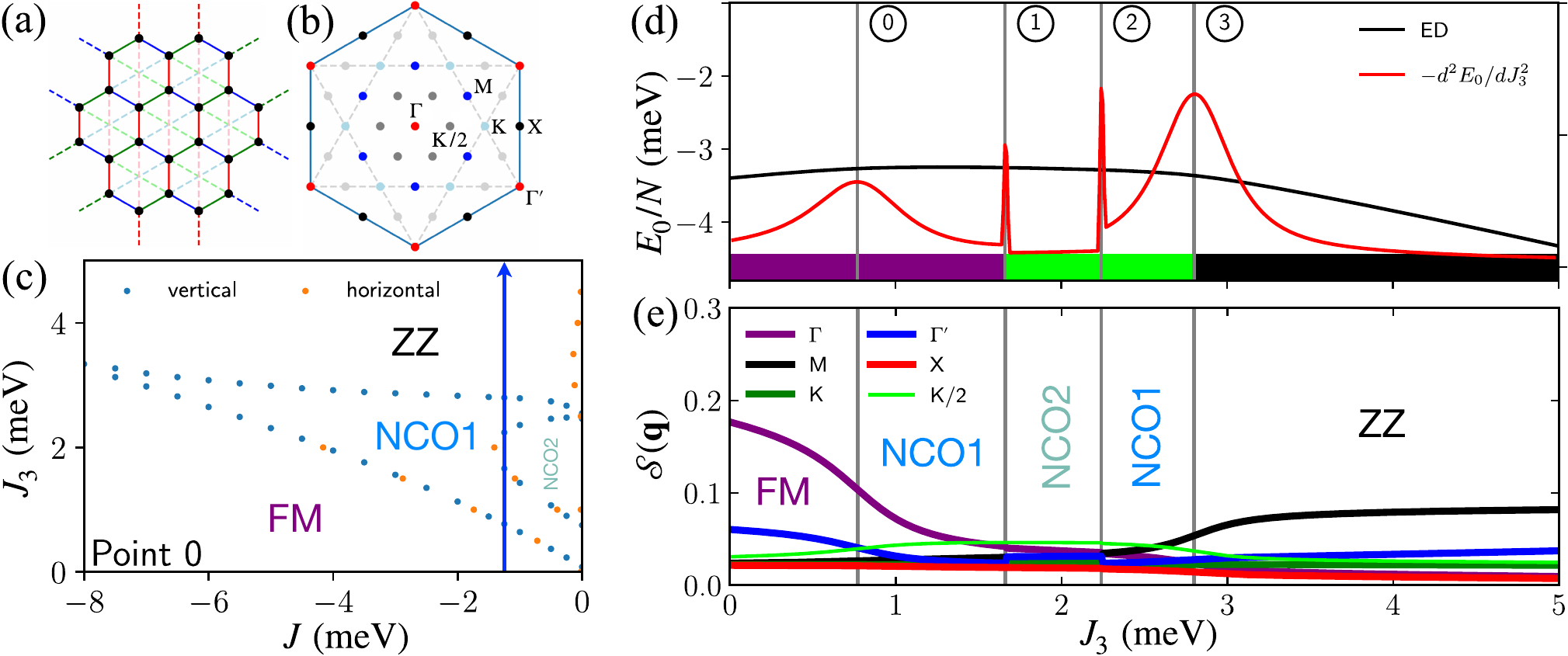} 
\caption{(a) The 24-site cluster used in the ED calculations; the outer bonds  visualize its periodicity. (b) The allowed momenta in the reciprocal space with the high-symmetry points. (c) Dots mark maxima in  $-\partial^2 E_0/\partial J_{(3)}^2$ in the vertical and horizontal sweeps for the Point~0 parameter set, see Figs.~\ref{fig_phaseDJ1J3_0}(b). (d) ED ground state energy per site and its second derivative vs $J_3$ for $J_1\!=\!-1.25$~meV and  Point~0 parameters. (e) Structure factor ${\cal S}({\bf q})$ at the high-symmetry points for the same sweep.}
\label{fig_ED1}
\end{figure*}

\subsection{Application to the ${\sf J_1}$--$\Delta$--${\sf J_{\pm\pm}}$--${\sf J_{z\pm}}$--$J_3$ and $K$--$J
$--$\Gamma$--$\Gamma'$--$J_3$ honeycomb lattice models}
\label{app_B2}

The honeycomb lattice considered in this work is bipartite  with the same spin length on each site. The couplings that are of interest for the model  (\ref{eq_Hij}) concern the first- and the third-nearest neighbors, providing connections only between the A and B sublattices, see Fig.~\ref{fig_axes}.  

With that, using the primitive vectors from  Fig.~\ref{fig_axes}(c), ${\bm \delta}_1 \!=\! a(1,0)$ and ${\bm \delta}_{2(3)}\!=\!a(-1/2,\pm\sqrt{3}/2)$, with $a$ being the lattice spacing, and  ${\bm \delta}^{(3)}_\alpha \!=\! -2{\bm \delta}_\alpha$, one can straightforwardly derive the elements of the Fourier transformed $3\times3$ exchange matrix $\hat{\rm \bf J}_{AB}({\bf q})$. 

For the model (\ref{HJpm})  in the ${\sf J_1}$--$\Delta$--${\sf J_{\pm\pm}}$--${\sf J_{z\pm}}$--$J_3$  representation they are
\begin{align}
J^{xx}_{AB}&={\sf J_1} \gamma_\mathbf{q}+2{\sf J}_{\pm\pm} \gamma'_\mathbf{q}+J_3 \gamma^{(3)}_\mathbf{q} ,\nonumber\\
J^{yy}_{AB}&={\sf J_1} \gamma_\mathbf{q}-2{\sf J}_{\pm\pm} \gamma'_\mathbf{q}+J_3 \gamma^{(3)}_\mathbf{q},\nonumber\\
\label{appB:Matrix1}
J^{zz}_{AB}&=\Delta {\sf J_1} \gamma_\mathbf{q}+J_3 \gamma^{(3)}_\mathbf{q},\\
J^{xy}_{AB}&=J^{yx}_{AB}=-2{\sf J}_{\pm\pm} \gamma''_\mathbf{q}, \nonumber\\
J^{xz}_{AB}&=J^{zx}_{AB}={\sf J}_{z\pm} \gamma''_\mathbf{q}, \nonumber\\ 
J^{yz}_{AB}&=J^{zy}_{AB}=-{\sf J}_{z\pm} \gamma'_\mathbf{q}, \nonumber
\end{align}
where   the hopping amplitudes are
\begin{align}
\label{gks}
\gamma_{\bf q}&=\sum_{\alpha} e^{i{\bf q}{\bm \delta}_\alpha},\ 
\gamma^{(3)}_{{\bf q}}=\sum_{\alpha} e^{i{\bf q}{\bm \delta}^{(3)}_\alpha},\\
\gamma'_{\bf q}&=\sum_{\alpha} \cos\tilde{\varphi}_\alpha e^{i{\bf q}{\bm \delta}_\alpha}, \ \ 
\gamma''_{\bf q}=\sum_{\alpha} \sin\tilde{\varphi}_\alpha e^{i{\bf q}{\bm \delta}_\alpha},
\nonumber
\end{align}
and the bond-dependent phases $\tilde{\varphi}_\alpha\!=\!\{0,2\pi/3,-2\pi/3\}$  for the $\alpha\!=\!\{1,2,3\}$ bonds  in Fig.~\ref{fig_axes} are the bond angles of the primitive vectors ${\bm \delta}_\alpha$ with the $x$ axis, as before.  Obviously, since the real-space exchanges are real, the  $\hat{\rm \bf J}_{AB}({\bf q})$ matrix is symmetric. 

The matrix elements of $\hat{\rm \bf J}_{AB}({\bf q})$ in the $K$--$J$--$\Gamma$--$\Gamma'$--$J_3$ parameterization (\ref{H_JKGGp}) are
\begin{align}
J^\text{xx}_{AB}&= J\gamma_\mathbf{q} +K e^{i\mathbf{q} {\bm \delta}_2}+J_3 \gamma^{(3)}_\mathbf{q}, \nonumber\\
J^\text{yy}_{AB}&= J\gamma_\mathbf{q} +K e^{i\mathbf{q} {\bm \delta}_3}+J_3 \gamma^{(3)}_\mathbf{q}, \nonumber\\
\label{appB:Matrix2}
J^\text{zz}_{AB}&= J\gamma_\mathbf{q} +K e^{i\mathbf{q} {\bm \delta}_1}+J_3 \gamma^{(3)}_\mathbf{q}, \\
J^\text{xy}_{AB}&=J^\text{yx}_{AB}= \Gamma e^{i\mathbf{q} {\bm \delta}_1} +\Gamma' \left(  e^{i\mathbf{q} {\bm \delta}_2}+ e^{i\mathbf{q} {\bm \delta}_3}\right), \nonumber\\
J^\text{xz}_{AB}&=J^\text{zx}_{AB}= \Gamma e^{i\mathbf{q} {\bm \delta}_3} +\Gamma' \left(  e^{i\mathbf{q} {\bm \delta}_1}+ e^{i\mathbf{q} {\bm \delta}_2}\right), \nonumber\\
J^\text{yz}_{AB}&=J^\text{zy}_{AB}= \Gamma e^{i\mathbf{q} {\bm \delta}_2} +\Gamma' \left(  e^{i\mathbf{q} {\bm \delta}_1}+ e^{i\mathbf{q} {\bm \delta}_3}\right). \nonumber
\end{align}
Using $\hat{\rm \bf J}_{BA}\!=\!\hat{\rm \bf J}_{AB}^{\dagger}$, numerical diagonalization together with the scan of the reciprocal space provide the ground state and the ordering vector ${\bf Q}$ for the given choice of the model parameters. Further analysis of the spin arrangement can be done using the inverse Fourier transform of the spin eigenvectors.

\section{ED Details}
\label{app_C}

The exact diagonalization (ED) calculations for the phase diagrams in Figs.~\ref{fig_phaseDJ1J3_B_ED},  \ref{fig_phaseDJ1J3_0}(b), \ref{fig_phaseDJ1J3_A}(b), and \ref{fig_phaseDpolar1}(b) were performed on the  high-symmetry  24-site honeycomb-lattice cluster with periodic boundary conditions, shown in Fig.~\ref{fig_ED1}(a).   It respects all space-group symmetries of the lattice. The allowed momenta of the reciprocal space in the first three Brillouin zones are shown in Fig.~\ref{fig_ED1}(b) together with the high symmetry points. 

The ED phase diagrams are obtained by  one-dimensional sweeps along various paths through the parameter space,  with the quantum model solved to find its ground state  at each point of the sweep along the varying parameter $X$~\cite{Wang17,Trebst_tr,Winter18,Trebst22,Hickey23}. The phase boundaries are identified with the positions of the maxima in the (negative) second derivative of the ground state energy with respect to the sweeping parameter, $-\partial^2 E_0/\partial X^2$. The momentum ${\bf q}$ from the set of the allowed momenta of the cluster with the maximal static spin-structure factor 
$${\cal S}({\bf q}) \!=\! \frac{1}{N^2} \sum_{\alpha} \sum_{\mathbf{r}, \mathbf{r'}} \langle S_{\mathbf{r}}^{\alpha} S_{\mathbf{r'}}^{\alpha}\rangle e^{i\mathbf{q}(\mathbf{r} - \mathbf{r'})},$$ 
provides a candidate for the ordering vector of the state in the thermodynamic limit. If no obvious dominant candidate can be found, hinting at either an incommensurate or more complicated form of ordering, the corresponding segment of the sweep is labeled as NCO for the non-commensurate order.

\begin{figure*}[t]
\includegraphics[width=1\linewidth]{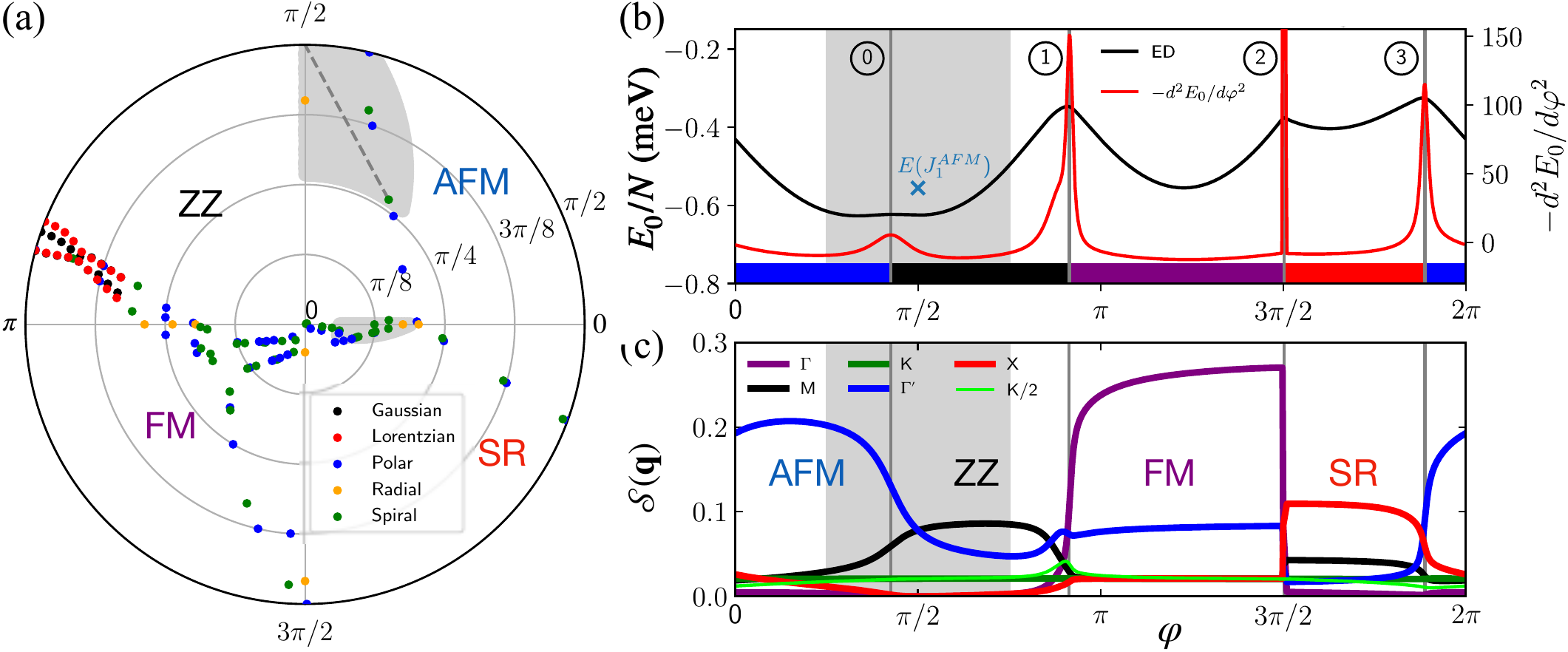} 
\caption{(a) Same as Fig.~\ref{fig_ED1}(c) obtained by the polar, radial, and spiral sweeps. The  Gaussian and Lorentzian fits of the close phase boundaries and the areas affected by the finite-size effects of the ED cluster, marked in gray,  are discussed in text. (b) and (c) Same as Figs.~\ref{fig_ED1}(d) and \ref{fig_ED1}(e), respectively, for $\theta\!=\!\pi/2$, see text for details.}
\label{fig_ED3}
\end{figure*}

\subsection{Cartesian ED phase diagrams}
\label{app_C1}

To obtain the phase diagrams for the Point~0, A, and B sets in Figs.~\ref{fig_phaseDJ1J3_0}(b), \ref{fig_phaseDJ1J3_A}(b), and \ref{fig_phaseDJ1J3_B_ED},  respectively, ED sweeps where carried out in the horizontal (fixed $J_3$, varying $J$) and vertical (fixed $J$, varying $J_3$) directions, with a discretization step of $0.01$~meV within the depicted range of the phase diagrams.  In Fig.~\ref{fig_ED1}(c), positions of the maxima of the $-\partial^2 E_0/\partial J_{(3)}^2$ are shown as dots with the color labeling the sweep direction.  The phase boundaries are then drawn by linearly interpolating these dots. Depending on the clarity of the course of the phase boundary, the distance between the sweeps was chosen to be $0.25$ or $0.5$~meV, with each sweep containing 500 points.

To demonstrate the process of obtaining the phase boundaries in more detail, we  discuss a representative example of such an ED sweep of the  phase diagram for the Point~0 parameter set in Fig.~\ref{fig_phaseDJ1J3_0}(b), shown in Figs.~\ref{fig_ED1}(d) and \ref{fig_ED1}(e).  It is a vertical sweep with $J_3$ as a varied parameter for the fixed $J \!=\! -1.25$~meV, which is shown in Fig.~\ref{fig_ED1}(c) by the vertical arrow. In Fig.~\ref{fig_ED1}(d), the ED ground state energy per site and its negative second derivative are shown. Fig.~\ref{fig_ED1}(e) shows the values of the static structure factor ${\cal S}({\bf q})$ at the color-coded high-symmetry points along the sweep, with the identified phases indicated. 

In the energy derivative, one can  see four peaks, with the two sharp inner ones, marked as \#1 and \#2, and two outer ones for the smaller and larger values of $J_3$, marked as \#0 and \#3, respectively, being considerably broader.  The sharp peaks can be taken as a signature of the singularities already in the first order derivatives of energy, implying first-order transitions, while the latter is suggestive  of the more continuous phase transitions in the thermodynamic limit~\cite{Trebst14}.

From the analysis of the structure factor in Fig.~\ref{fig_ED1}(e),  the small $J_3$ region of the sweep spans the FM,  while the large $J_3$ region of it belongs to the ZZ phase. The three intermediate regions have no dominant indicators of the ordering vector aside from the proximities to the FM and ZZ transitions in the two sectors marked as NCO1.  They are considered as belonging to the same phase because of their continuity that is obvious from Fig.~\ref{fig_ED1}(c). The NCO2 phase is distinct and is carving itself a separate area of the phase diagram. 

While the small size of the ED cluster should not be able to accommodate the IC1 and IC2 spiral phases discussed in Sec.~\ref{Sec_IC}, it is curious that it is still able to detect  sharp transitions between some phases that must be reminiscent of them. It is also notable that the ED results clearly indicate a prominent intermediate region between the FM and ZZ phases, which is also in a close qualitative agreement with the LT and DMRG results.   

\subsection{Polar ED phase diagram}
\label{app_C2}

The polar ED phase diagram of the ${\sf J^{\it XY}_1}$--${\sf J_{z\pm}}$--$J_3$ model discussed in Sec.~\ref{Sec_polarPD} is shown in Fig.~\ref{fig_phaseDpolar1}(b).  It is obtained using the grid of the radial, polar, and spiral sweeps, with the summary of them shown in Fig.~\ref{fig_ED3}(a). Each dot corresponds to a maximum in the Gaussian or Lorentzian fit of the $-\partial^2 E_0/\partial X^2$, with $X\!=\!\{\theta,\varphi\}$ and the color labeling the sweep direction, as before.

The polar sweeps covered the entire range, $0\!\leq\!\varphi\!<\!2\pi$, of the polar angle, encoding the relative values of the  ${\sf J_1}$ and $J_3$ terms, for the circles of  different radius, which is controlled by the angle $0\!\leq\!\theta\!\leq\!\pi/2$, encoding the ${\sf J_{z\pm}}$ term. The radii were fixed at $\theta/\pi\!=\!\{1/80,1/32,1/16,3/40,7/80,1/8,1/5,1/4,3/8,1/2\}$ with the $\varphi$ step of $\pi/500$, i.e., 1000 points in a circle. In order to closely explore the region near the ZZ-FM boundary, which is of most interest, additional polar sweeps have been performed for the smaller range of $\varphi$, $3\pi/4\!\leq\!\varphi\!<\!5\pi/4$, for a finer mesh of radii, $\theta/\pi\!=\!\{0.34, 0.36, 0.38,0.4,0.42,0.44,0.46,0.48,0.5\}$, and also with the higher discretization of $\pi/2000$, corresponding to 1000 points for the sweep range.

The  radial sweeps with $\theta$ were done along the radial lines of fixed $\varphi\!=\!\{0,\pi/2,\pi,3\pi/2\}$ with a step  of $\pi/1000$ for the entire range of $0\!\leq\!\theta\!\leq\!\pi/2$. In addition, two  spiral sweeps, which had both $\theta$ and $\varphi$ varying along the sweep, have been performed. Both had four full rotations in $\varphi$, $0\!\leq\!\varphi\!<\!8\pi$,  with the first varying $\theta$ from 0 to $\pi/2$, and the second from 0 to $\pi/5$, aimed at a closer investigation of the central region of the phase diagram. Both spiral sweeps had 1000 points. 

Generally, because of the finite size of the ED cluster, the indicators of the transitions between different phases provided by the peaks in the energy derivative often have a finite width. This feature requires additional analysis in the cases of the two transitions coming close and peaks overlapping, such as the case shown in Fig.~\ref{fig_ED3}(b) and discussed below. The simple criterion for the transition as the derivative's maximum fails in these cases, while the finer step of the sweep cannot help with increasing the resolution as the width of the peak is the ED cluster property. The  solution is to fit the more complicated structures in the energy derivatives by a pair of the  Gaussian or Lorentzian peaks, identifying their individual maxima with the phase boundaries. 

The fitting procedure by the two overlapping peaks using FindFit routine in Wolfram Mathematica~\cite{mathematica} 
reproduces the ED energy derivatives very well, but it leads to slightly different positions of the phase boundaries in the FM-ZZ region for the  Gaussian and Lorentzian fits, see Fig.~\ref{fig_ED3}(a).  For the phase diagram in Fig.~\ref{fig_phaseDpolar1}, the Lorentzian border has been chosen.

In Figs.~\ref{fig_ED3}(b)  and \ref{fig_ED3}(c), we present the results for the representative polar sweep that corresponds to the circumference of the circle in Fig.~\ref{fig_ED3}(a),  $\theta\!=\!\pi/2$, for which ${\sf J}_{z\pm}$ term vanishes and the model reduces to the simpler ${\sf J^{\it XY}_1}$--$J_3$ model, much studied recently~\cite{J1J3us}. Notations are the same as in Figs.~\ref{fig_ED1}(d) and \ref{fig_ED1}(e), respectively.

We first discuss the asymmetric peak \#1 in the energy derivative in Fig.~\ref{fig_ED3}(b), which corresponds to the proximity of the FM-ZZ boundary and is indicative of the two transitions, not one. Moreover, the structure factor in Fig.~\ref{fig_ED3}(c) also shows a combination of the contributions from the AFM $\Gamma'$ and  K$/2$ points  in this region, suggesting an intermediate NCO phase. Applying the analysis of the overlapping peaks discussed above produces a clear intermediate region between the FM and ZZ phases.  Its existence is also supported by a continuity argument with the ${\sf J}_{z\pm}\!>\!0$ sweeps, for which the two peaks for the transitions into the NCO phase from the neighboring FM and ZZ phases become distinct. This scenario is also in a close accord with the LT phase diagram in Fig.~\ref{fig_phaseDpolar1}(a). 

Two additional features of the sweep in Figs.~\ref{fig_ED3}(b)  and \ref{fig_ED3}(c) and of the ED phase diagram in Fig.~\ref{fig_ED3}(a) should be noted. Both concern the regions of the dominant $J_3$, i.e., proximities of the $\varphi\!=\!\pi/2$ and $\varphi\!=\!3\pi/2$ points with ${\sf J_1}\!=\!0$, for which the  24-site  cluster splits into three independent clusters containing only 8 sites each connected by the sparse  $J_3$ network. 

Because of this finite-size effects, various artificial degeneracies emerge. The energy of the pure AFM $J_1$ model, shown in Fig.~\ref{fig_ED3}(b) by the cross,  should be equal to that of the pure $J_{3}$ model ($\varphi\!=\!\pi/2$) in the thermodynamic limit, demonstrating strong  finite-size effects for this region of $\varphi$. The affected regions are marked with gray color and the AFM-ZZ sector of the phase diagram was clarified by DMRG. 

For the  ferromagnetic $J_{3}$ limit ($\varphi\!=\!3\pi/2$), the FM state is purely classical. The abrupt transition between the FM and stripe phases is also clearly reflected in the jump of the ordering vector associated with the two phases. 

Altogether, the analysis provided here underscores a close quantitative unity of the results between the ED, LT, and DMRG approaches.  

\begin{figure}[t]
\includegraphics[width=\linewidth]{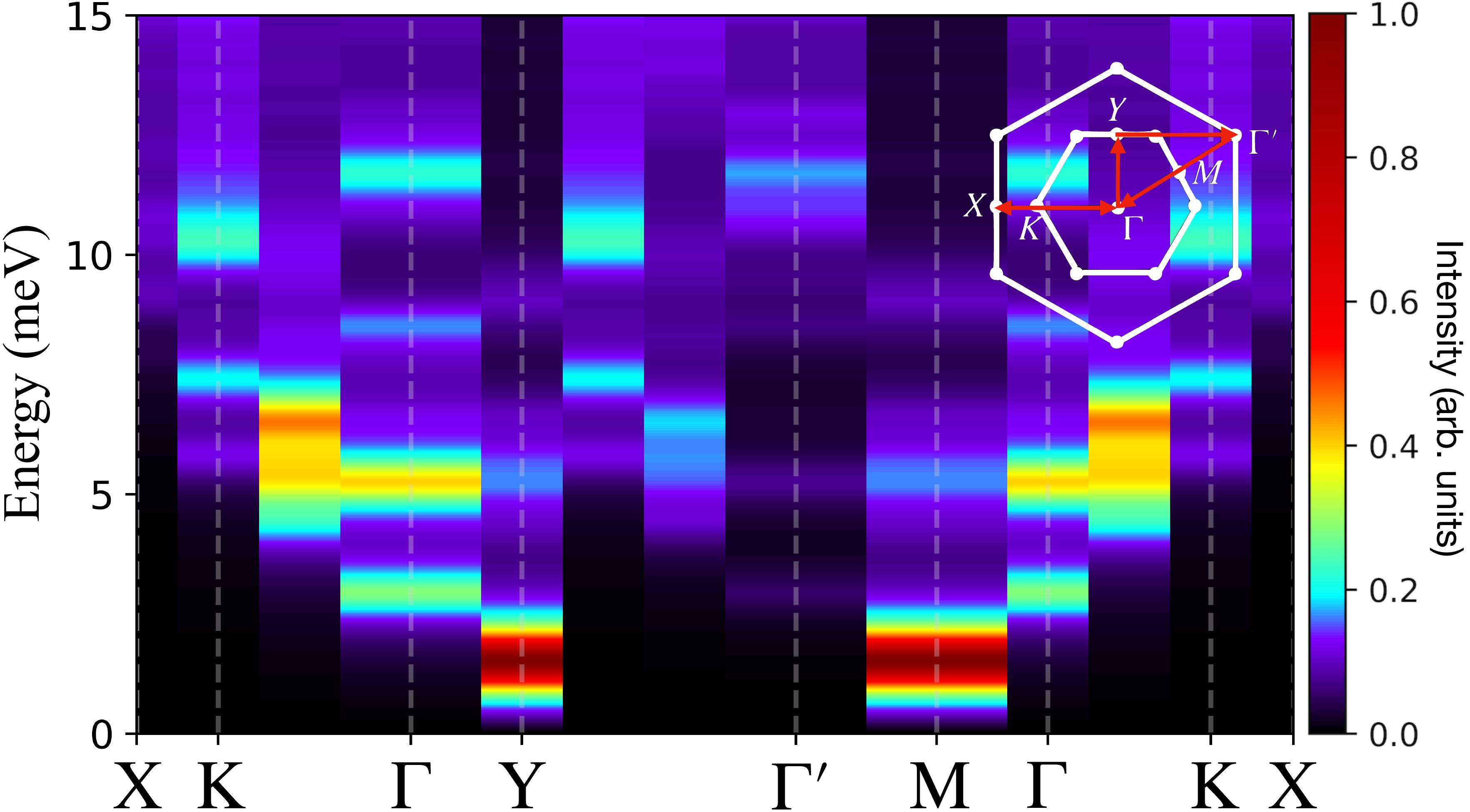}
\caption{The intensity map of ${\cal S}(\mathbf{q},\omega)$ from Eq.~(\ref{eq_App_C3_Sqw2}) for  the Point~$\bigstar$ parameter set in zero field along the ${\bf q}$-path shown in the inset.}
\label{fig_DynSF_ED}
\end{figure}

\subsection{Dynamical spin-spin correlation function}
\label{app_C3}

The dynamical spin-spin correlation function 
\begin{equation}
\label{eq_App_C3_Sqw1}
{\cal S}^{\mu \nu}(\mathbf{q},\omega) = \frac{1}{2\pi} \int_{-\infty}^{\infty} e^{i\omega t} \langle S^{\mu}_{-\mathbf{q}}(t)  S^{\nu}_{\mathbf{q}}(0)\rangle \, dt\, ,
\end{equation}
can be related to  various forms of   dynamical structure factors, which are the subjects of   ESR and  INS  spectroscopic experiments. 

The correlation functions  (\ref{eq_App_C3_Sqw1}) can be calculated numerically using ED in finite clusters, such as the one discussed in this work; see also Refs.~\cite{Winter18,winter17,CsCeSe}. ESR measures  the ${\bf q}\!=\!0$  sector of the dynamical structure factor, and for the results discussed in Sec.~\ref{Sec_DMRGchecks}, the components perpendicular to the applied field  are of interest. For a field in the $a$-direction, the in-plane ${\cal S}^{bb}(0,\omega)$ component was used  in Fig.~\ref{fig_ED_ESR} as a proxy for the true ESR structure factor. 

In Figure~\ref{fig_DynSF_ED} we provide an intensity plot of the  sum of the diagonal components of the spin-spin correlation function from (\ref{eq_App_C3_Sqw1}) 
\begin{equation}
\label{eq_App_C3_Sqw2}
{\cal S}(\mathbf{q},\omega) = \sum_{\mu}  {\cal S}^{\mu \mu}(\mathbf{q},\omega) ,
\end{equation}
for  the Point~$\bigstar$ parameter set of model (\ref{eq_Hij}) in zero field. It can serve as a rough proxy for the INS dynamical structure factor~\cite{Stot_tri_13}. Because the $C_3$ symmetry of the spin ground state is maintained in the ED calculations, the results in Fig.~\ref{fig_DynSF_ED} also correspond to an effective averaging over the three distinct zigzag domains~\cite{winter17} for the Point~$\bigstar$ set ground state. 

As in the ESR comparison in Sec.~\ref{Sec_DMRGchecks}, the ED calculations for Fig.~\ref{fig_DynSF_ED} used the Lanczos algorithm~\cite{lanczos1950iteration} and the continued fraction method~\cite{dagotto1994correlated}, with a Krylov dimension of $150$ and a Lorentzian broadening of $0.5$~meV. 

As described in Sec.~\ref{Sec_Outlook}, these preliminary results require a more detailed analysis to be  compared with the available INS data. However, certain improvements relative to  similar comparisons using different parameter sets~\cite{kaib19,winter17} can be pointed out, such as a better quantitative matching of the lowest-energy excitations at the M- and $\Gamma$-points and a wider spectral weight distribution in the magnetic spectrum across the BZ,  in accord with  experimental expectations.   

\section{LT polar phase diagram with ${\sf J_{\pm\pm}}\!>\!0$}
\label{app_D}

\begin{figure}[t]
\includegraphics[width=\linewidth]{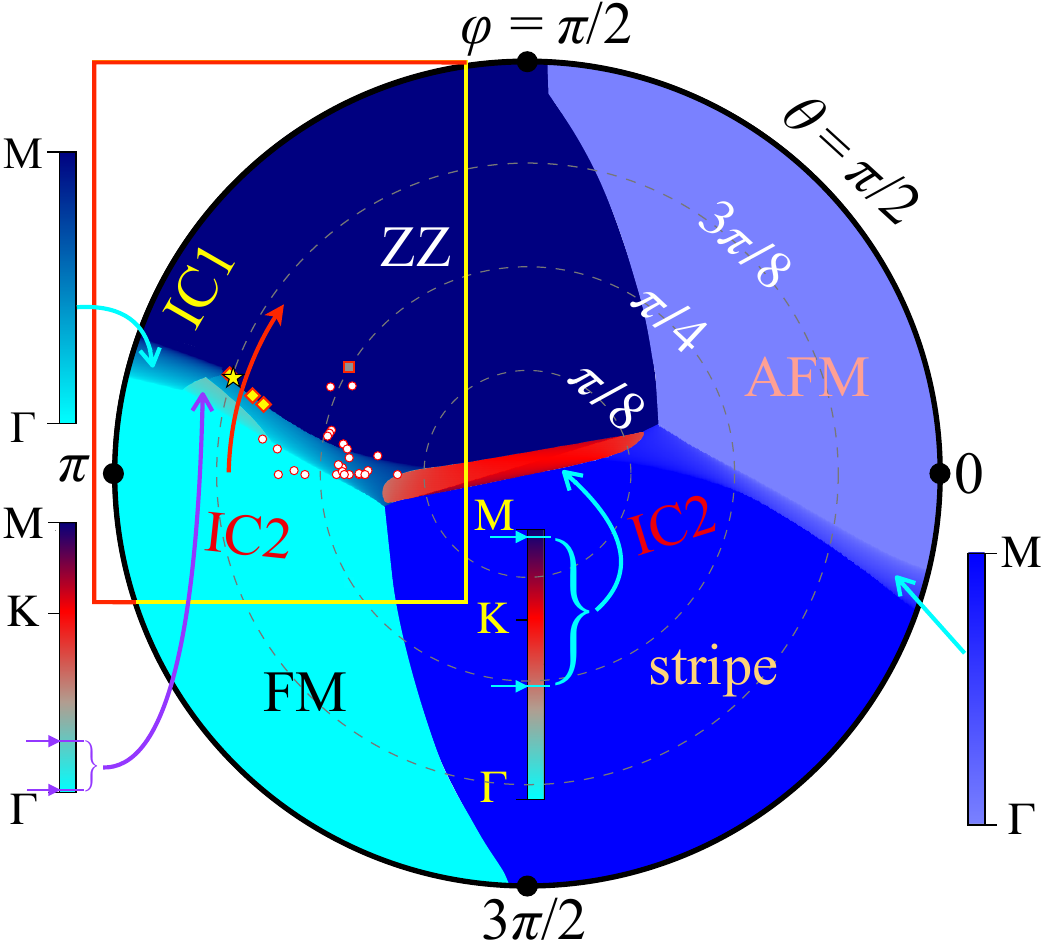}
\caption{Polar phase diagram of the model (\ref{HJpm}) for $\Delta\!=\!0$, ${\sf J_{\pm\pm}}\!=\!0.05$, with ${\sf J_1}$ and $J_3$ encoding the polar and ${\sf J_{z\pm}}$ the radial coordinates in units of ${\sf J^{\rm 2}_1}+{\sf J^{\rm 2}_{z\pm}}+J_3^2\!=\!1$. The box shows the cutout shown in Fig.~\ref{fig_phaseDpolar1}(c). For details, see Sec.~\ref{Sec_PhDs}.}
\label{fig_phaseDpolarJpp}
\end{figure}

Our Figure~\ref{fig_phaseDpolarJpp} shows the effect of ${\sf J_{\pm\pm}}$ in the full polar phase diagram obtained by the LT method, with the cutout of  the FM-ZZ region from it shown in the main text in Fig.~\ref{fig_phaseDpolar1}(c). The phase diagram is of the model (\ref{HJpm}) for the $XY$ limit of the $XXZ$ term, $\Delta\!=\!0$, and for a small ${\sf J_{\pm\pm}}\!=\!0.05$, with ${\sf J_1}$ and $J_3$ encoding the polar and ${\sf J_{z\pm}}$ the radial coordinates in units of ${\sf J^{\rm 2}_1}+{\sf J^{\rm 2}_{z\pm}}+J_3^2\!=\!1$, as before. The choice of ${\sf J_{\pm\pm}}\!=\!0.05$ is motivated by its relative value for the physical region of $\alpha$-RuCl$_3$ parameters discussed in Sec.~\ref{Sec_Jpp_Jzp_space}.  

Compared to the LT phase diagram in Fig.~\ref{fig_phaseDpolar1}(a) for ${\sf J_{\pm\pm}}\!=\!0$, which shows a complete symmetry between FM and AFM and ZZ and stripe phases, respectively, such a symmetry is lost in Fig.~\ref{fig_phaseDpolarJpp}, suggesting a similarity between the effects of the finite ${\sf J_{\pm\pm}}$ to the quantum effects observed in ED phase diagram in Fig.~\ref{fig_phaseDpolar1}(b). 

Another important effect of the finite and positive ${\sf J_{\pm\pm}}$ is the protrusion of the IC2 region in the IC1 phase within the FM-ZZ quadrant, bringing a consistency of the polar phase diagram consideration with that of  the Cartesian phase diagrams in the generalized KH parametrization (\ref{H_JKGGp}) in Figs.~\ref{fig_phaseDJ1J3_0} and \ref{fig_phaseDJ1J3_A}, which feature both IC phases prominently. It is also relevant to the consideration of these IC phases, which are proximate to the  physical parameter space of $\alpha$-RuCl$_3$, given in Sec.~\ref{Sec_IC}.

In addition to the symbols representing  projections of all individual $\alpha$-RuCl$_3$ parameter sets listed in Table~\ref{table3} onto the $\Delta\!=\!0$ plane of the phase diagram, Fig.~\ref{fig_phaseDpolarJpp} shows the group of three diamonds, which are the sets  listed in Table~\ref{table2rethinking} that were proposed in Ref.~\cite{rethinking}. They were put forward using phenomenological constraints on the $\alpha$-RuCl$_3$ parameters, but were deliberately not included in Table~\ref{table3} for the sake of not skewing  independent distribution of parameters from the prior literature, exposed and discussed in Sec.~\ref{Sec_compilation}. 

\begin{table*}
  \begin{tabular}{ | l | c || c | c | c | c | c | c | c | c |}
    \hline
\text{Reference} 
& \text{Method} 
& $K$ & $\Gamma$ & $\Gamma'$ &  $J$ & $J_3$  & ${\sf J_{\pm \pm}}$ & ${\sf J_{z\pm}}$ & ${\sf J}_1(1-\Delta)$ \\ \hline\hline
\multirow{3}{3.1cm}{Maksimov et al. \cite{rethinking}}  
&  point 1 & -4.8          & 4.08 & 2.5   &  -2.56 & 2.42         &  0.3 & -3.0  & { 9.08}              \\ \cline{2-10}
& point 2  & -10.8        & 5.2   & 2.9   & { -4.0}    & { 3.26}  &  1.0  & -6.2  & { 11.0}                \\ \cline{2-10}
&   point 3 & -14.8     & 6.12 & 3.28 & {  -4.48} & {  3.66}  &  1.5  &  -8.3 & {  12.7}                  \\ \hline
\end{tabular}
\caption{Same as Table~\ref{table2} for the sets in Ref.~\cite{rethinking}, also showing $J$ and $J_3$.}
\label{table2rethinking}
\vskip -0.2cm
\end{table*}

\section{Classical helix ansatz}
\label{app_E}

One can study incommensurate spiral phases using the general classical spin-helix ansatz, which, unlike the LT method, respects the spins' length
\begin{align}
\mathbf{S}_{i,\gamma}=\mathbf{u} \cos \theta_{i,\gamma}+\mathbf{v} \sin\theta_{i,\gamma},
\label{app_helix}
\end{align}
where vectors $\mathbf{u}$ and $\mathbf{v}$ of length $|{\bf u}|\!=\!|{\bf v}|\!=\!S$ define the plane of spins' rotation, $\gamma$ is the sublattice index, and phases are $\theta_{i,\gamma}\!=\!\mathbf{Q}_\gamma\mathbf{r}_i+\varphi_\gamma$, with the propagation vector of the spiral $\mathbf{Q}_\gamma$ and   its phase shift $\varphi_\gamma$.  

For instance, the $y$-$z$ and $x$-$z$ helical states, such as the ones associated with, respectively, the IC1 and IC2 phases of the main text, are given by
\begin{align}
&\mathbf{S}_{i,\gamma}=\hat{\bf y}S \cos\theta_{i,\gamma}+\hat{\bf z}S \sin\theta_{i,\gamma},
\label{eq_yz}
\\
&\mathbf{S}_{i,\gamma}=\hat{\bf x}S \cos\theta_{i,\gamma}+\hat{\bf z}S \sin\theta_{i,\gamma},
\label{eq_xz}
\end{align}
see Eqs.~(\ref{ansatz}) and (\ref{ansatz1}) for comparison. 

In addition to the spiral states, the form in Eq.~(\ref{app_helix}) can also  describe {\it all} commensurate-${\bf Q}$ states that are discussed in this work, such as the ferromagnetic, antiferromagnetic N\'{e}el, stripe, and zigzag collinear orders.

 Specifically, the choice of ${\bf Q}\!=\!0$ with $\varphi_A\!=\!\varphi_B$ corresponds to a FM state, while an AFM state is associated with the same ${\bf Q}$ and phases $\varphi_A\!=\!\varphi_B+\pi$. The zigzag and stripe states correspond to  ${\bf Q}\!=\!{\bf G}/2$=M-point, where ${\bf G}$ is the nearest-neighbor reciprocal-lattice vector. The phases are $\varphi_A\!=\!\varphi_B$ for the stripe and $\varphi_A\!=\!\varphi_B+\pi$ for zigzag states. The values of  $\varphi_{A(B)}$ phases are determined from the energy minimization and define the direction of the ordered moments in the $\mathbf{u}$--$\mathbf{v}$ plane.

Using Eq.~(\ref{eq_yz}) and crystallographic parametrization of the model (\ref{HJpm}), after some  algebra, one can obtain the energies of the FM, AFM-z, and zigzag states, per  the honeycomb-lattice unit cell and in units of $S^2$, as
\begin{align}
E_\text{FM}&=3{\sf J_1}+3 J_3\nonumber\\
E_\text{AFM-z}&=-3\Delta {\sf J_1}-3 J_3
\label{eq_commensurate}\\
E_\text{zz}&=\frac{1}{2} \Big[{\sf J_1}+ \Delta {\sf J_1} - 6 J_3 + 4 {\sf J_{\pm \pm}} +\nonumber\\
&({\sf J_1} - \Delta {\sf J_1} + 4 J_{\pm\pm}) \cos 2\varphi_A + 
   4{\sf J_{z \pm}}  \sin 2\varphi_A\Big],\nonumber
\end{align} 
where $\varphi_A\!=\!\alpha$ is  the out-of-plane tilt angle of the spins in Eq.~\eqref{eq_alpha}.

Motivated by the results analyzed in Sec.~\ref{Sec_helices}, we have studied the incommensurate single-${\bf Q}$ helical states with the co-rotating, $\mathbf{Q}_B\!=\!\mathbf{Q}_A$, and counter-rotating, $\mathbf{Q}_B\!=\!-\mathbf{Q}_A$ ($\varphi_B\!\Rightarrow\!-\varphi_B$), helices defined by Eq.~\eqref{app_helix}. In addition to the $y$-$z$ and $x$-$z$ helical states defined above, Eqs.~(\ref{eq_yz}) and (\ref{eq_xz}), we have also considered the energetics of the $x$-$y$ spirals that are coplanar with the ${\bf Q}$ vector. 

Of the considered states, the energies of the $y$-$z$ and $x$-$z$ counter-rotating helical states, which are close analogues of the IC1 and IC2 states discussed in Sec.~\ref{Sec_helices}, are of most interest. They are given by 
\begin{align}
E_\text{yz}=& \sum_{{\bm \delta}_\alpha}\Big\{\Big( \frac{{\sf J_1}}{2}(1-\Delta)-{\sf J_{\pm \pm}} \tilde{c}_\alpha \Big)
\cos \left( \mathbf{Q} {\bm \delta}_\alpha+\Delta\varphi_{AB}\right) \nonumber\\
&\quad\quad-{\sf J_{z \pm}} \tilde{c}_\alpha \sin \left( \mathbf{Q} {\bm \delta}_\alpha+\Delta\varphi_{AB}\right)\Big\},
\label{eq_en_counter1}\\
E_\text{xz}=& \sum_{{\bm \delta}_\alpha}\Big\{\Big( \frac{{\sf J_1}}{2}(1-\Delta)+{\sf J_{\pm \pm}} \tilde{c}_\alpha \Big)
\cos \left( \mathbf{Q} {\bm \delta}_\alpha+\Delta\varphi_{AB}\right) \nonumber\\
&\quad\quad-{\sf J_{z \pm}} \tilde{s}_\alpha \sin \left( \mathbf{Q} {\bm \delta}_\alpha+\Delta\varphi_{AB}\right)\Big\},
\label{eq_en_counter2}
\end{align}
in the crystallographic parametrization of the model (\ref{HJpm}), per  the honeycomb-lattice unit cell, and in units of $S^2$, with $\Delta\varphi_{AB}\!=\!\varphi_{A}-\varphi_{B}$ and ${\bm \delta}_\alpha$ being the nearest-neighbor vectors of the honeycomb lattice. An important feature of the results in Eqs.~(\ref{eq_en_counter1}) and (\ref{eq_en_counter2}) is their independence of the isotropic terms of the model  (\ref{eq_Hij}). That is, if rewritten in the generalized KH parametrization, the energies of the counter-rotating spirals depend only on the $\{K,\Gamma,\Gamma'\}$ parameter sets, the feature noticed in the discussion of such states in  Sec.~\ref{Sec_helices}.

\begin{figure}[b]
\includegraphics[width=\linewidth]{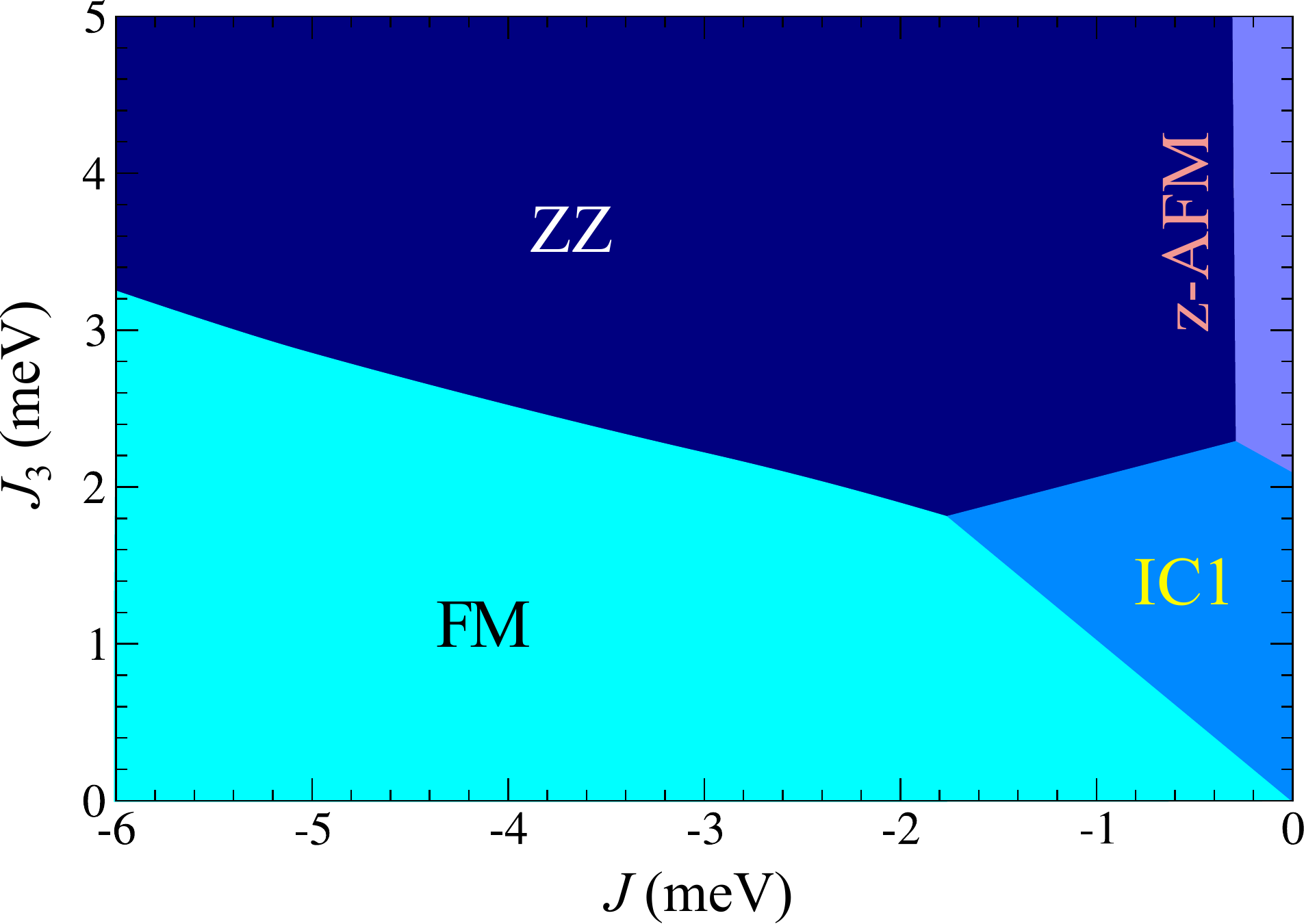}
\caption{Phase diagram of the model \eqref{eq_Hij} using the classical helical ansatz~\eqref{app_helix} for the Point~0 set, c.f.,  Fig.~\ref{fig_phaseDJ1J3_0}.}
\label{fig_point0}
\end{figure}

Minimizing Eqs.~(\ref{eq_en_counter1}) and (\ref{eq_en_counter2})  with respect to $\mathbf{Q}$ and  $\Delta\varphi_{AB}$, doing the same for the similar expressions for the energies of the helical states  with different spin planes and co-rotating configurations, and comparing them to the energies of the commensurate states in Eq.~(\ref{eq_commensurate}) allows one to obtain the purely classical phase diagrams of the model (\ref{eq_Hij}). 

Our  Fig.~\ref{fig_point0} shows a representative phase diagram for the Point~0 parameter set in the  Cartesian $J$--$J_3$  axes of the KH parametrization, to be compared with the LT and ED phase diagrams in Fig.~\ref{fig_phaseDJ1J3_0} for the same parameters. One can see that while identical to LT  regarding the commensurate FM, ZZ, and AFM phases, of the classical helical states  considered here it is only   IC1 ($y$-$z$) counter-rotating state that is stabilized. It is also stable in a smaller region of  $J$ and $J_3$. These results underscore the importance of  quantum fluctuations in stabilizing IC phases in a significantly  extended parameter space.

However, we note that the classical solutions for the $y$-$z$ and $x$-$z$ counter-rotating helical states exhibit  great similarities  to the findings presented in Sec.~\ref{Sec_helices}. First of all, in both cases, the propagation vector of the helix is found to be orthogonal to the plane of spins, ${\bf Q}\!\parallel\!\hat{\bf x}\!\parallel\! \Gamma$M line for the $y$-$z$ (IC1) and ${\bf Q}\!\parallel\!\hat{\bf y}\!\parallel\! \Gamma$K line for the $x$-$z$ (IC2) helices. Then, the phase differences are found in a complete accord with the LT and DMRG results in Sec.~\ref{Sec_helices}, with $\Delta\varphi_{AB}$ being zero for the IC2 state and close to 41$\degree$ for the representative Point~0 parameter set in the IC1 state. Last, but not the least are  close values of the IC1 and IC2 pitches of the spirals, discussed in the main text.


\bibliography{Saga_aRu3bib}

\end{document}